\documentclass[traditabstract]{aa}
\usepackage{lmodern}
\pdfoutput=1
\usepackage{graphicx}
\usepackage{subfig}
\usepackage{natbib}
\usepackage{subfig}
\usepackage{tabularx}
\usepackage{lscape}
\bibpunct{(}{)}{;}{a}{}{,}

\usepackage[colorlinks=true, citecolor=blue, linkcolor=blue, breaklinks=true]{hyperref}

\usepackage[fleqn]{amsmath} 
\usepackage{amsfonts} 
\usepackage{amstext}
\usepackage{amssymb} 
\newcommand{\hi } {{\rm H}\,{\small\rm I} \,}

\newcommand{\hiiA} {{\rm H}\,{\small\rm II}}

\begin{document}

\title{Dynamics of starbursting dwarf galaxies. III.}
\subtitle{A \hi study of 18 nearby objects.}
\author{Federico Lelli\inst{1,}\inst{2}
\and Marc Verheijen\inst{1}
\and Filippo Fraternali\inst{3,}\inst{1}}

\institute{Kapteyn Astronomical Institute, University of Groningen, Postbus 800, 9700 AV, Groningen, The Netherlands
\and Department of Astronomy, Case Western Reserve University, 10090 Euclid Avenue, Cleveland, OH 44106, USA \\
\email{federico.lelli@case.edu}
\and Department of Physics and Astronomy, University of Bologna, viale Berti Pichat 6/2, 40127, Bologna, Italy}

\date{}

\abstract{
We investigate the dynamics of starbursting dwarf galaxies, using both new and
archival \hi observations. We consider 18 nearby galaxies that have been resolved
into single stars by HST observations, providing their star formation history and
total stellar mass. We find that 9 objects have a regularly rotating \hi disk,
7 have a kinematically disturbed \hi disk, and 2 show unsettled \hi distributions.
Two galaxies (NGC~5253 and UGC~6456) show a velocity gradient along the minor
axis of the \hi disk, which we interpret as strong radial motions. For galaxies
with a regularly rotating disk we derive rotation curves, while for galaxies
with a kinematically disturbed disk we estimate the rotation velocities in
their outer parts. We derive baryonic fractions within about 3 optical scale
lengths and find that, on average, baryons constitute at least 30$\%$ of the
total mass. Despite the star formation having injected $\sim$10$^{56}$ ergs in
the ISM in the past $\sim$500~Myr, these starbursting dwarfs have both baryonic
and gas fractions similar to those of typical dwarf irregulars, suggesting that
they did not eject a large amount of gas out of their potential wells.}

\titlerunning{Dynamics of starbursting dwarf galaxies. III.}
\authorrunning{Lelli, Verheijen, and Fraternali}

\maketitle

\section{Introduction}\label{sec:intro}

Starburst activity is thought to strongly affect the evolution of dwarf galaxies.
Both observations and theoretical models suggest that massive star formation can
alter the morphology and kinematics of the gas in dwarf galaxies \citep[e.g.,][]
{MacLow1999, Cannon2011b}, as well as their chemical properties \citep[e.g.,][]
{Recchi2004, Romano2006}. Moreover, models of galaxy formation in a $\Lambda$
cold dark matter ($\Lambda$CDM) cosmology require strong feedback from star
formation to explain several observational facts, such as i) the existence of
bulgeless disk galaxies by removing low angular-momentum gas from the galaxy
center \citep[e.g.,][]{Governato2010, Brook2011}; ii) the ``cored'' DM profiles
observed in dwarfs by flattening the presumed central ``cusps'' \citep[e.g.,]
[]{Oh2011b, Governato2012}; iii) the slope of the baryonic Tully-Fisher relation
by reducing the baryonic fraction in galaxies \citep[e.g.,][]{McGaugh2012,
Stringer2012}, and iv) the number density of low-luminosity galaxies by
suppressing star formation in low-mass DM halos \citep[e.g.,][]{Okamoto2010,
Sawala2013}. Detailed dynamical studies of nearby starbursting dwarfs are
necessary to determine the actual efficiency of these processes.

Starbursting dwarfs can be identified by i) their blue colors and high surface
brightness, such as the blue compact dwarfs (BCDs) (e.g., \citealt{GilDePaz2003});
ii) their strong emission-lines, such as the \hiiA-galaxies (e.g., \citealt{Terlevich1991,
Taylor1995}); and iii) their peculiar morphologies, such as the ``amorphous dwarfs''
(e.g., \citealt{Gallagher1987, Marlowe1999}). Hereafter, we refer to any
starbursting dwarf as a BCD. 

To date, detailed studies of the \hi kinematics of BCDs have been focused either
on individual galaxies \citep[e.g.,][]{Viallefond1983, Hunter1996, Wilcots1998,
Matthews2008} or on small galaxy samples with four to five objects \citep[e.g.,]
[]{vanZee1998b, vanZee2001, Thuan2004, Ramya2011}. These studies show that some
BCDs have regularly rotating \hi disks \citep[e.g.,][]{vanZee1998b, vanZee2001},
whereas others have complex \hi kinematics \citep[e.g.,][]{Cannon2004, Kobulnicky2008}.
The relative fraction of BCDs with ordered or disturbed \hi kinematics remains
unclear, as does the possible relation between the gas kinematics and the
starburst. The DM content of starbursting dwarfs is also poorly constrained.
\citet{Elson2010, Elson2013} argue that the BCDs NGC~1705 and NGC~2915 are
dominated by DM at all radii \citep[see also][]{Meurer1996, Meurer1998}. In
contrast, \citet{Walter1997} and \citet{Johnson2012} studied the starbursting
dwarfs II~Zw~33 and NGC~1569, respectively, and conclude that there is no
need for DM to explain their inner kinematics.

\begin{table*}[t]
\caption{Galaxy Sample}
\centering
\setlength{\tabcolsep}{3pt}
\resizebox{18.5cm}{!}{
\begin{tabular}{l l c c c c c c c c c c}
\hline
\hline
Name     & Alternative  & Dist & $M_{*}$              & $M_{\rm R}$ & $M_{*}/L_{\rm{R}}$ & $b$ & SFR$_{\rm p}$ & $t_{\rm p}$ & $\Sigma_{\rm SFR}(t_{\rm p})$ & 12+log(O/H) & Ref. \\
         & Name         & (Mpc)&(10$^{7}$ M$_{\odot}$)&              &(M$_{\odot}$/L$_{\odot}$)&    &(10$^{-3}$M$_{\odot}$ yr$^{-1}$)&(Myr)&(10$^{-3}$M$_{\odot}$ yr$^{-1}$ kpc$^{-2}$) &             &     \\
\hline
NGC 625  & ESO 297-G005 & 3.9$\pm$0.4 & 26$\pm$10   & -17.25$\pm$0.24 & $>$0.6      & 3.0$\pm$0.1 & 86$\pm$20  & 820$\pm$180 & 2.5$\pm$0.6 & 8.08$\pm$0.12& a, g, l  \\
NGC 1569 & UGC 3056     & 3.4$\pm$0.2 & 70$\pm$7    & -17.14$\pm$0.25 & 1.7$\pm$0.2 & 21$\pm$1    & 240$\pm$10 & 40$\pm$10   & 8.5$\pm$0.3 & 8.19$\pm$0.02& a, h, m  \\
NGC 1705 & ESO 158-G013 & 5.1$\pm$0.6 & $>$20       & -16.35$\pm$0.26 & $>$1        & $\sim$6     & 314$\pm$78 & $\sim$3     & 44$\pm$11   & 8.21$\pm$0.05& b, i, l \\
NGC 2366 & UGC 3851     & 3.2$\pm$0.4 & 26$\pm$3    & -16.64$\pm$0.27 & 1.0$\pm$0.1 & 5.6$\pm$0.4 & 160$\pm$10 & 450$\pm$50  & 2.6$\pm$0.2 & 7.91$\pm$0.05& a, h, l  \\
NGC 4068 & UGC 7047     & 4.3$\pm$0.1 & 22$\pm$3    & -15.67$\pm$0.05 & 2.0$\pm$0.3 & 4.7$\pm$0.3 & 42$\pm$3   & 360$\pm$40  & 4.5$\pm$0.3 & ...          & a, h \\
NGC 4163 & UGC 7199     & 3.0$\pm$0.1 & 10$\pm$3    & -14.81$\pm$0.10 & 2.0$\pm$0.6 & 2.9$\pm$0.6 & 12$\pm$3   & 450$\pm$50  & 3.8$\pm$0.9 & 7.56$\pm$0.14& a, h, l \\
NGC 4214 & UGC 7278     & 2.7$\pm$0.2 & $>$28       & -17.77$\pm$0.24 & $>$0.4      & 3.1$\pm$0.9 & 130$\pm$40 & 450$\pm$50  & 8.5$\pm$2.6 & 8.22$\pm$0.05& a, h, l  \\
NGC 4449 & UGC 7592     & 4.2$\pm$0.5 & 210$\pm$35  & -18.88$\pm$0.26 & 1.0$\pm$0.2 & 6.0$\pm$0.5 & 970$\pm$70 &   5$\pm$3   & 28$\pm$2    & 8.26$\pm$0.09& a, h, l  \\
NGC 5253 & Haro 10      & 3.5$\pm$0.4 & 154$\pm$21  & -17.61$\pm$0.27 & 2.4$\pm$0.3 & 9.0$\pm$0.9 & 400$\pm$40 & 450$\pm$50  & 29$\pm$3    & 8.12$\pm$0.05& a, g, k \\
NGC 6789 & UGC 11425    & 3.6$\pm$0.2 & 7$\pm$2     & -15.09$\pm$0.14 & 1.1$\pm$0.3 & 3.8$\pm$1.3 & 15$\pm$5   & 565$\pm$65  & 9.7$\pm$3.2 & ...          & a, i \\
UGC 4483 & ...          & 3.2$\pm$0.2 & 1.0$\pm$0.2 & -12.97$\pm$0.19 & 1.1$\pm$0.3 &  14$\pm$3   & 11$\pm$2   & 565$\pm$65  & 8.8$\pm$3.5 & 7.56$\pm$0.03& a, i, l \\
UGC 6456 & VII Zw 403   & 4.3$\pm$0.1 & 5$\pm$2     & -14.41$\pm$0.05 & 1.5$\pm$0.6 & 7.6$\pm$1.1 & 23$\pm$3   &  16$\pm$8   & 5.1$\pm$0.7 & 7.69$\pm$0.01& a, j, n \\
UGC 6541 & Mrk 178      & 4.2$\pm$0.2 & $>$0.8      & -14.61$\pm$0.10 & $>$0.2      & $\sim$3     & $\sim$3    &  ...        & 1.2$\pm$0.6 & 7.82$\pm$0.06& c, j, l \\
UGC 9128 & DDO 187      & 2.2$\pm$0.1 & 1.3$\pm$0.2 & -12.82$\pm$0.12 & 1.6$\pm$0.2 & 6.3$\pm$1.4 & 5$\pm$1    & 150$\pm$50  & 4.4$\pm$0.9 & 7.75$\pm$0.05& a, h, l \\
UGCA 290 & Arp 211      & 6.7$\pm$0.4 & $>$1.0      & -14.09$\pm$0.18 & $>$0.4      & $\sim$3     & 42$\pm$15  &  $\sim$15   & 16$\pm$6    & ...          & d, i \\
I Zw 18  & Mrk 116      &18.2$\pm$1.4 & $>$1.7      & -14.99$\pm$0.26 & $>$0.3      & $\sim$30    & $\sim$100  &  $\sim$10   & $\sim$127   & 7.20$\pm$0.01& e, j, k \\
I Zw 36  & Mrk 209      & 5.9$\pm$0.5 & $>$0.8      & -14.88$\pm$0.23 & $>$0.1      & $\sim$7     & $\sim$25   &  ...        & $\sim$9.8   & 7.77$\pm$0.01& f, i, k \\
SBS 1415+437 & ...      &13.6$\pm$1.4 & 17$\pm$3    & -15.90$\pm$0.25 & 1.3$\pm$0.2 & 12$\pm$2    & 150$\pm$10 & 450$\pm$50  & 8.3$\pm$0.5 & 7.62$\pm$0.03& a, i, o \\
\hline
\hline
\end{tabular}
}
\tablefoot{Distances are derived from the TRGB. Stellar masses are calculated integrating the SFHs and assuming
i) a Salpeter IMF from 0.1 to 100~$M_{\odot}$, and ii) a gas-recycling efficiency of 30$\%$. The birthrate
parameter $b$ is defined as $b = \rm{SFR_{p}}/\overline{\rm{SFR}}_{0-6}$, where SFR$_{\rm{p}}$ is the peak
SFR over the past 1 Gyr and $\overline{\rm{SFR}}_{0-6}$ is the mean SFR over the past 6 Gyr \citep[see][]
{McQuinn2010a}. $t_{\rm p}$ is the look-back time at $\rm{SFR_{p}}$. $\Sigma_{\rm SFR}(t_{\rm p})$ is the
SFR surface density at $t_{\rm p}$, calculated as SFR$_{\rm{p}}$/$(\pi R_{\rm opt}^{2})$ where $R_{\rm opt}$
is the optical radius (see Table \ref{tab:extent}). The last column provides references for the HST studies
of the resolved stellar populations, the integrated photometry, and the ionized gas metallicity, respectively.}
\tablebib{(a)~\citet{McQuinn2010a}; (b)~\citet{Annibali2003}; (c)~\citet{SchulteLadbeck2000}; (d)~\citet{Crone2002};
(e)~\citet{Annibali2013}; (f)~\citet{SchulteLadbeck2001}; (g)~\citet{Lauberts1989}; (h)~\citet{Swaters2002b};
(i)~\citet{GilDePaz2003}; (j)~\citet{Papaderos2002}; (k)~\citet{Izotov1999}; (l)~\citet{Berg2012};
(m)~\citet{Kobulnicky1997}; (n)~\citet{Thuan2005}; (o)~\citet{Guseva2003}.}
\label{tab:sample}
\end{table*}
To clarify these issues, we considered a sample of 18 starbursting dwarfs, for which
we collected both new and archival \hi observations. We selected objects that
have been resolved into single stars by the \textit{Hubble Space Telescope} (HST),
providing a direct estimate of the recent star-formation history (SFH) and of the
total stellar mass \citep[e.g.,][]{Annibali2003}. The latter information allows us
to break the ``disk-halo degeneracy'' \citep{vanAlbada1986} and to estimate baryonic
fractions. In \citet{Lelli2012a, Lelli2012b}, we presented our results for two show-case
galaxies: I~Zw~18 and UGC~4483. For both objects, we showed that i) the \hi gas
forms a compact, rotating disk; ii) the rotation curve rises steeply in the inner
parts and flattens in the outer regions; and iii) old stars and atomic gas are
dynamically important, since they constitute at least $\sim30\%$ of the total
dynamical mass within the last measured point of the rotation curve. Here we
present a dynamical study of the remaining 16 objects.

\section{The sample}\label{sec:sample}

Table \ref{tab:sample} summarizes the main properties of our sample of starbursting
dwarfs. For these 18 galaxies, the studies of the resolved stellar populations
provide i) the galaxy distance from the tip of the red giant branch (TRGB), ii)
the spatial distribution of the different stellar populations, iii) the recent
SFH (ages $\lesssim$1~Gyr) by modeling the color-magnitude diagrams (CMDs),
iv) the energy produced during the burst by supernovae and stellar winds, and
v) the stellar mass in young and old stars. For 13 objects, we adopt the SFHs
derived by \citet{McQuinn2010a} using archival HST images. The remaining 5 objects
(I~Zw~18, I~Zw~36, NGC~1705, UGC~6541, and UGCA~290) have not been studied
by \citet{McQuinn2010a} because the HST observations have a relatively shallow
photometric depth ($\lesssim$1~mag below the TRGB). We use the SFHs derived by
other authors \citep{Annibali2003, Annibali2013, Crone2002, SchulteLadbeck2000,
SchulteLadbeck2001}, although they are more uncertain due to the limited
photometric depth. Note that all these 18 objects are well-defined starburst
galaxies, as their star-formation rates (SFRs) show an increase in the recent
SFH by a factor $\gtrsim$~3 with respect to the past, average SFR. The sample
covers a broad range in luminosities ($-19\lesssim M_{\rm{R}}\lesssim -13$),
stellar masses ($10^{7}\lesssim M_{*}/M_{\odot}\lesssim10^{9}$), and metallicities
($0.3 \lesssim Z/Z_{\odot}\lesssim 0.03$).

For all these galaxies, we collected both new and archival \hi data. We obtained new
\hi observations for SBS~1415+437 and UGCA~290 using the \textit{Jansky Very Large Array}
(VLA, during its upgrade period), and of NGC~6789 using the \textit{Westerbork Synthesis
Radio Telescope} (WSRT). We analyzed raw data from the VLA archive for I~Zw~18
\citep{Lelli2012a}, UGC~4483 \citep{Lelli2012b}, UGC~6456 and NGC~625 (this work).
The \hi datacubes of NGC~1705 \citep{Elson2013} and NGC~5253 \citep{LopezSanchez2012}
were kindly provided by Ed Elson and Angel R. Lopez-Sanchez, respectively. For the
remaining 9 galaxies, we used \hi cubes from 3 public surveys: WHISP \citep{Swaters2002a},
THINGS \citep{Walter2008}, and LITTLE-THINGS \citep{Hunter2012}. For 4 galaxies
(NGC~2366, NGC~4163, NGC~4214, and UGC~9128), \hi cubes are available from both
WHISP and THINGS/LITTLE-THINGS; we used the VLA data as they have higher spatial
resolution than the WSRT observations.

\begin{table*}[t]
\caption{New 21 cm-line observations}
\centering
\begin{tabular}{l c c c c c c c c}
\hline
\hline
Galaxy       & Array & Project     & Observing Dates          & Time on Source   & Calibrators \\
\hline
UGCA~290     & VLA/B & 10C-200 & 12, 26, 28 Mar. 2011        &  7.6 h & 3C286, 1227+365 \\
             & VLA/C & 12A-246 & 25, 27, 28 Apr. 2012        &  4.6 h & 3C286, 1227+365 \\
             & VLA/D & 11B-075 & 2, 13 Nov. 2011             & 0.28 h & 2C295, 1227+365 \\
SBS~1415+345 & VLA/B & 10C-200 & 4, 27 Mar., and 8 Apr. 2011 & 7.6 h  & 3C286, 1400+621 \\
             & VLA/C & 12A-246 & 26, 28, 30 Apr. 2012        & 4.6 h  & 3C286, 1400+621 \\
             & VLA/D & 11B-075 & 6, 9 Oct. 2011              & 0.28 h & 3C295, 1400+621 \\
NGC 6789     & WSRT  & R11A007 & 20 May. 2011                & 12   h & 3C286, 3C48     \\
\hline
\hline
\end{tabular}
\label{tab:obs}
\end{table*}
\section{Data Reduction \& Analysis\label{sec:dataAnal}}

\subsection{\hi data}

In the following, we outline the main steps of the \hi data analysis. We first
describe the new 21 cm-line observations and the data reduction. For the latter,
we followed procedures similar to \citet{Lelli2012a, Lelli2012b} and refer to
these papers for further details. For the existing \hi datacubes, we refer to
the original papers (see Table~\ref{tab:data}). Finally, we describe the derivation
of total \hi maps and velocity fields.

NGC~6789 was observed in May 2011 with the WSRT in a standard 12 h session.
The correlator was used in dual-polarization mode, with a total bandwidth of
10 MHz and 1024 spectral channels, providing a velocity resolution of $\sim$2.5
km~s$^{-1}$. SBS~1415+437 and UGCA~290 were observed with the B, C, and D
arrays of the VLA between March 2011 and April 2012. The correlator was used in
dual-polarization WIDAR mode with a total bandwidth of 2.0 MHz and 256 spectral
line channels, providing a velocity resolution of $\sim$1.9 km~s$^{-1}$. Between
20 Sept. 2011 and 3 Dec. 2011, the VLA correlator back-end, by mistake, integrated
for only 1 sec per record, thus the D-array observations have a time on source
of only $\sim$16 mins instead of the expected 2 hours. The new \hi observations
are summarized in Table \ref{tab:obs}. We also reduced archival VLA
observations of NGC~625 and UGC~6456.

The raw UV data were flagged, calibrated, and combined using the AIPS package
and following standard procedures. We Fourier-transformed the UV data using a
robust weighting technique \citep{Briggs1995}. After various trials, we chose
the value of the robust parameter $\Re$ (either $-1$, $-0.5$, or 0) that minimizes
sidelobes and wings in the beam profile. After the Fourier transform, we continued
the data analysis using the Groningen Imaging Processing SYstem (GIPSY)
\citep{vanderHulst1992}. The channel maps were continuum-subtracted using
line-free channels and then cleaned \citep{Hogbom1974} down to 0.3$\sigma$
using a mask to define the search areas.

A detailed study of the \hi kinematics requires a combination of spatial resolution,
spectral resolution, and sensitivity that varies from object to object, depending
both on the quality of the \hi observations and on the intrinsic properties of the
galaxy (e.g., angular size, rotation velocity, mean \hi column density). We used the
following approach. For every galaxy, we first analyzed the \hi datacube at the highest
spatial and spectral resolutions available. This cube is typically obtained using
$\Re \simeq 0$ and has relatively low column-density sensitivity, but the synthesized beam
profile is close to a Gaussian and does \textit{not} have the broad wings that are
typical for natural-weighted UV-data. Then, we smoothed the cube both in the image plane
and in velocity using various Gaussian tapers, until we found the optimal compromise
between resolution and sensitivity. The properties of both original and final datacubes
are summarized in Appendix~\ref{app:tables} (Table \ref{tab:data}). The spatial and
spectral resolutions range between 5$''$ to 30$''$ and 5 to 10 km~s$^{-1}$, respectively.

Total \hi maps were obtained by summing the masked channel maps. The masks were
constructed by first smoothing the datacubes in the image plane to 30$''$ or 60$''$
(depending on the angular extent of the galaxy) and in velocity to $\sim$10 or
$\sim$20 km~s$^{-1}$ (depending on the \hi line-width of the galaxy), and subsequently
clipping the channel maps at 3$\sigma_{\rm{s}}$ ($\sigma_{\rm{s}}$ is the rms noise
in the smoothed cube). A pseudo-3$\sigma$ contour in the total \hi map was calculated
following \citet{Verheijen2001}. The \hi datacube of NGC~1569 is strongly affected
by Galactic emission that we have interactively blotted out; the resulting \hi map
is rather uncertain. Velocity fields (VFs) were derived by fitting a Gaussian function
to the \hi line profiles. Fitted Gaussians with a peak intensity less than 3$\sigma$
were discarded. For most galaxies, the \hi line profiles are very broad and
asymmetric, thus the VFs only provide a rough description of the galaxy kinematics.
As a consequence, our kinematical analysis is mostly based on channel maps,
Position-Velocity (PV) diagrams, and 3-dimensional (3D) disk models.

\begin{figure*}
\centering
\includegraphics[width=17 cm]{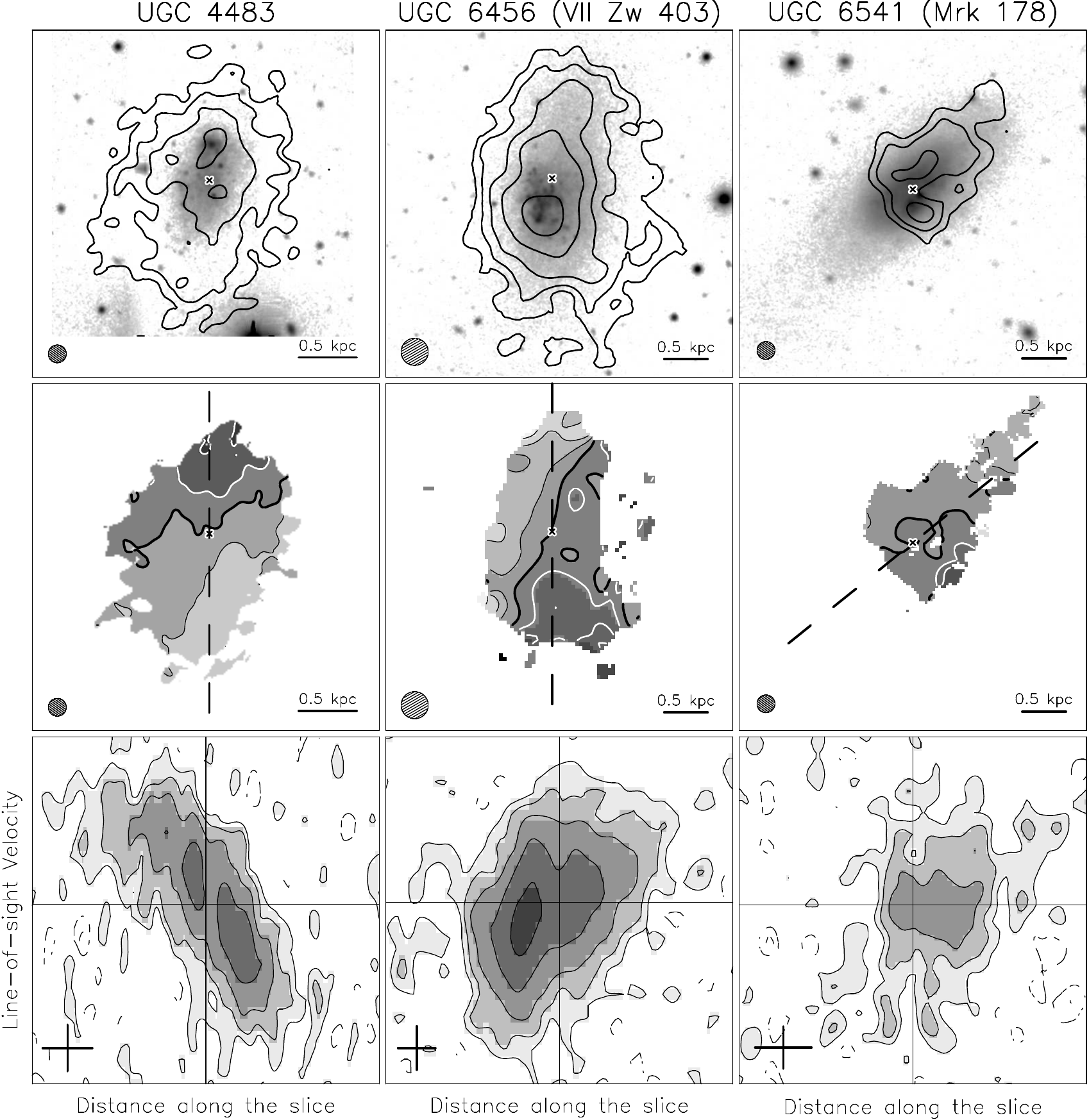}
\caption{Examples of BCDs with different \hi distribution and kinematics:
UGC~4483 (left) has a regularly rotating \hi disk, UGC~6456 (center) has a
kinematically disturbed \hi disk, and UGC~6456 (right) has an unsettled \hi
distribution. \textit{Top}: optical image superimposed with the total \hi
map (contours). The contours are the same as in Appendix~\ref{app:Atlas}.
The cross shows the optical center; the circle shows the \hi beam. 
\textit{Middle:} \hi velocity field. Light and dark shading indicate
approaching and receding velocities, respectively. The thick, black
line shows the systemic velocity. The isovelocity contours are the
same as in Appendix~\ref{app:Atlas}. The dashed line indicates the
\hi major axis. The cross and the circle are the same as in the top panel.
\textit{Bottom:} position-velocity diagrams taken along the major axis.
The cross corresponds to 0.5 kpc $\times$ 10 km~s$^{-1}$. The horizontal
and vertical lines indicate the systemic velocity and the galaxy center,
respectively.}
\label{fig:examples}
\end{figure*}
\subsection{Optical data}

In order to compare the relative \hi and stellar distributions of the BCDs,
we collected optical images via the NASA/IPAC Extragalactic Database
(NED\footnote{The NASA/IPAC Extragalactic Database is operated by the
Jet Propulsion Laboratory, California Institute of Technology, under
contract with the National Aeronautics and Space Administration.}).
When available, we used $R$-band images, otherwise we used $V$-band
ones. The images come from the following studies: \citet{Kuchinski2000},
\citet{GilDePaz2003}, \citet{Taylor2005}, \citet{Hunter2006}, and
\citet{Meurer2006}. In I~Zw~18, the nebular emission line dominates
the optical morphology, thus we have also used the H$\alpha$-subtracted,
$R$-band image from \citet{Papaderos2002}.

The images were analyzed as follows. We determined the sky level by
masking the sources in the frame and fitting a 2D polynomial to the
masked image. Then, we created sky-subtracted images and isophotal maps,
using the calibration parameters provided in the original papers. To
improve the signal-to-noise ratio in the outer regions, the isophotal
maps were pixel-averaged with a 3$\times$3 box; this preserves the
resolution of the images as the pixel sizes were typically $\sim$3
times smaller than the seeing.

The images were interactively fitted with ellipses to determine the
optical center ($\alpha_{\rm_{opt}}, \beta_{\rm_{opt}}$), position angle
(PA$_{\rm{opt}}$), and ellipticity $\epsilon_{\rm{opt}}$. Foreground
stars and scattered light in the frames were masked out by hand. Since
the inner isophotes of BCDs are usually irregular due to the starburst,
we estimated ($\alpha_{\rm_{opt}}, \beta_{\rm_{opt}}$), PA$_{\rm{opt}}$,
and $\epsilon_{\rm{opt}}$ by taking the mean value over the outer isophotes.
We also estimated the optical inclination $i_{\rm{opt}}$ using the formula
\begin{equation}
 \cos^{2} (i_{\rm{opt}}) = \dfrac{(1-\epsilon_{\rm{opt}})^{2} - q_{0}^{2}}{1 - q_{0}^{2}}
\end{equation}
where $q_{0}$ is the intrinsic thickness of an oblate stellar disk.
We assumed $q_{0} = 0.3$, as indicated by statistical studies of
the ellipticities of dwarf galaxies \citep{Binggeli1995, Sanchez2010}.
The orientation parameters of different isophotes often show
relatively large variations with radii, that we used to estimate the
uncertainties in PA$_{\rm opt}$ and $i_{\rm opt}$ ($\sim$2$^{\circ}$ to
$\sim$6$^{\circ}$, except for NGC~4214 that is very close to face-on).
The resulting orientation parameters are provided in Appendix~\ref{app:tables}
(Table \ref{tab:param}). The sky-subtracted images and isophotal
maps are presented in Appendix \ref{app:Atlas}. 

\begin{table}[t]
\caption{Structural properties of the \hi disk.}
\centering
\resizebox{9cm}{!}{
\begin{tabular}{l c c c c c}
\hline
\hline
Name        & $M_{\hi}$          & $R_{\hi}$ & $R_{\rm{opt}}$ & $R_{\hi}/R_{\rm{opt}}$ & Ref.\\
            &($10^{7} M_{\odot}$)& (kpc)     & (kpc)          &                        &     \\
\hline
\multicolumn{5}{l}{\textit{Galaxies with a regularly rotating \hi disk}}\\
NGC 1705    & 11.1$\pm$2.9 & 2.1 & 1.5 & 1.4 & a\\
NGC 2366    & 62$\pm$17    & 6.8 & 4.4 & 1.5 & b\\
NGC 4068    & 14.9$\pm$1.6 & 3.1 & 1.8 & 1.7 & b\\
NGC 4214    & 43$\pm$8     & 5.5 & 2.2 & 2.5 & b\\
NGC 6789    & 1.8$\pm$0.3  & 1.0 & 0.7 & 1.4 & a\\
UGC 4483    & 2.9$\pm$0.5  & 1.4 & 0.6 & 2.3 & a\\
I Zw 18     & 21$\pm$3     & 3.3 & 0.5 & 6.6 & c\\
I Zw 36     & 6.7$\pm$1.3  & 1.9 & 0.9 & 2.1 & a\\
SBS 1415+437& 20.1$\pm$4.6 & 4.3 & 2.4 & 1.8 & a\\
\multicolumn{5}{l}{\textit{Galaxies with a kinematically disturbed \hi disk}}\\
NGC 625     & 9.7$\pm$2.2  & 2.6 & 3.3 & 0.8 & a\\
NGC 1569    & 29.1$\pm$4.5 & 3.9 & 3.0 & 1.3 & b\\
NGC 4163    & 1.5$\pm$0.2  & 1.1 & 1.0 & 1.1 & b\\
NGC 4449    & 300$\pm$77   & 8.9 & 3.3 & 2.7 & b\\
NGC 5253    & 13.8$\pm$3.4 & 3.1 & 2.1 & 1.5 & a\\
UGC 6456    & 4.5$\pm$0.5  & 1.8 & 1.2 & 1.5 & c\\
UGC 9128    & 1.3$\pm$0.2  & 0.9 & 0.6 & 1.5 & b\\
\multicolumn{5}{l}{\textit{Galaxies with unsettled \hi distribution}}\\
UGC 6541    & 1.2$\pm$0.2  & ... & 0.9 & ... & c\\
UGCA 290    & 1.4$\pm$0.2  & ... & 0.9 & ... & a\\
\hline
\hline
\end{tabular}
}
\tablefoot{The \hi radius $R_{\hi}$ is defined as the radius where
the \hi surface density profile reaches 1 $M_{\odot}$~pc$^{-2}$.
The optical radius $R_{\rm opt}$ is defined as 3.2 $R_{\rm d}$,
where $R_{\rm d}$ is the exponential scale length. The last column
provides references for $R_{\rm d}$.}
\tablebib{a) this work, b) \citet{Swaters2002b}, c) \citet{Papaderos2002}.}
\label{tab:extent}
\end{table}
For 11 galaxies in our sample, \citet{Swaters2002b} and \citet{Papaderos2002}
derived $R$-band luminosity profiles that were used to estimate the scale
length $R_{\rm{d}}$ and the central surface brightness $\mu_{0}$ of the old
stellar component by fitting an exponential law to their outer parts. For
the remaining 7 objects, we derived luminosity profiles both for the whole
galaxy and for the approaching/receding sides separately, by azimuthally averaging
the $R$-band images over concentric ellipses. We did not correct the profiles
for internal extinction, as BCDs are usually metal-poor (see Table \ref{tab:sample})
and the dust content is expected to be low. Finally, we estimated the central
surface brightness $\mu_{0}$ and scale length $R_{\rm{d}}$ of the old stellar
component by fitting an exponential law to the side of the disk that is least
affected by the starburst (as in \citealt{Lelli2012b} for UGC~4483).

\section{\hi distribution and kinematics\label{sec:HIdistr}}

In the following, we discuss the distribution and kinematics of the
high-column-density gas associated with the stellar component of BCDs.
In Appendix~\ref{app:indi} we discuss individual galaxies in detail
and compare our results to previous studies, while in Appendix~\ref{app:Atlas}
we present an atlas with optical images, isophotal maps, total \hi maps,
\hi surface density profiles, \hi velocity fields, and PV-diagrams.

Based on the \hi morphology and kinematics, we classify starbursting
dwarfs into three main families:
\begin{itemize}
 \item \textit{BCDs with a regularly rotating \hi disk} ($\sim$50$\%$). \\
The PV-diagram along the \hi major axis shows a regular velocity gradient 
(see Fig.~\ref{fig:examples}, left). The VF displays a pattern that is
typical of a rotating disk, although minor asymmetries caused by non-circular
motions may be present. In some cases, the kinematical center and PA may
not coincide with the optical ones. For these galaxies, we derive rotation
curves (see Sect.~\ref{sec:RegRotBCD}).
 \item \textit{BCDs with a kinematically disturbed \hi disk} ($\sim$40$\%$). \\
The \hi distribution resembles a disk, but the PV-diagrams and the VF
are irregular and asymmetric (see Fig.~\ref{fig:examples}, center).
For these galaxies, it is not possible to derive a reliable rotation
curve, but we obtain rough estimates of the kinematical parameters
and of the rotation velocity in the outer parts (see Sect.~\ref{sec:DistRotBCD}).
 \item \textit{BCDs with unsettled \hi distributions} ($\sim$10$\%$).\\
Both the \hi distribution and kinematics are irregular and asymmetric
(see Fig.~\ref{fig:examples}, right), and they are inconsistent with
a rotating disk.
\end{itemize}

For the 16 galaxies with a \hi disk, we derived \hi surface density profiles
by azimuthally averaging the \hi maps over ellipses with a width of 1 beam-size.
We assumed the orientation parameters defined by the \hi kinematics (see
Table \ref{tab:param}). Following \citet{Swaters2002a}, we calculated i)
the \hi radius $R_{\hi}$, defined as the radius where the \hi surface density
profile (corrected for inclination) reaches 1 $M_{\odot}$~pc$^{-2}$, and ii)
the optical radius $R_{\rm opt}$, defined as 3.2 scale lengths $R_{\rm d}$.
The latter definition allows us to compare the sizes of galaxies with
different central surface brightnesses: for an exponential disk with
$\mu_{0}(B) = 21.65$ mag~arcsec$^{-2}$ \citep{Freeman1970}, $R_{\rm{opt}}
= 3.2 R_{\rm{d}}$ corresponds to the isophotal radius $R_{25}$. In our
sample of BCDs, the ratio $R_{\hi}/R_{\rm{opt}}$ ranges from $\sim$1
to $\sim$3 (see Table~\ref{tab:extent}), which is typical for gas-rich
spirals and irregulars (Irrs) \citep[e.g.,][]{Verheijen2001}. The only exception
is I~Zw~18 with $R_{\hi}/R_{\rm{opt}}\simeq 6.6$, as the \hi distribution
extends towards a secondary stellar body (see Appendix~\ref{app:Atlas}
and \citealt{Lelli2012a}). Excluding I~Zw~18, the mean value of
$R_{\hi}/R_{\rm{opt}}$ is 1.7$\pm$0.5, in close agreement with the
values found by \citet{Swaters2002a} for 73 gas-rich dwarfs (1.8$\pm$0.8)
and by \citet{Broeils1997} for 108 gas-rich spirals (1.7$\pm$0.5).

\begin{figure*}
\centering
\includegraphics[width=17cm]{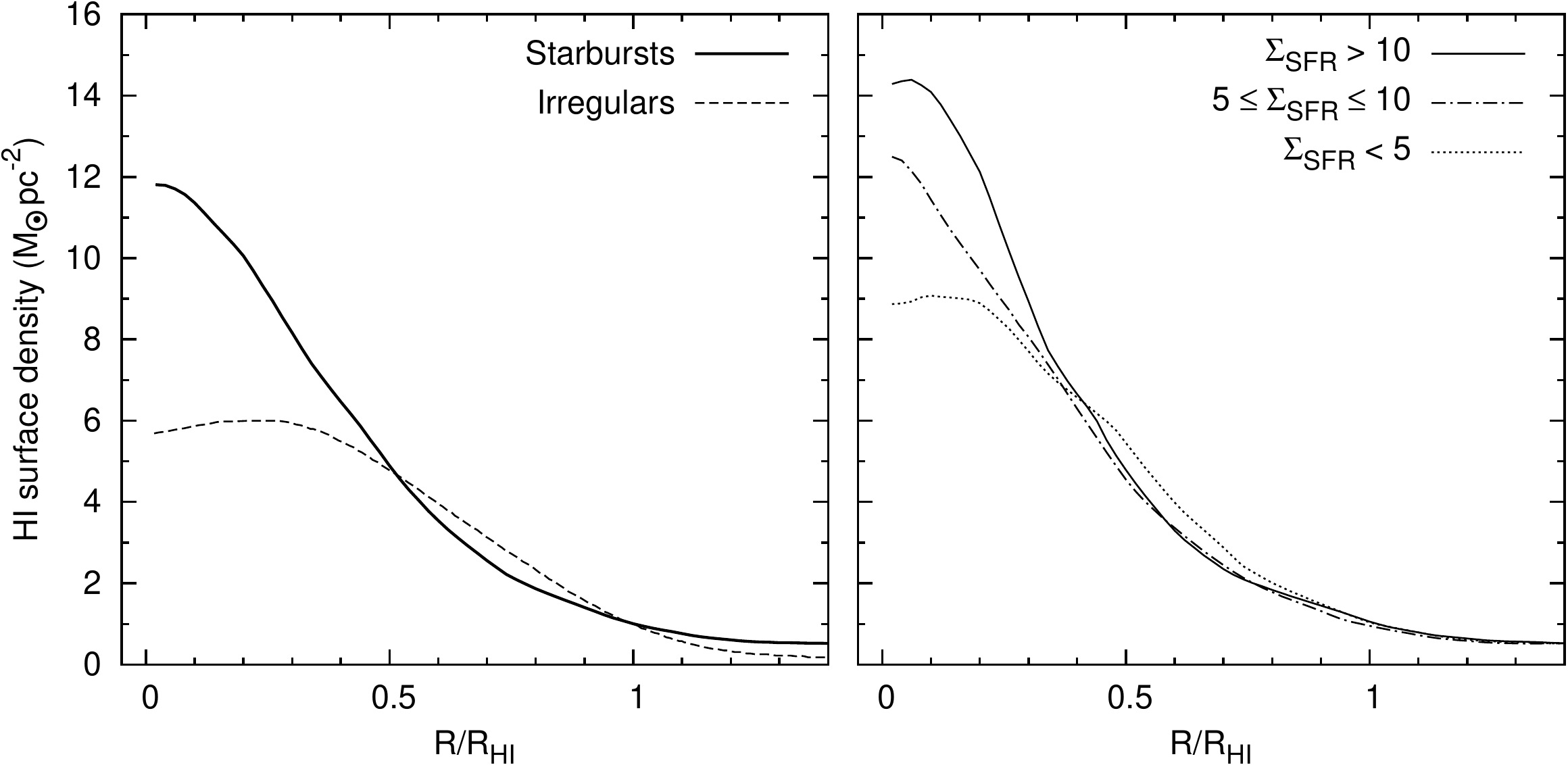}
\caption{\textit{Left:} average \hi surface density profile for
starbursting dwarfs (solid line) and for typical irregulars (dashed
line, from \citealt{Swaters2002a}). \textit{Right:} average \hi
surface density profiles for starbursting dwarfs binned by $\Sigma
_{\rm SFR}$ (in units of $10^{-3}$~$M_{\odot}$~yr$^{-1}$~kpc$^{-2}$)
as indicated in the legend to the top-right.}
\label{fig:hiprof}
\end{figure*}
Previous \hi studies have shown that BCDs have centrally-concentrated
\hi distributions with higher surface densities than typical Irrs
\citep{Taylor1994, vanZee1998b, vanZee2001, Simpson2000}. We confirm
this result for a larger sample of galaxies. In Fig.~\ref{fig:hiprof}
(left), we compare the mean \hi surface density profile of late-type
dwarfs obtained by \citet[][averaging over 73 objects]{Swaters2002a}
with the mean \hi surface density profile of our starbursting dwarfs
(averaging over the 16 objects with a \hi disk). Note that the sample
of \citet{Swaters2002a} contains a few BCDs that are also part of
our sample, but it is fully dominated by typical, non-starbursting Irrs.
Fig.~\ref{fig:hiprof} (left) shows that BCDs, on average, have
central \hi surface densities a factor of $\sim$2 higher than typical
Irrs. In several cases, the central, azimuthally averaged \hi surface
densities can be as high as $\sim$20 $M_{\odot}$~pc$^{-2}$ (see e.g.,
NGC~1569 and NGC~1705 in Appendix \ref{app:Atlas}). The actual peak
\hi column densities can reach even higher values, up to $\sim$50-100
$M_{\odot}$~pc$^{-2}$ in I~Zw~18 at a linear resolution of $\sim$200
pc \citep[][]{Lelli2012a}.

When comparing the \hi surface densities of different galaxies, a
possible concern is the different linear resolutions (in kpc) of
the 21 cm-line observations. The 16 \hi cubes used here have linear
resolutions ranging from $\sim$0.2 to $\sim$0.7 kpc (see Table
\ref{tab:data}), while the 73 cubes used by \citet{Swaters2002a}
have linear resolutions ranging from $\sim$0.4 to $\sim$2 kpc
(apart for 13 cases where the linear resolution is $\gtrsim$2 kpc).
To quantify the effect of beam smearing, we derived \hi surface density
profiles by smoothing our data to $30''$ \citep[as in][]{Swaters2002a},
and checked that the distribution of the ratio $R_{\hi}$/beam-size
is comparable for the two samples. We found that the \textit{azimuthally
averaged} \hi surface density of BCDs decreases by only $\sim$20$\%$
in the inner parts. Thus, we conclude that the difference between the
mean \hi surface density profile of BCDs and Irrs is not due to
observational effects.

Fig.~\ref{fig:hiprof} (right) shows the mean \hi surface density
profiles of BCDs binned by SFR surface density $\Sigma_{\rm SFR}=
{\rm SFR_{\rm p}}/(\pi R_{\rm opt}^{2})$ (see Table~\ref{tab:sample}).
We used three $\Sigma_{\rm SFR}$ bins (in units of $10^{-3}$~$M_{\odot}$
yr$^{-1}$~kpc$^{-2}$): $\Sigma_{\rm SFR}< 5$ (five objects), $5 \leq 
\Sigma_{\rm SFR} \leq 10$ (seven objects), and $\Sigma_{\rm SFR}>10$
(four objects). It is clear that galaxies with higher values of
$\Sigma_{\rm SFR}$ have higher central \hi surface densities, as
one may expect from the Kennicutt-Schmidt law \citep[e.g.,]
[]{Kennicutt1998}. We also derived mean \hi surface density
profiles binned by the value of the birthrate parameter $b$
(see Table~\ref{tab:sample}), but found no strong relation
between $b$ and the shape of the \hi surface density profile. 
The relation between star formation, gas surface density,
and internal dynamics is further investigated and discussed
in \citet{Lelli2014}.

Finally, we compare the overall \hi kinematics of BCDs and typical
Irrs. A detailed comparison between the rotation curves of BCDs
and Irrs is presented in \citet{Lelli2014}. Irrs generally have
regularly rotating \hi disks. For example, \citet{Swaters2009}
studied the \hi kinematics of 69 late-type dwarfs and could derive
rotation curves for 62 objects ($\sim$90$\%$). In contrast, for our
sample of 18 starbursting dwarfs, rotation curves could be derived
for only 50$\%$ of the galaxies, as the other objects have either a
kinematically disturbed \hi disk or an unsettled \hi distribution.
This suggests that complex \hi kinematics are much more common
in BCDs than in typical Irrs. This may be related to the starburst
trigger (e.g., interactions/mergers or disk instabilities) and/or
be a consequence of feedback from supernovae (SN) and stellar winds.

\begin{figure*}
\centering
\includegraphics[width=17 cm]{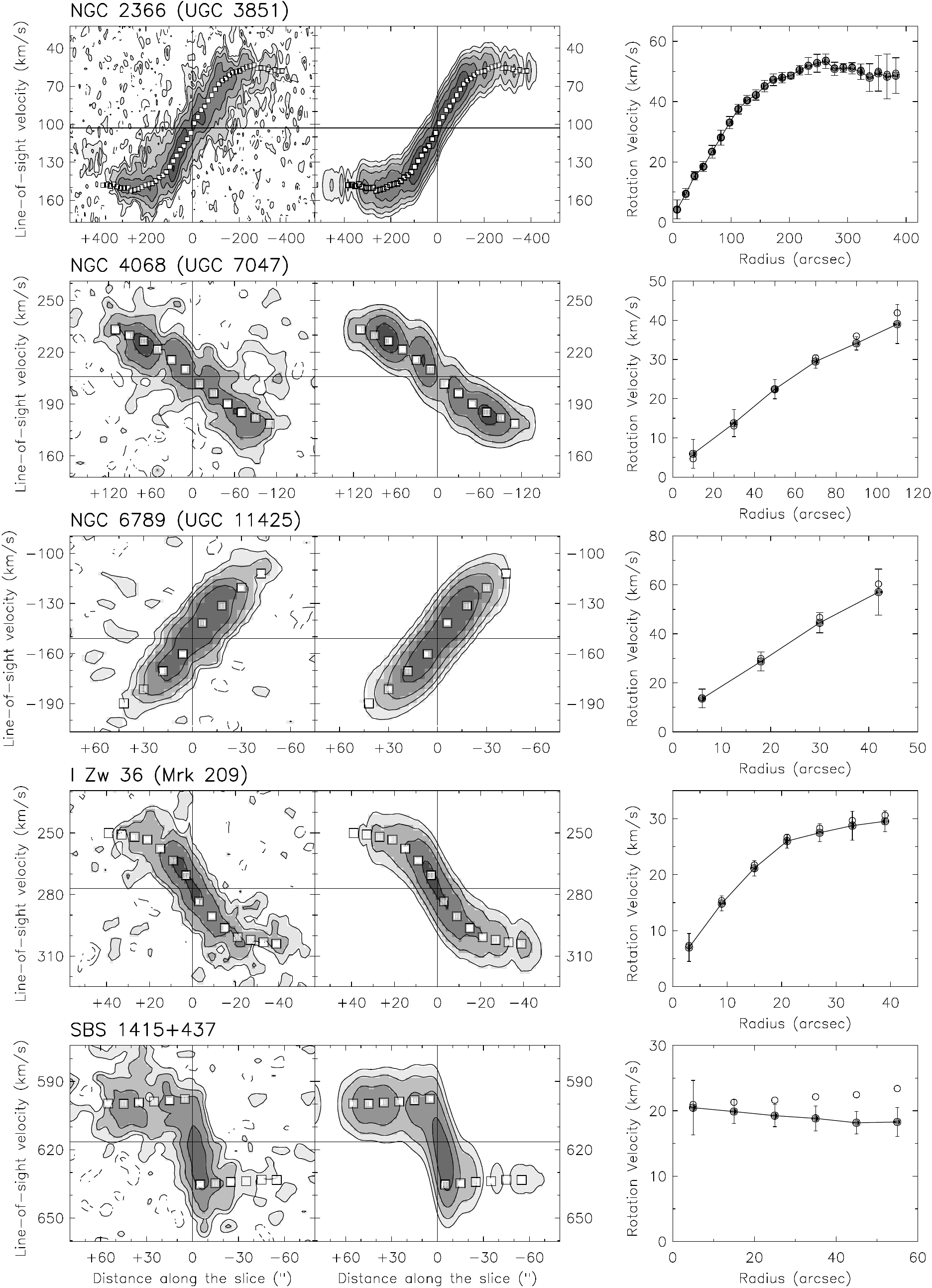}
\caption{Kinematical analysis of five BCDs with a regularly rotating \hi disk.
\textit{Left}: PV-diagrams obtained along the disk major axis from the observed
cube and the model-cube. Contours are at -1.5 (dashed), 1.5, 3, 6, 12 $\sigma$.
The squares show the rotation curve used to build the models, projected along
the line of sight. \textit{Right:} observed rotation curve (filled circles) and
asymmetric-drift-corrected rotation curve (open circles). See Sect.~\ref{sec:RegRotBCD}
for details.}
\label{fig:PVmodels}
\end{figure*}
\begin{figure*}
\centering
\includegraphics[width=18 cm]{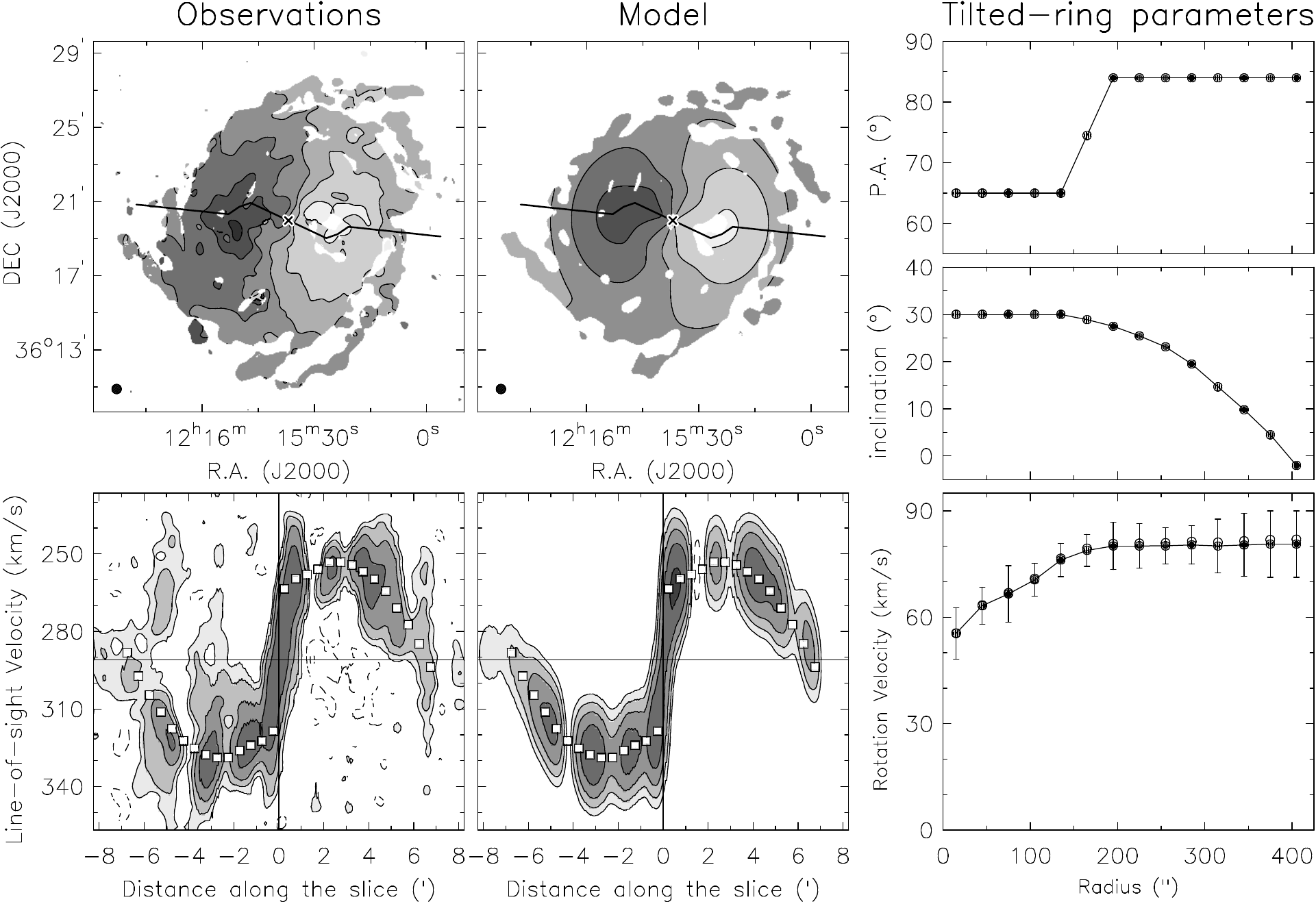}
\caption{The warped galaxy NGC~4214.
\textit{Left}: velocity fields (top) and PV-diagrams (bottom) obtained from the observed
cube and the model-cube. In the velocity fields, the contours range from 257 to 337
km~s$^{-1}$ with steps of 16 km~s$^{-1}$; the thick line shows the line of nodes; the
circle to the bottom-left shows the beam. In the PV-diagrams, the contours are at -1.5
(dashed), 1.5, 3, 6, 12 $\sigma$; the squares show the rotation curve projected along
the line of sight. \textit{Right:} tilted-ring parameters for the disk model. The open
circles show the asymmetric-drift-corrected rotation curve. See Sect.~\ref{sec:RegRotBCD}
for details.}
\label{fig:n4214}
\end{figure*}
\begin{figure*}
\centering
\includegraphics[width=18 cm]{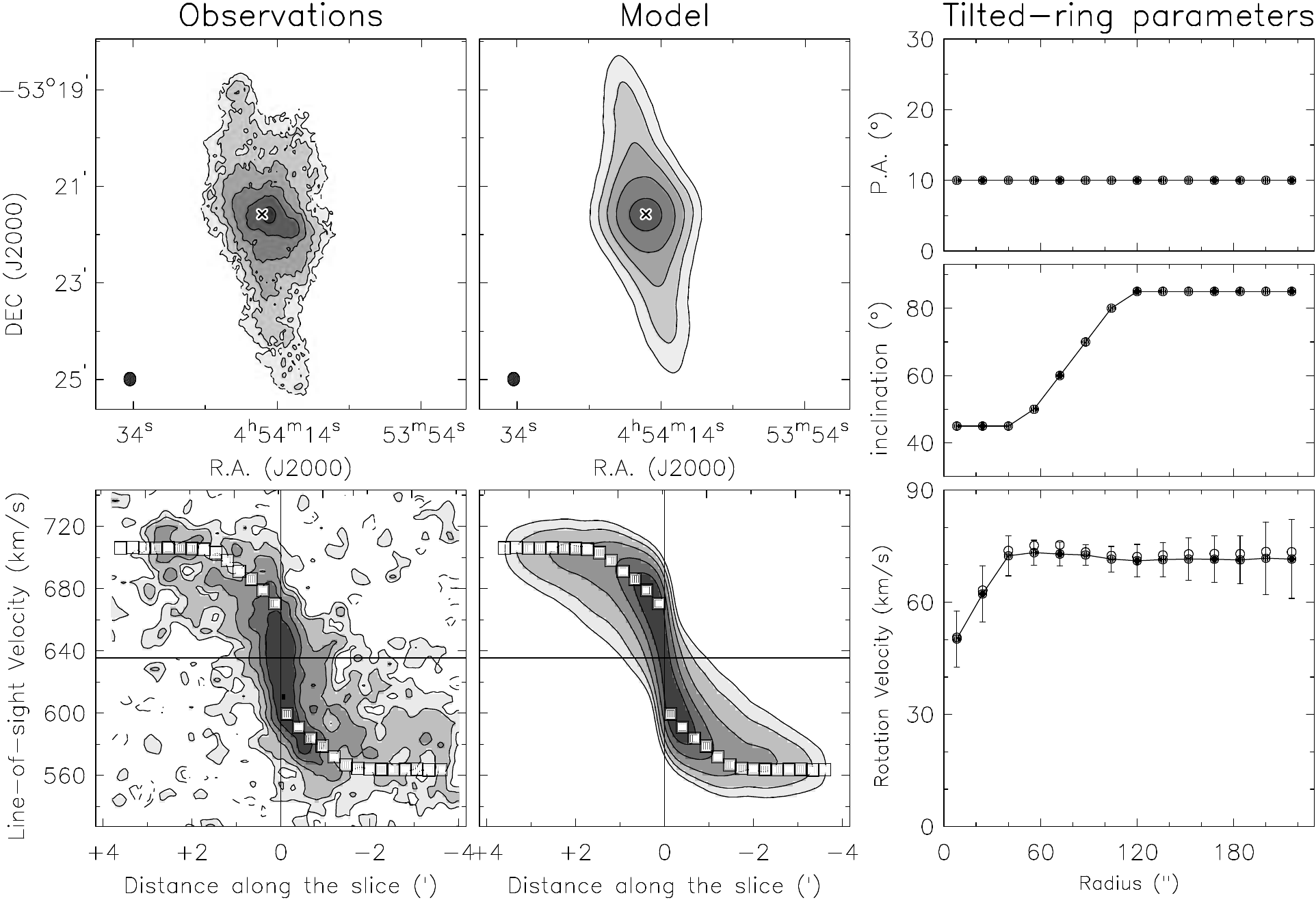}
\caption{The warped galaxy NGC~1705.
\textit{Left}: total \hi map  at $\sim$16$''$ resolution (top) and PV-diagrams at 20$''$
resolution (bottom) obtained from the observed cube and the model-cube. In the total
\hi map, contours are at 1.1, 2.2, 4.4, 8.8, 17.6 $\times$10$^{20}$ atoms cm$^{-2}$;
the circle to the bottom-left shows the beam. In the PV-diagrams, the contours are at
-1.5 (dashed), 1.5, 3, 6, 12, 24 $\sigma$; the squares show the rotation curve projected
along the line of sight. \textit{Right:} tilted-ring parameters for the disk model.
The open circles show the asymmetric-drift-corrected rotation curve.
See Sect.~\ref{sec:RegRotBCD} for details.}
\label{fig:n1705}
\end{figure*}
\section{Galaxies with a regularly rotating \hi disk \label{sec:RegRotBCD}}

\subsection{Derivation of the rotation curves \label{sec:rotcur}}

Nine galaxies in our sample (50$\%$) have a regularly rotating \hi disk:
NGC~1705, NGC~2366, NGC~4068, NGC~4214, NGC~6789, UGC~4483, I~Zw~18, I~Zw~36,
and SBS~1415+437. For these objects, we derived rotation curves following a
procedure similar to \citet{Swaters2009}. As a first step, we obtained initial
estimates of the geometrical parameters and of the rotation curve by fitting
a tilted-ring model to the VF \citep{Begeman1987}. These initial estimates
were then used as input to build a 3D kinematic model, which was subsequently
corrected by trial and error to obtain a model-cube that matches the observations.
The model-cubes consider the spatial and spectral resolution of the observations,
the observed gas distribution, the velocity dispersion, the disk thickness,
and possibly non-circular motions. For I~Zw~18 and UGC~4483, the derivation
of the rotation curve is described in detail in \citet{Lelli2012a, Lelli2012b}.
In the following, we briefly describe the derivation of the rotation curves
for the remaining galaxies.

The VF was fitted with a tilted-ring model using a ring width of one beam
and weighting every point by $\cos^{2}(\theta)$, where $\theta$ is the
azimuthal angle in the plane of the galaxy from the major axis. The parameters
of the fit are the kinematical center $(\alpha_{\rm kin}, \delta_{\rm kin})$,
the systemic velocity $V_{\rm sys}$, the position angle PA$_{\rm kin}$, the
inclination $i_{\rm kin}$, and the rotation velocity $V_{\rm rot}$. We first
left all the parameters free, and determined $(\alpha_{\rm kin}, \delta_{\rm kin})$
and $V_{\rm sys}$ by taking the average value within the innermost rings.
Then, we fixed $(\alpha_{\rm kin}, \delta_{\rm kin})$ and $V_{\rm sys}$,
and determined PA$_{\rm kin}$ and $i_{\rm kin}$. Finally, we determined
$V_{\rm rot}$ at every radius, keeping all the other parameters fixed.
The errors on $V_{\rm rot}$ have been estimated by considering the
differences in the rotation curves derived from the approaching and
receding sides separately \citep[see e.g.,][]{Lelli2012b}.

The 3D disk models were built assuming that the \hi kinematics is axisymmetric
while the \hi distribution is clumpy, i.e. the surface density varies with 
position as in the observed \hi map (see \citealt{Lelli2012a, Lelli2012b}
for details). We also assumed that i) the \hi disk has an exponential vertical
distribution with a constant scale height $z_{0}$ of $\sim 100$~pc, and
ii) the velocity dispersion $\sigma_{\hi}$ is constant with radius.
The actual value of $z_{0}$ does not affect our results, since kinematic
models with different scale heights (up to 1~kpc) show no significant
differences in their channel maps and PV-diagrams. For the mean velocity
dispersion $\sigma_{\hi}$, we estimated values between 8 and 12 km~s$^{-1}$
(see Table \ref{tab:asym}) by comparing several PV-diagrams obtained
from models and observations; values higher than 12 km~s$^{-1}$ are
generally ruled out. Possible variations of $\sigma_{\hi}$ with radius
of $\sim$2-3 km~s$^{-1}$ would generally produce no appreciable differences
in the model-cubes. An exception is NGC~4214, which seems to have $\sigma
_{\hi}\simeq15$ km~s$^{-1}$ for $R\lesssim1'$.

Figs.~\ref{fig:PVmodels}, \ref{fig:n4214}, and \ref{fig:n1705} compare
PV-diagrams obtained from both the observations and the models along the
disk major axis. We generally found a good agreement between models and
observations when increasing the values of $V_{\rm rot}$ derived with
the tilted-ring fit to the VF by $\sim$2 to 3 km~s$^{-1}$ at the
innermost radii. Exceptions are NGC~4214 and SBS~1415+437, as their
inner velocity-points require a correction of $\sim$10 km~s$^{-1}$
due to severe beam-smearing effects. For NGC~1705, the corrections to
the velocity-points are even larger ($\sim$20 to 30 km~s$^{-1}$) because
the \hi disk appears close to edge-on in the outer parts and, thus, the
VF only provides a poor representation of the rotation velocities. We
also used PV-diagrams at the full resolution (see Table \ref{tab:data})
to further check that beam-smearing effects do not significantly affect
our rotation curves. In two cases (NGC~4068 and SBS~1415+437), the 3D disk
models also suggest that the position of the kinematical center should
be shifted by about one beam with respect to the results of the tilted-ring
fit to the VF.

The observed PV-diagram of NGC~2366 shows \hi emission close to the systemic
velocity that cannot be reproduced by our disk model (see Fig.~\ref{fig:PVmodels}).
This is likely due to extra-planar \hi emission that is rotating at a lower
velocity than the disk (a so-called ``lagging \hi halo'', see e.g.,
\citealt{Fraternali2002}), as it is observed in several nearby spiral
galaxies \citep[e.g.,][]{Sancisi2008}. The modeling of a lagging \hi
halo is beyond the scope of this paper.

The galaxies NGC~4214 and NGC~1705 deserve special attention, as their \hi
disks are close to face-on in the inner parts and strongly warped in the
outer parts. A tilted-ring fit to the VF, therefore, poses severe limitations
when determining the dependence of PA$_{\rm kin}$ and $i_{\rm kin}$ on
radius \citep[cf.][]{Begeman1987}. We built a series of disk models assuming
different types of warps. In the following, we only discuss our best models
(shown in Figs.~\ref{fig:n4214} and \ref{fig:n1705}). For both galaxies,
the warp is slightly asymmetric between the approaching and receding
sides of the disk, and the actual dependence of $i_{\rm kin}$ on radius
remains uncertain.

For NGC~4214, we found that PA$_{\rm kin}\simeq65^{\circ}$ and $i_{\rm kin}
\simeq30^{\circ}$ for $R \lesssim 3'$, in agreement with the optical values
within the uncertainties, while at larger radii PA$_{\rm kin}\simeq84^{\circ}$
and $i_{\rm kin}$ gradually decreases (see Fig.~\ref{fig:n4214}, right panels).
This model provides a good match to the observed VF and PV-diagram taken
along the line of nodes (see Fig.~\ref{fig:n4214}, left panels); minor
discrepancies ($\sim$20 km~s$^{-1}$) are observed, possibly due to non-circular
motions caused by the inner stellar bar and/or streaming motions along the
prominent \hi spiral arms (see the \hi map in Appendix~\ref{app:Atlas}).

The \hi disk of NGC~1705 appears highly inclined in the outer parts, thus
the VF does not provide useful information regarding the dependence of
$i_{\rm kin}$ on radius. For this galaxy, we built 3D models using an
\textit{axisymmetric} \hi distribution in each ring, and determined
the values of $i_{\rm kin}$ by comparing total \hi maps and PV-diagrams
obtained from the observed cube and the model-cubes. Similarly to
\citet{Elson2013}, we found that PA$_{\rm kin}\simeq10^{\circ}$ while
$i_{\rm kin}$ abruptly changes at $R\simeq1.5'$ (see Fig.~\ref{fig:n1705},
right panels). We adopted, however, a higher value of $i_{\rm kin}$ in the
outer parts than \citet{Elson2013} (85$^{\circ}$ instead of 65$^{\circ}$).
This is necessary to reproduce the tails of \hi emission towards
$V_{\rm sys}$ that are clearly visible in PV-diagrams at 20$''$ resolution
(see Fig.~\ref{fig:n1705}, bottom panels). To fully reproduce these broad
\hi profiles, especially the \hi emission in the ``forbidden'' quadrants
of the PV-diagram, we also had to include radial motions of $\sim$15
km~s$^{-1}$ (cf. with NGC~2915, \citealt{Elson2011}). As a consequence
of the high value of $i_{\rm kin}$ in the outer parts, the thickness
of the \hi disk must be $\sim 500$~pc in order to reproduce the
observed total \hi map (see Fig.~\ref{fig:n1705}, top panels).

\subsection{Asymmetric-drift correction}

\begin{table}[t]
\caption{Parameters for the asymmetric-drift correction}
\centering
\resizebox{9cm}{!}{
\begin{tabular}{l c c c c c}
\hline
\hline
Galaxy              & Funct. & $\Sigma_{0, \hi}$     & $R_{0, \hi}$ & $s$ & $\sigma_{\hi}$\\
                    &        &(M$_{\odot}$ pc$^{-2}$)& (kpc)        & (kpc) & (km s$^{-1}$)\\
\hline
NGC~1705 ($R<80''$) & Exp    & 24.0                  & 0.7          & ...  & 12\\
NGC~1705 ($R>80''$) & Exp    & 1.5                   & 2.8          & ...  & 12\\
NGC~2366            & Gauss  & 9.1                   & 0.8          & 6.31 & 10\\
NGC~4068            & Gauss  & 9.0                   & 1.0          & 0.92 & 8\\
NGC~4214            & Exp    & 13.6                  & 2.5          & ...  & 10\\
NGC~6789            & Gauss  & 16.4                  & 0.0          & 0.37 & 8\\
I~Zw~36             & Exp    & 11.9                  & 0.7          & ...  & 9\\
SBS~1415+357        & Exp    & 11.8                  & 1.4          & ...  & 9\\
\hline
\hline
\end{tabular}
}
\label{tab:asym}
\end{table}
For several galaxies in our sample, the observed rotation velocity $V_{\rm{rot}}$
is only a factor $\sim$2 or 3 larger than the \hi velocity dispersion $\sigma_{\hi}$.
In order to trace the underlying mass distribution, the observed rotation curves
have to be corrected for pressure support. We calculated the asymmetric-drift
correction following \citet{Meurer1996}. We assumed that i) the \hi kinematics
is axisymmetric, ii) the \hi velocity dispersion is isotropic, iii) the velocity
dispersion and the scale-height of the \hi disk are constant with radius, and
iv) the \hi surface density profile can be approximated either by an exponential
function $\Sigma_{0, \hi} \times \exp(-R/R_{0, \hi})$ or by a Gaussian function
$\Sigma_{0, \hi} \times \exp[-(R-R_{0, \hi})^{2}/(2 s^{2})]$. The circular
velocity $V_{\rm{circ}}$, corrected for asymmetric-drift, is thus given by
\begin{equation}
 V_{\rm{circ}} = \sqrt{V_{\rm{rot}}^{2} + \sigma_{\hi}^{2} (R/R_{0, \hi})}
\end{equation}
in the case of an exponential surface density profile, and by
\begin{equation}
 V_{\rm{circ}} = \sqrt{V_{\rm{rot}}^{2} + \sigma_{\hi}^{2} R (R-R_{0,\hi})/s^{2}}
\end{equation}
in the case of a Gaussian surface density profile. For NGC~1705, the first
assumption is \textit{not} valid because the \hi disk is strongly warped
with an abrupt change of the inclination by $\sim$40$^{\circ}$ (see
Fig.~\ref{fig:n1705}), thus we calculated the asymmetric-drift correction
separately for the inner, face-on disk and for the outer, edge-on disk.

\begin{figure*}
\centering
\includegraphics[width=18 cm]{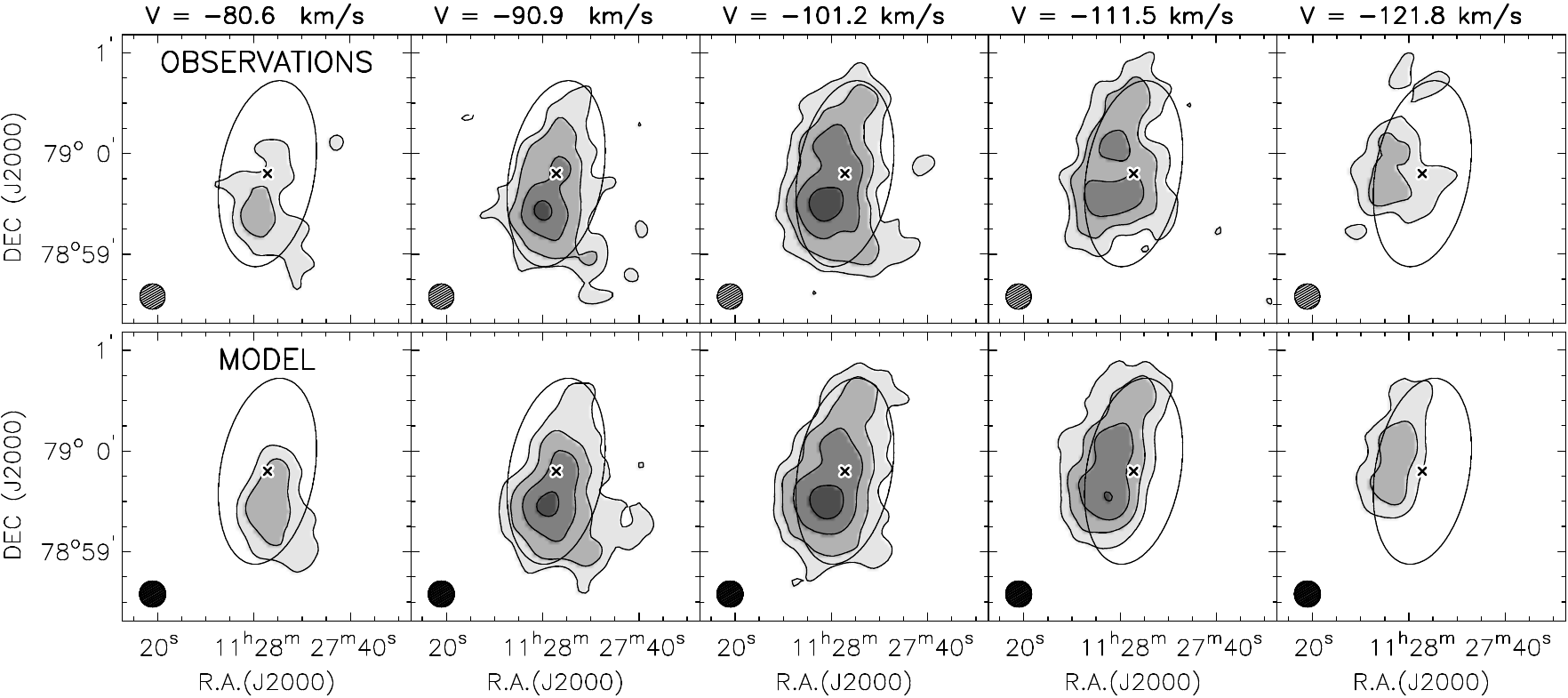}
\caption{Channel maps for UGC~6456 (VII Zw 403). \textit{Top}: observations.
\textit{Bottom}: a 3D disk model with $V_{\rm{rot}}\simeq V_{\rm{rad}} \simeq
\sigma_{\hi} \simeq 10$ km~s$^{-1}$. Contours are at 3, 6, 12, 24 $\sigma$.
The cross shows the optical center. The ellipse corresponds to the $R$-band
isophote at 25 mag arcsec$^{-2}$.}
\label{fig:u6456}
\end{figure*}
Table \ref{tab:asym} provides the values of $\sigma_{\hi}$, derived by building 3D
disk models, and the parameters for the exponential/Gaussian functions, obtained by
fitting the observed \hi surface density profiles. For NGC~1705, we fitted the \hi
surface density profile with 2 different exponential functions for the inner face-on
disk and the outer edge-on disk. The open-circles in Figs.~\ref{fig:PVmodels},
\ref{fig:n4214}, and \ref{fig:n1705} (right panels) show the asymmetric-drift-corrected
rotation curves. The correction is generally smaller than the error bars, except
for SBS~1415+437 that is rotating at only $\sim$20 km~s$^{-1}$ (see also UGC~4483
in \citealt{Lelli2012b}).

\subsection{Non-circular motions}

The PV-diagrams in Figs.~\ref{fig:PVmodels}, \ref{fig:n4214}, and \ref{fig:n1705}
clearly show that the \hi kinematics of these galaxies is dominated by rotation.
In several cases, however, a simple rotating-disk model cannot reproduce all the
features of the observed cube. In \citet{Lelli2012a, Lelli2012b}, we showed that
the \hi disks of I~Zw~18 and UGC~4483 likely have radial motions of $\sim$15 and
$\sim$5 km~s$^{-1}$, respectively. We did not find such regular radial motions
in the 7 BCDs analyzed here, with the possible exception of NGC~1705 that shows
double-peaked \hi profiles near the center and may have radial motions of $\sim$15
km~s$^{-1}$. Several galaxies, however, do show kinematically-anomalous components
that deviate from the main rotation pattern of the disk. We briefly discuss 3
interesting cases: NGC~1705, NGC~2366, and NGC~4214.

NGC~1705 has a \hi ``spur'' to the North-West \citep[see][]{Meurer1998, Elson2013},
that may be interpreted as an \hi outflow associated with the H$\alpha$ wind
\citep{Meurer1992}. The VF of NGC~2366 shows a strong distortion to the North-West
(see Appendix~\ref{app:Atlas}), that \citet{Oh2008} interpreted as non-circular
motions of $\sim20$ km~s$^{-1}$ (see their Figure 3). Finally, NGC~4214 shows
non-circular motions of $\sim$20 to $\sim$50 km~s$^{-1}$ (corrected for $i$)
that are visible in several PV-diagrams taken across the disk (not shown here,
but some kinematically-anomalous gas can be seen in the bottom panels of
Fig.~\ref{fig:n4214}); these non-circular motions are likely associated
with the \hi spiral arms and/or the inner stellar bar.

\begin{figure*}
\centering
\includegraphics[width=18 cm]{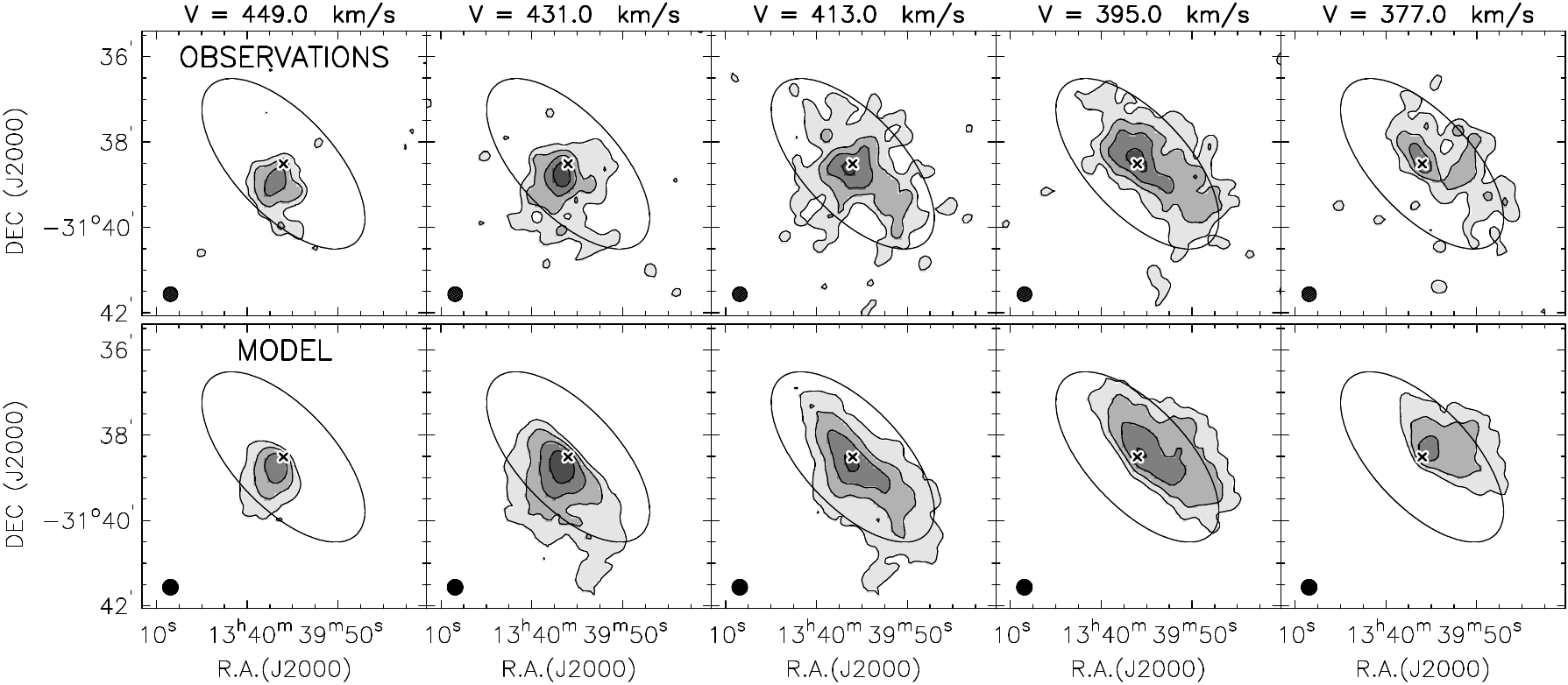}
\caption{Channel maps for NGC~5253. Contours are at 3, 6, 12, 24 $\sigma$.
\textit{Top}: observations. \textit{Bottom}: a 3D disk model with
$V_{\rm{rot}}=0$, $V_{\rm{rad}}= 25$, and $\sigma_{\hi}=$ 15 km~s$^{-1}$.
The cross shows the optical center. The ellipse corresponds to the $R$-band
isophote at 24 mag arcsec$^{-2}$.}
\label{fig:n5253}
\end{figure*}
\section{Galaxies with a kinematically disturbed \hi disk\label{sec:DistRotBCD}}

Seven galaxies in our sample have a kinematically disturbed \hi disk:
NGC~625, NGC~1569, NGC~4163, NGC~4449, NGC~5253, UGC~6456, and
UGC~9128. For these objects, we used channels maps and PV-diagrams
to estimate $(\alpha_{\rm{kin}}, \delta_{\rm{kin}})$, $V_{\rm sys}$,
PA$_{\rm{kin}}$, and the rotation velocity in the outer parts. For the
inclination, we assumed the optical value. The values of the kinematic
parameters have been tested by building 3D disk models with constant
$V_{\rm{rot}}$. These models cannot reproduce the observations in details,
but indicate that the observed \hi kinematics are consistent with a rotating
disk. The resulting values for the kinematic parameters are rather uncertain.

Two galaxies (NGC~5253 and UGC~6456) show a regular velocity gradient
approximately along the \hi minor axis (see the VFs in Appendix~\ref{app:Atlas}).
Velocity gradients along the \hi minor axis have also been observed in
other BCDs and interpreted as unusual rotation around the major axis
\citep[e.g.,][]{Thuan2004, BravoAlfaro2004}. This peculiar kinematic behavior,
however, can also be interpeted as a \hi disk with strong \textit{radial}
motions (e.g., \citealt{LopezSanchez2012}). We built 3D disk models assuming
the PA suggested by the total \hi map (consistent with the optical
value within the errors) and tried different combinations of circular
and radial motions. Our best-models are shown in Figs ~\ref{fig:u6456}
and \ref{fig:n5253}. For UGC~6456, the \hi emission can be reproduced
by a combination of circular and radial motions ($V_{\rm{rot}} \simeq
V_{\rm{rad}} \simeq 10$ km~s$^{-1}$, see Fig.~\ref{fig:u6456}).
The case of NGC~5253 is even more extreme, as the radial component
is $\sim$25 km~s$^{-1}$ while the rotation is constrained to be
$\lesssim$5 km~s$^{-1}$ (see Fig.~\ref{fig:n5253}). In our opinion,
strong radial motions are a more likely interpretation than rotation
around the major axis, since gas inflows/outflows are expected in
starburst galaxies.

For UGC~6456, it is not possible to distinguish between inflow and
outflow, as it is unknown which side of the disk is nearest to the
observer. For NGC~5253, instead, shadowing of the X-ray emission
suggests that the southern side of the disk is the nearest one
\citep{Ott2005a}, implying that the radial motion is an inflow.
For both galaxies, the inflow timescale $t_{\rm in}= R_{\hi}/V_{\rm rad}$
is $\sim$100-200 Myr, thus comparable with the starburst duration
\citep[see][]{McQuinn2010b}. We also calculated the gas inflow
rates $\dot{M}_{\rm in} = 1.33 \, M_{\hi}/t_{\rm in}$
(the factor 1.33 takes the contribution of Helium into account)
and found that they are about 1 order of magnitude higher than
the current SFRs \citep[from][]{McQuinn2010a}: for NGC~5253
$\dot{M}_{\rm in}\simeq1.5$ M$_{\odot}$~yr$^{-1}$ and SFR$\simeq$0.16
M$_{\odot}$~yr$^{-1}$, while for UGC~6456 $\dot{M}_{\rm in}\simeq0.3$
M$_{\odot}$~yr$^{-1}$ and SFR$\simeq$0.02 M$_{\odot}$~yr$^{-1}$.
Similar results can be derived for I~Zw~18 and UGC~4483, which
show a global radial motion superimposed on a regularly rotating
\hi disk \citep[][]{Lelli2012a, Lelli2012b}: for I~Zw~18
$\dot{M}_{\rm in}\simeq1$ M$_{\odot}$~yr$^{-1}$ and SFR$\simeq$0.1
M$_{\odot}$~yr$^{-1}$, while for UGC~4483 $\dot{M}_{\rm in}\simeq0.1$
M$_{\odot}$~yr$^{-1}$ and SFR$\simeq$0.01 M$_{\odot}$~yr$^{-1}$.
If the hypothesis of a radial inflow is correct, these results
would imply that only $\sim$10$\%$ of the inflowing gas is converted
into stars, in line with several estimates of the star-formation
efficiencies in dwarf galaxies \citep[e.g.,][]{Leroy2008}.

\section{Mass models}

\subsection{Preliminary considerations \label{sec:prelimi}}

In Sect.~\ref{sec:RegRotBCD} we derived rotation curves for nine BCDs
with a regularly rotating \hi disk, while in Sect.~\ref{sec:DistRotBCD}
we estimated the outer rotation velocities of seven BCDs with a
kinematically disturbed \hi disk. In several cases, we found that
the values for the kinematical center, PA, and $i$ do not coincide
with the optical ones (see Table~\ref{tab:param}). Non-circular
motions are also present in the \hi disks of several galaxies.
This raises the question as to whether the \hi disks are in a
fully-stable configuration and the observed rotation velocities
are suitable to investigate the mass distributions in these galaxies.

\begin{figure}
\centering
\includegraphics[width=8.5 cm]{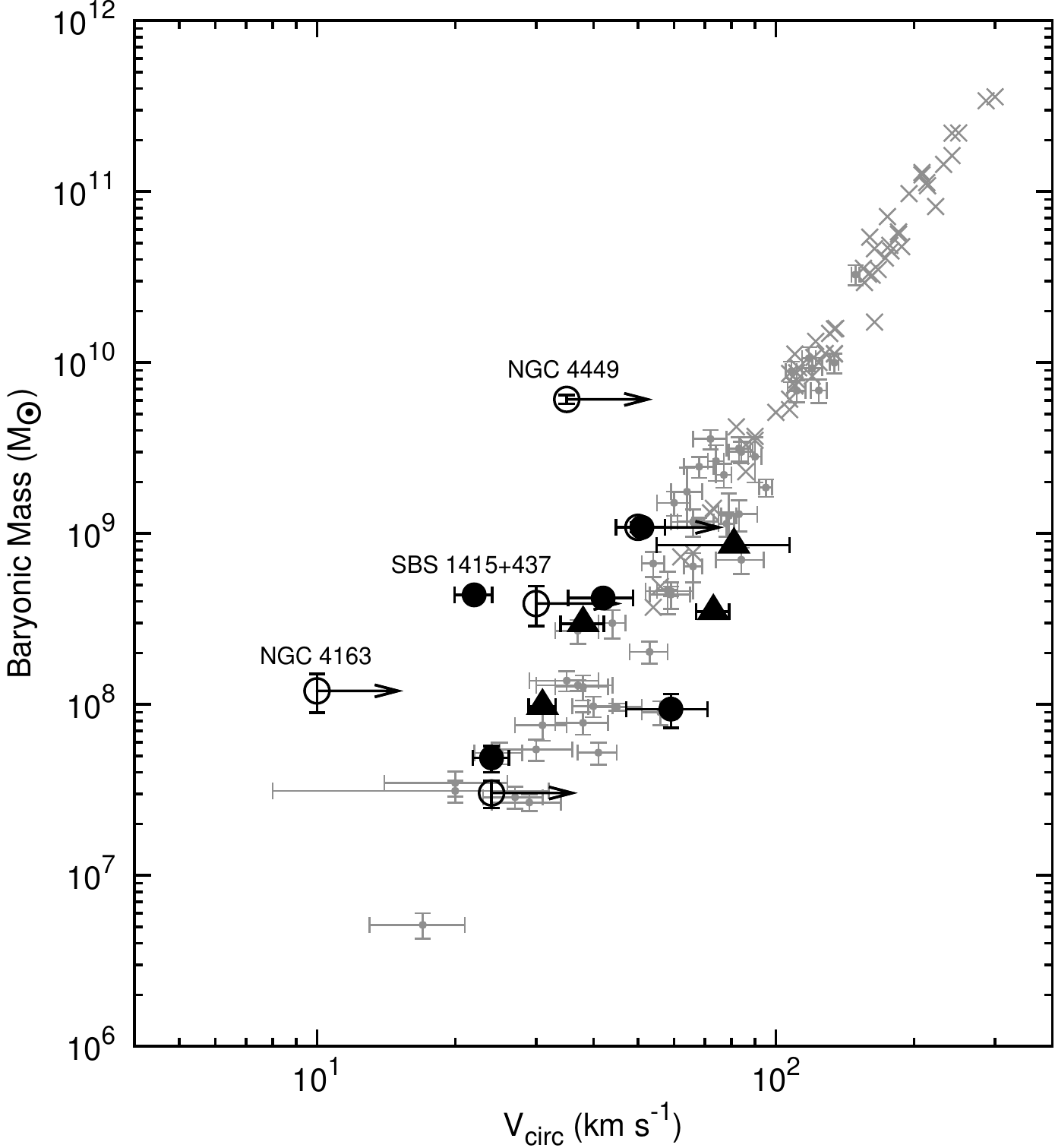}
\caption{The location of starbursting dwarfs on the baryonic Tully-Fisher
relation. Grey-crosses indicate star-dominated galaxies from \citet{McGaugh2005},
while grey-dots indicate gas-dominated galaxies from \citet{McGaugh2011}.
Filled and open symbols indicate, respectively, BCDs with a regularly rotating
\hi disk and BCDs with a kinematically disturbed \hi disk. For the latter
ones, the circular velocity may be underestimated. The triangles indicate
galaxies for which the baryonic mass is a lower limit. Galaxies that
significantly deviate from the relation are labeled.}
\label{fig:TFrel}
\end{figure}
\begin{figure*}
\centering
\includegraphics[width=18 cm]{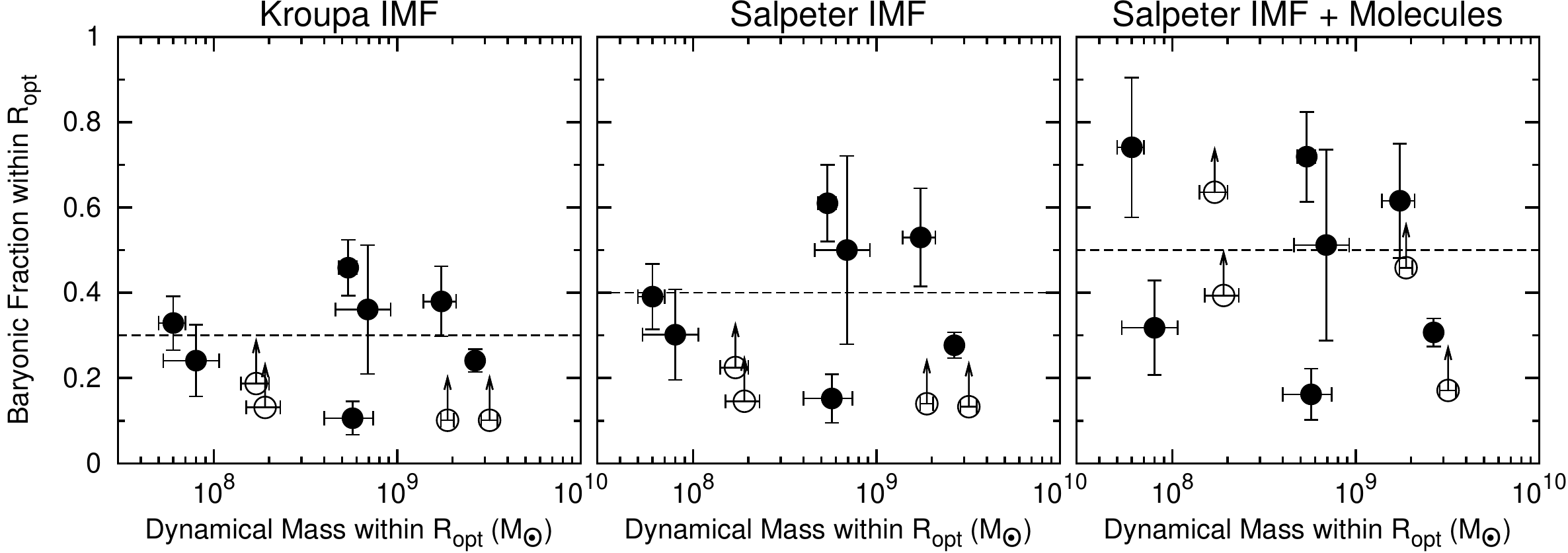}
\caption{Baryonic fractions $f_{\rm bar} = M_{\rm bar}/M_{\rm dyn}$
versus $M_{\rm dyn}$ calculated within the optical radius $R_{\rm opt}$.
The open circles indicate lower limits for $f_{\rm{bar}}$. The dashed
lines indicate the average value of $f_{\rm bar}$ for the different
assumptions on $M_{\rm bar}$. See Sect.~\ref{sec:barFrac} for details.}
\label{fig:bar}
\end{figure*}
In Fig.~\ref{fig:TFrel}, we consider the location of the starbursting
dwarfs in our sample on the baryonic Tully-Fisher relation \citep[BTFR,]
[]{McGaugh2000}. We exclude UGC~6456 and NGC~5253 as their \hi kinematics
seems to be dominated by radial motions (Figs.~\ref{fig:u6456} and
\ref{fig:n5253}), and UGCA~290 and UGC~6541 as they have unsettled
\hi distributions. We use data from \citet{McGaugh2005} for star-dominated
galaxies (grey crosses) and \citet{McGaugh2012} for gas-dominated
ones (grey dots). Most BCDs with a regularly rotating \hi disk have
flat rotation curves in the outer parts (Figs.~\ref{fig:PVmodels},
\ref{fig:n4214}, and \ref{fig:n1705}), thus we use the mean value
of $V_{\rm circ}$ along the flat part \citep{Verheijen2001b}.
In two cases (NGC~4068 and NGC~6789), however, the rotation curve does
not reach the flat part, thus we use the maximum observed value of
$V_{\rm circ}$. One would expect these galaxies to deviate from
the BTFR if $V_{\rm circ}$ keeps rising after the last measured
point. For BCDs with a kinematically disturbed \hi disk, our
estimates of $V_{\rm circ}$ are considered as lower limits in
Fig.~\ref{fig:TFrel} because i) they may not correspond to the
flat part of the rotation curve, and ii) they have not been corrected
for pressure support. As in \citet{McGaugh2005, McGaugh2012}, the
total baryonic mass is calculated as $M_{*}+1.33 M_{\hi}$; the
possible contributions of molecular and ionized gas are neglected.
Note that \citet{McGaugh2005, McGaugh2012} estimated $M_{*}$ using
integrated colors and synthetic stellar population models, while our
estimate of $M_{*}$ is based on the HST studies of the resolved
stellar populations.

Fig.~\ref{fig:TFrel} shows that both BCDs with a regularly rotating
\hi disk (filled symbols) and BCDs with a kinematically disturbed
\hi disk (open symbols) follow the BTFR within the uncertainties, 
except for NGC~4163, NGC~4449, and SBS~1415+437. Thus, for the
majority of galaxies in our sample, the observed rotation velocity
is a reasonable proxy for the total dynamical mass within the
\hi radius. The three objects that strongly deviate from the BTFR
may be unusual for the following reasons. NGC~4163 has a disturbed
\hi distribution with tails and plumes; it is unclear whether the
observed velocity gradient (of only $\sim$10 km~s$^{-1}$) is really
due to rotation. NGC~4449 is characterized by two counter-rotating
gas systems (see VF in Appendix \ref{app:Atlas} and \citealt{Hunter1999b});
we estimated the rotation velocity of the \textit{inner} \hi disk,
that possibly does not correspond to the asymptotic velocity along
the flat part of the rotation curve. It is unclear whether the
\textit{outer} gas system forms a rotating disk; intriguingly, its
inferred circular velocity would be consistent with the BTFR.
Finally, SBS~1415+437 has a kinematic center that is strongly
off-set ($\sim$800 pc) with respect to the optical center and to
the centroid of the \hi distribution; the lopsided \hi morphology
and kinematics may be explained by a pattern of elliptical orbits
viewed almost edge-on \citep[cf.][]{Baldwin1980}. A detailed
investigation of these models is beyond the scope of this paper,
but it is clear that the observed rotation curve might not be
a reliable tracer of the dynamical mass.

Considering the uncertainties involved, we proceed as follows.
In Sect.~\ref{sec:barFrac} we estimate global baryonic fractions
within $R_{\rm{opt}}$ for the 14 objects in Fig.~\ref{fig:TFrel},
while in Sect.~\ref{sec:rotDec} we build detailed mass models
for four galaxies with a regularly rotating disk (NGC~2366, NGC~6789,
NGC~4068, and UGC~4483). In \citet{Lelli2012a, Lelli2012b},
we presented similar mass models for I~Zw~18 and UGC~4483,
respectively. For I~Zw~18, the stellar mass from the HST studies
is very uncertain, thus we built mass models assuming a maximum-disk.
For UGC~4483, instead, we used the HST information on the stellar
mass to break the ``disk-halo degeneracy'' \citep{vanAlbada1986};
here we extend our previous analysis on UGC~4483 by making different
assumptions about the IMF and including the gravitational contribution
of molecules. We do not decompose the rotation curves of NGC~1705,
I~Zw~36, and SBS~1415+437 because the optical and kinematical centers
show a strong off-set ($\gtrsim 1$ disk scale length), thus it is not
possible to calculate the gravitational contributions of stars, gas,
and DM using a common dynamical center. We also exclude NGC~4214
because the \hi disk is close to face-on and strongly warped.

\subsection{Baryonic fractions \label{sec:barFrac}}

As we mentioned in Sect. \ref{sec:sample}, the HST studies of
the resolved stellar populations provide a direct estimate of
the total stellar mass of a galaxy. For seven objects, however,
these stellar masses are \textit{lower limits} because either
the HST field of view only covers a small portion of the galaxy
(NGC~1705, NGC~4214, and NGC~625), or the ancient SFH ($>$1 Gyr)
is uncertain due to the limited photometric depth (I~Zw~18 and
UGCA~290), or both (I~Zw~36 and UGC~6541). Moreover, the values
of $M_{*}$ depend on the assumed IMF and on the mass returned
to the ISM by stellar ejecta. The stellar masses in
Table~\ref{tab:sample} are calculated assuming a Salpeter
IMF from 0.1 to 100 M$_{\odot}$ and a gas-recycling efficiency
of 30$\%$; a Kroupa IMF would give stellar masses lower by a
factor of 1.6 \citep{McQuinn2012b}. In Appendix~\ref{app:tables}
(Table~\ref{tab:dynamics}), we provide three different estimates
for the baryonic mass within $R_{\rm opt}$ (in order of increasing mass):
i) $M^{\rm Kr}_{\rm bar} = M^{\rm Kr}_{*} + 1.33 \, M_{\hi}(R_{\rm opt})$,
where $M^{\rm Kr}_{*}$ is the stellar mass assuming a Kroupa IMF
and $M_{\hi}(R_{\rm opt})$ is the \hi mass within $R_{\rm opt}$;
ii) $M^{\rm Sal}_{\rm bar} = M^{\rm Sal}_{*} + 1.33 \, M_{\hi}
(R_{\rm opt})$, where $M^{\rm Sal}_{*}$ is the stellar mass assuming
a Salpeter IMF; and iii) $M^{\rm mol}_{\rm bar} = M^{\rm Sal}_{*}
+ 1.33 \, M_{\hi}(R_{\rm opt}) + M_{\rm mol}$, where we also include
an indirect estimate of the molecular mass $M_{\rm mol}$.

The molecular content of dwarf galaxies is very uncertain
as they usually have low metallicities and the CO-line,
which traces the molecular hydrogen, is often undetected
\citep[e.g.,][]{Taylor1998}. Moreover, even when the CO-line
is detected, the conversion factor from CO luminosity to
H$_{2}$ mass is poorly constrained, as it may differ from
the Milky-Way value and vary with metallicity and/or other
physical conditions \citep[e.g.,][]{Boselli2002}. Thus, we
chose to indirectly estimate the molecular mass by using
the correlation between SFR surface density $\Sigma_{\rm{SFR}}$
and H$_{2}$ surface density $\Sigma_{\rm{H}_{2}}$ \citep[e.g.,]
[]{Bigiel2008}. In particular, we use Eq.~28 of \citet{Leroy2008},
which assumes that the star-formation efficiency in dwarf galaxies
is the same as in spirals \citep[but see also][]{Roychowdhury2011}.
This equation can be written as:
\begin{equation}\label{eq:mol}
M_{\rm{H_{2}}} [M_{\odot}] = 1.9 \times 10^{9} \, \rm{SFR} [M_{\odot} yr^{-1}].
\end{equation}
We use the average SFR over the last $\sim$10 Myr as obtained
by the HST studies, and assume a systematic uncertainty of
30$\%$ on $M_{\rm{H_{2}}}$. The real uncertainty, however,
may be larger as starbursting dwarfs may deviate from the
$\Sigma_{\rm{H}_{2}}-\Sigma_{\rm{SFR}}$ relation. 

\begin{table*}[t]
\caption{Results of the rotation curve decompositions.}
\centering
\begin{tabular}{l |c c c |c c c |c c c}
\hline
\hline
Galaxy    & \multicolumn{3}{|c|}{$F_{\rm bar}(2.2\,R_{\rm{d}})$} & \multicolumn{3}{c|}{$\rho_{0}$}              & \multicolumn{3}{c}{$r_{\rm{c}}$}\\
          & \multicolumn{3}{|c|}{}                       & \multicolumn{3}{c|}{(10$^{-3}$ M$_{\odot}$ pc$^{-3}$)} & \multicolumn{3}{c}{(kpc)}\\
          & Kroupa & Sal. & Sal.+ Mol.                   & Kroupa   & Sal.     & Sal.+ Mol.                    & Kroupa & Sal. & Sal.+ Mol.\\
\hline
NGC~2366  & 0.5    & 0.6  & 0.6                         & 37       & 34       & 31                            & 1.2    & 1.3  & 1.3\\
NGC~4068  & 0.9    & 1.2  & 1.3                         & 14       & 7        & 3                             & \textit{1.8}& \textit{1.8}&\textit{1.8}\\
NGC~6789  & 0.4    & 0.5  & 0.5                         & 450      & 397      & 384                           & \textit{0.7}& \textit{0.7}&\textit{0.7}\\
UGC~4483  & 0.6    & 0.7  & 0.9                         & 122      & 87       & 15                            & 0.3    & 0.4  & 2.0\\
\hline
\hline
\end{tabular}
\label{tab:rotDec}
\tablefoot{For NGC~4068 and NGC~6789, the values of $r_{\rm c}$ are unconstrained, thus we assumed $r_{\rm c} = R_{\rm opt}$ (in italics).}
\end{table*}
We used the three different estimates of the total baryonic mass to
calculate baryonic fractions $f_{\rm bar} = M_{\rm bar}/M_{\rm dyn}$
within $R_{\rm opt}$. Assuming a spherical mass distribution,
the dynamical mass within $R_{\rm opt}$ is given by $M_{\rm dyn}
(R_{\rm opt}) = V^{2}_{\rm circ}(R_{\rm opt}) \times R_{\rm opt}/G$,
where $V_{\rm circ}(R_{\rm opt})$ is the asymmetric-drift-corrected
circular velocity at $R_{\rm opt}$. We estimated the errors on
$M_{\rm dyn}(R_{\rm opt})$ considering i) the random errors $\delta
_{V_{\rm circ}}$ on $V_{\rm circ}$ due to possible kinematic asymmetries
and/or non-circular motions, and ii) the errors $\delta_{i}$ on the
inclination angle, which would cause a systematic shift of all the
velocity-points but preserve the overall shape of the rotation curve
(unless the \hi disk is warped). Considering that $V_{\rm circ} \simeq
V_{\rm los}/\sin(i)$, the error on $M_{\rm dyn}$ is thus given by
\begin{equation}
\delta_{M_{\rm dyn}} = \bigg\{\dfrac{2 V_{\rm circ} R}{G}\sqrt{
\delta_{V_{\rm circ}}^{2} + \bigg[\dfrac{V_{\rm circ}\delta_{i}}{\tan(i)} \bigg]^{2}}\bigg\}_{R=R_{\rm opt}}
\end{equation}
The resulting dynamical masses and baryonic fractions are listed in
Table~\ref{tab:dynamics}. The errors on $f_{\rm bar}$ have been
calculated by propagating the various uncertainties on $M_{*}$,
$M_{\rm gas}(R_{\rm opt})$, $M_{\rm mol}$, and $M_{\rm dyn}(R_{\rm opt})$.
The three galaxies that deviate from the BTFR (NGC~4163, NGC~4449,
and SBS~1415+437) have unphysically large baryonic fractions
($f_{\rm bar}\gtrsim$1), further suggesting that the observed
circular velocities are not adequate to trace the dynamical
mass. For the remaining 11 objects, we plot $f_{\rm bar}$
against $M_{\rm dyn}$ in Fig.~\ref{fig:bar}. We find no clear
trend with $M_{\rm dyn}$ or with other physical parameters
such as $M_{*}$, $M_{\hi}$, SFR$_{\rm p}$, and $t_{\rm p}$
(see Table~\ref{tab:sample}). The mean baryonic fractions
within the stellar component are relatively high: $\sim$0.3
for a Kroupa IMF, $\sim$0.4 for a Salpeter IMF, and $\sim$0.5
for a Salpeter IMF plus molecules. Old stars (with ages $>$1~Gyr)
generally provide the major contribution to these baryonic
fractions, except for a few cases where either the atomic gas
dominates (NGC~2366) or the molecular gas may be very important
(NGC~1705 and UGC~4483, see Table~\ref{tab:dynamics}).

\subsection{Rotation curve decompositions \label{sec:rotDec}}

\begin{figure*}
\centering
\includegraphics[width=18.5 cm]{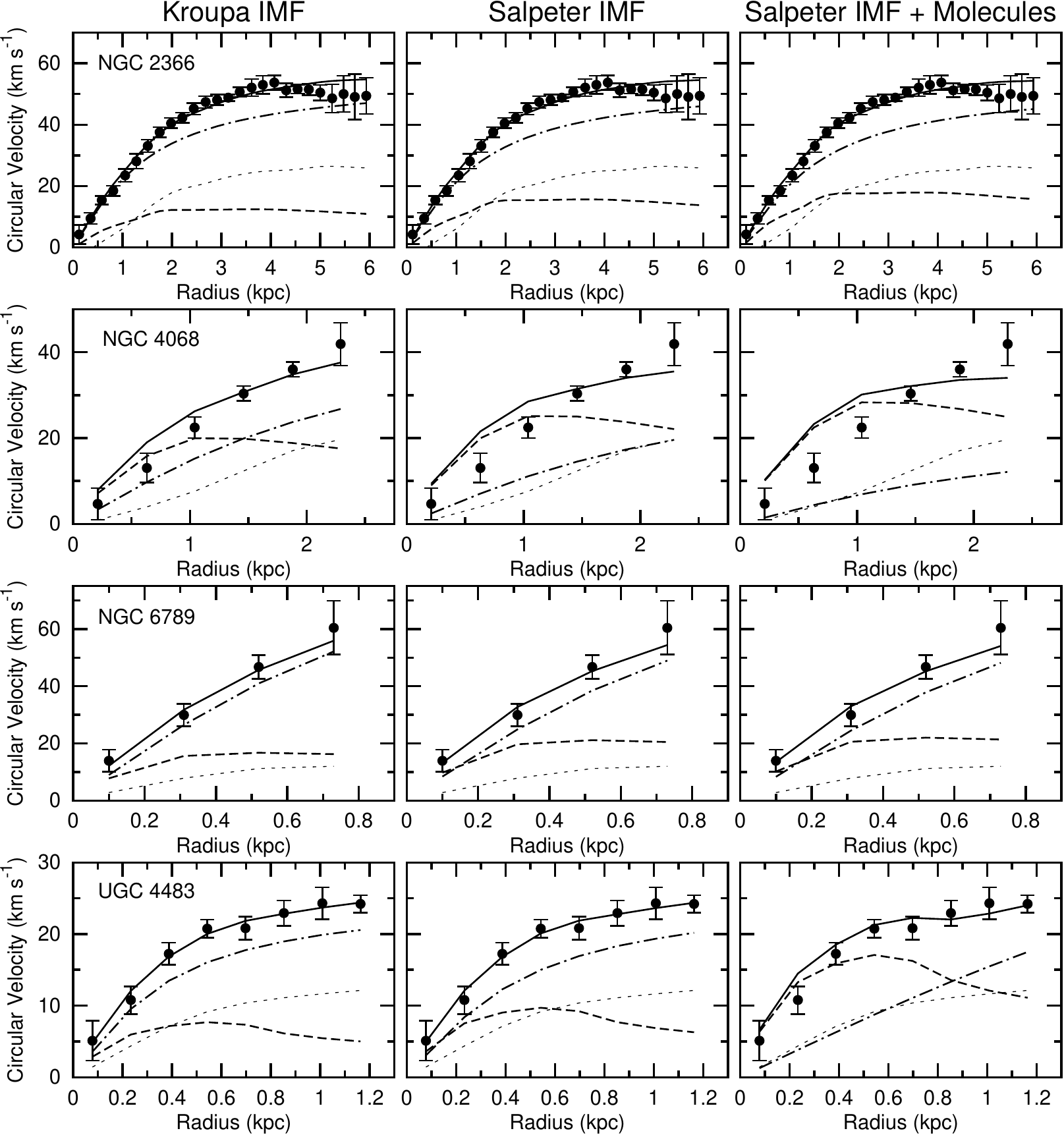}
\caption{Rotation curve decompositions. Dots show the
observed rotation curve (corrected for asymmetric-drift).
Long-dashed, short-dashed, and dot-dashed lines show
the gravitational contribution due to stars, atomic gas,
and dark matter, while the solid line shows the resulting 
total contribution to the rotation curve.
See Sect.~\ref{sec:rotDec} for details.}
\label{fig:rotDec}
\end{figure*}

In this section, we decompose the rotation curves of four galaxies
(NGC~2366, NGC~4068, NGC~6789, and UGC~4483), which have a
regularly rotating \hi disk centered on the stellar component.
We follow standard procedures described by \citet[][]{Begeman1987}.
Similarly to Sect.~\ref{sec:barFrac}, we compute three different
mass models which assume i) a Kroupa IMF; ii) a Salpeter IMF;
and iii) a Salpeter IMF plus the molecular mass inferred by
Eq.~\ref{eq:mol}. These mass models are shown in Fig.~\ref{fig:rotDec}.

The gravitational contribution of the atomic gas was
calculated using the \hi surface density profiles and
scaled to the total atomic gas mass as $\sqrt{1.33 M_{\hi}}$.
In agreement with the models in Sect.~\ref{sec:rotcur},
we assume that the gaseous disk has an exponential
vertical density distribution with a scale height
of 100~pc.

The gravitational contribution of the stellar component
$V_{*}(R)$ was calculated using $R$-band surface brightness
profiles and assuming that the stars are located in a disk
with a vertical density distribution given by ${\rm sech}^{2}(z/z0)$
\citep{vanderKruit1981} with $z_{0} = 0.3 R_{\rm{d}}$.
If one assumes that $z_{0}/R_{\rm d} = 0.2$ or 0.4,
$V_{*}(R)$ would change by a few percent in the inner
parts ($R \lesssim R_{\rm opt}$). The amplitude of $V_{*}$
was scaled to the total stellar mass as $\sqrt{M_{*}}$
(assuming either a Kroupa or a Salpeter IMF). This is
equivalent to using a stellar mass-to-light ratio that
is constant with radius (see Tab.~\ref{tab:sample}).

To include the possible contribution of molecular gas,
we simply assume that molecules are distributed in the
same way as the stars and thus scale the amplitude of $V_{*}$
by $\sqrt{M^{\rm Sal}_{*}+M_{\rm mol}}$. We also tried
to estimate the shape of the molecular gas contribution
$V_{\rm mol}$ using H$\alpha$ and 24$\mu$m images
(from \citealt{Dale2009} and \citealt{GilDePaz2003}).
These images show very clumpy and asymmetric distributions,
that are completely dominated by strong star-forming regions
and shell-like structures. We thus prefer to include the
molecular gas contribution in $V_{*}(R)$. We note that
$B$-band and UV images may be better tracers of the
molecular gas distribution than the $R$-band images
used here. However, since the $B-R$ color profiles of
BCDs typically show a sharp gradient at small radii
and a flat part in the outer regions \citep[e.g.,][]
{Papaderos2002, GilDePaz2005}, the use of $B$-band or
UV surface brightness profiles would lead to appreciable
differences in $V_{*}(R)$ only at the innermost radii
and have no significant effects on our general results.

For the DM distribution, we assume a pseudo-isothermal halo
described by equation
\begin{equation} \label{eq:ISO}
 \rho_{\rm{ISO}}(r) = \frac{\rho_{\rm{0}}}{1 + (r/r_{\rm{c}})^{2}} ,
\end{equation}
where the central density $\rho_{\rm{0}}$ and the core radius
$r_{\rm{c}}$ are the free parameters of the mass model. This
is motivated by the fact that observed rotation curves of
low-mass galaxies are well-reproduced by cored DM density
profiles \citep[e.g.,][]{Gentile2005, Oh2011a}. Since the
\hi rotation curves are not well sampled (except for NGC~2366),
the parameters of the DM halo cannot be determined with high
accuracy. In particular, for NGC~6789 and NGC~4068, the core
radius is completely unconstrained, thus we assumed $r_{\rm c}
= R_{\rm opt}$. Given these uncertainties, we did not explore
other DM density profiles than the pseudo-isothermal one.

A baryonic disk is usually defined to be maximum if $F_{\rm bar}
=V_{\rm bar}/V_{\rm circ} = 0.85 \pm 0.10$ at 2.2 disk scale
lengths \citep{Sackett1997, Bershady2011}, where $V_{\rm{bar}}$
is the contribution to the rotation curve given by the baryons.
The BCDs in Fig.~\ref{fig:rotDec} have $F_{\rm bar} \simeq 0.4$
to 0.6, except for NGC~4068 that has $F_{\rm bar} \simeq 0.9$
for a Kroupa IMF and $F_{\rm bar} > 1$ for a Salpeter IMF (see
Table~\ref{tab:rotDec}). The latter result may suggest that a
Salpeter IMF implies an unphysical, over-maximal disk for NGC~4068;
however, if one assumes a nearly spherical stellar body with
a scale height of $\sim$600~pc, a Salpeter IMF would give
acceptable results with $F_{\rm bar} \lesssim 1$. The sub-maximal
disks of NGC~2366, NGC~6789, and UGC~4483 are in line with the
results of the DiskMass survey \citep[e.g.,][]{Bershady2011,
Westfall2011, Martinsson2011}, who measured the stellar velocity
dispersions of a sample of spiral galaxies and found that
$F_{\rm bar} \simeq 0.5$. Note, however, that baryons are still
dynamically significant, as $F_{\rm bar} \simeq 0.4$ to 0.6
correspond to baryonic fractions $f_{\rm bar} = M_{\rm bar}/M_{\rm dyn}
\simeq 0.2$ to 0.4 within 2.2~$R_{\rm d}$, in line with the
results in Sect.~\ref{sec:barFrac}.

\section{Discussion}

\subsection{Comparison with other dwarf galaxies \label{sec:compa}}

In Sect.~\ref{sec:HIdistr}, we compared the \hi distribution
and kinematics of BCDs and Irrs. In agreement with previous
studies \citep[e.g.,][]{vanZee1998b, vanZee2001}, we found that
BCDs have central \hi densities a factor of $\sim$2 higher
than typical Irrs. The average extent of the \hi disk with
respect to the stellar component, instead, is similar for
BCDs, Irrs, and spirals ($R_{\hi}/R_{\rm opt}\simeq1.7$
with $R_{\rm opt}$ defined as 3.2 exponential scale lengths
$R_{\rm d}$). We also found that complex \hi kinematics are much
more common in BCDs ($\sim$50$\%$) than in typical Irrs ($\sim$10$\%$),
likely due to the effects of stellar feedback and/or of the
triggering mechanism (interactions/mergers or disk instabilities).
In \citet{Lelli2014}, we present a comparison between the
rotation curves of BCDs and those of Irrs, and discuss the
link between the starburst, the gas concentration, and the
gravitational potential \citep[see also][]{Lelli2012a,
Lelli2012b}. Here we compare the baryonic fractions of BCDs
with those of gas-rich Irrs and gas-poor spheroidals (Sphs).

For gas-rich Irrs, maximum-disk decompositions of \hi rotation
curves usually result in high values of the stellar mass-to-light
ratio, up to $\sim$15 in the $R$-band \citep[e.g.,][]{Swaters2011}.
These high values of $M_{*}/L_{\rm R}$ are difficult to explain
using stellar population synthesis models \citep[e.g.,][]{Zibetti2009},
suggesting that Irrs are dominated by DM at all radii. The detailed
baryonic fractions of Irrs, however, remain uncertain because they
depend on the assumed value of $M_{*}/L_{\rm R}$. For the BCDs in
our sample, instead, we can directly calculate the values of $M_{*}
/L_{\rm R}$ using the stellar masses from the HST studies of
the resolved stellar populations. Assuming a Salpeter IMF from
0.1 to 100 M$_{\odot}$ and a gas-recycling efficiency of 30$\%$,
we find that the mean value of $M_{*}/L_{\rm R}$ is $\sim$1.5
(see Table~\ref{tab:sample}). Photometric studies of BCDs \citep[e.g.,]
[]{Papaderos1996} suggest that the starburst typically increases
the total luminosity by a factor of $\sim$2, whereas studies of
the SFHs \citep[e.g.,][]{McQuinn2010b} indicate that the burst
only produces a small fraction of the total stellar mass ($\sim$10$\%$).
Thus, it is reasonable to assume that Irrs have, on average,
$M_{*}/L_{\rm R}\simeq3$ for a Salpeter IMF and $M_{*}/L_{\rm R}
\simeq2$ for a Kroupa IMF.

Similarly to Sect.~\ref{sec:barFrac}, we calculated the baryonic
fractions of 30 gas-rich dwarfs using the $R$-band luminosities,
\hi surface density profiles, and \hi rotation curves from
\citet{Swaters2002a, Swaters2009}, and assuming $M_{*}/L_{\rm R}=2$.
We only considered galaxies with $i>$30$^{\circ}$ and high-quality
rotation curves ($q>2$, see \citealt{Swaters2009}), that are
traced out to $\sim$3~$R_{\rm d}$ and have $V_{\rm rot}<100$
km~s$^{-1}$ at the last measured point. We also calculated the
baryonic fractions of several gas-poor dwarfs using the $V$-band
luminosities and dynamical masses from \citet{Wolf2010}, and
assuming $M_{*}/L_{\rm V}=2.0$. For the nearby Sphs Sculptor
and Fornax, the stellar masses from \citet{deBoer2012a,
deBoer2012b} imply, respectively, $M_{*}/L_{\rm V}\simeq2.2$
and $M_{*}/L_{\rm V}\simeq1.7$ (assuming a Kroupa IMF
and a gas-recycling efficiency of 30$\%$). We only considered
nine Sphs that have accurate estimates of the stellar velocity
dispersion: the eight ``classical'' satellites of the Milky Way
(Carina, Draco, Fornax, Leo~I, Leo~II, Sculptor, Sextans,
and Ursa Minor) and NGC~185, which is a satellite of M31.
For these Sphs, the baryonic fractions are computed at the
3D deprojected half-light radius $r_{1/2}$ (for an exponential
density profile $r_{1/2} \simeq 2.2 R_{\rm d}$, see \citealt{Wolf2010}),
thus they may be slightly overestimated with respect to those
of Irrs and BCDs (computed at 3.2 $R_{\rm d}$), since the DM
contribution is expected to increase at larger radii.

\begin{figure}
\centering
\includegraphics[width=8.5 cm]{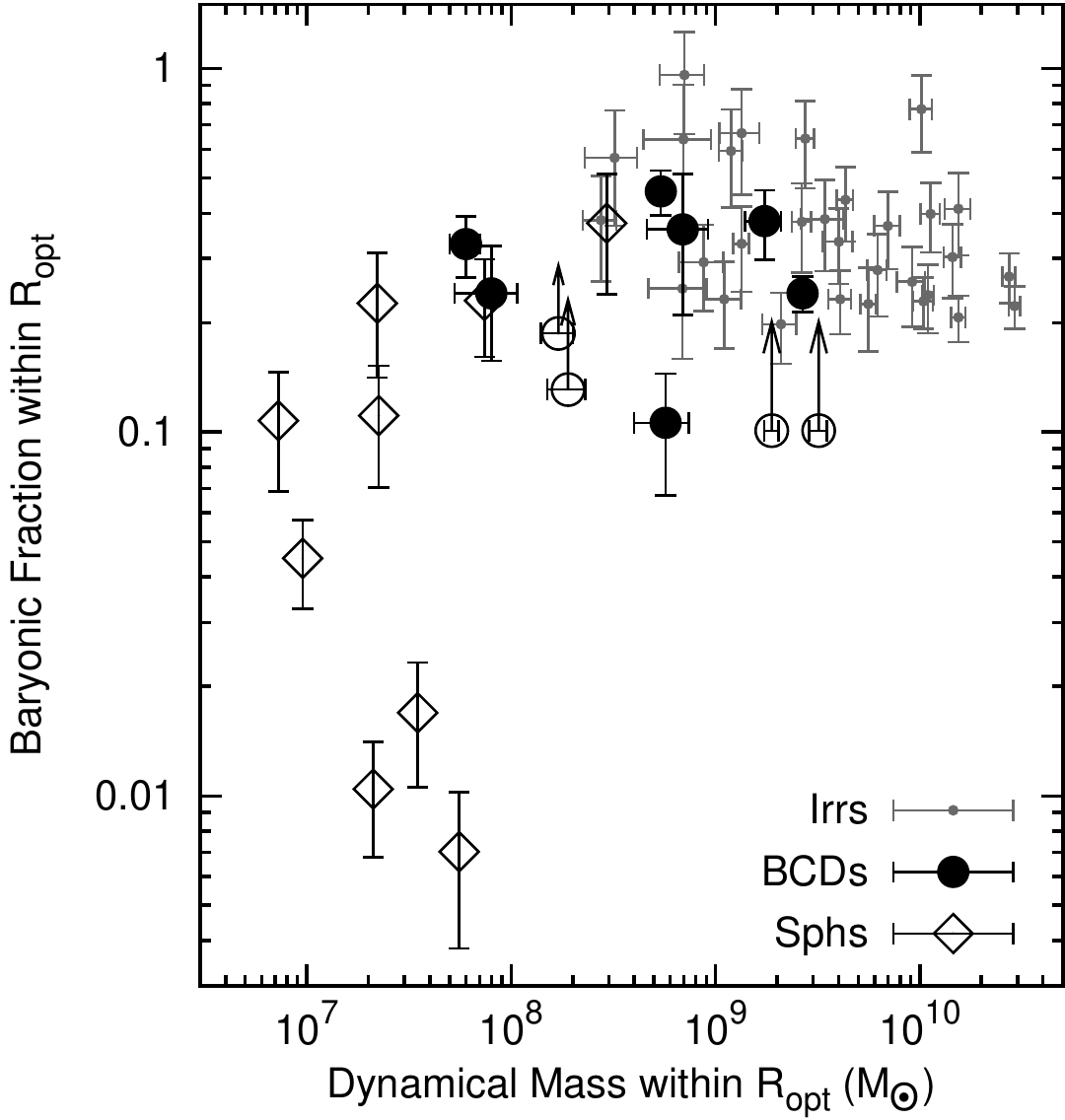}
\caption{Comparison between the baryonic fractions of BCDs (open
and filled circles, same as Fig.~\ref{fig:bar}, left panel), Irrs
(grey dots), and Sphs (open diamonds). The data for Irrs and Sphs are
taken from \citet{Swaters2009} and \citet{Wolf2010}, respectively.}
\label{fig:barTot}
\end{figure}
Figure~\ref{fig:barTot} shows that starbursting dwarfs (open and
filled circles) have baryonic fractions comparable with those of
typical Irrs (grey dots) and of some Sphs (open diamonds).
The Sphs with extremely low baryonic fractions ($f_{\rm bar}<0.1$)
are Carina, Draco, Sextans and Ursa Minor, that are very close
to the Milky Way and may have suffered from environmental
effects \citep[e.g.,][]{Mayer2006, Gatto2013}. We conclude that
the baryonic content of BCDs is similar to that of other types
of dwarf galaxies, except for some low-luminosity satellities of
the Milky Way.

\begin{figure*}
\centering
\includegraphics[width=17.5 cm]{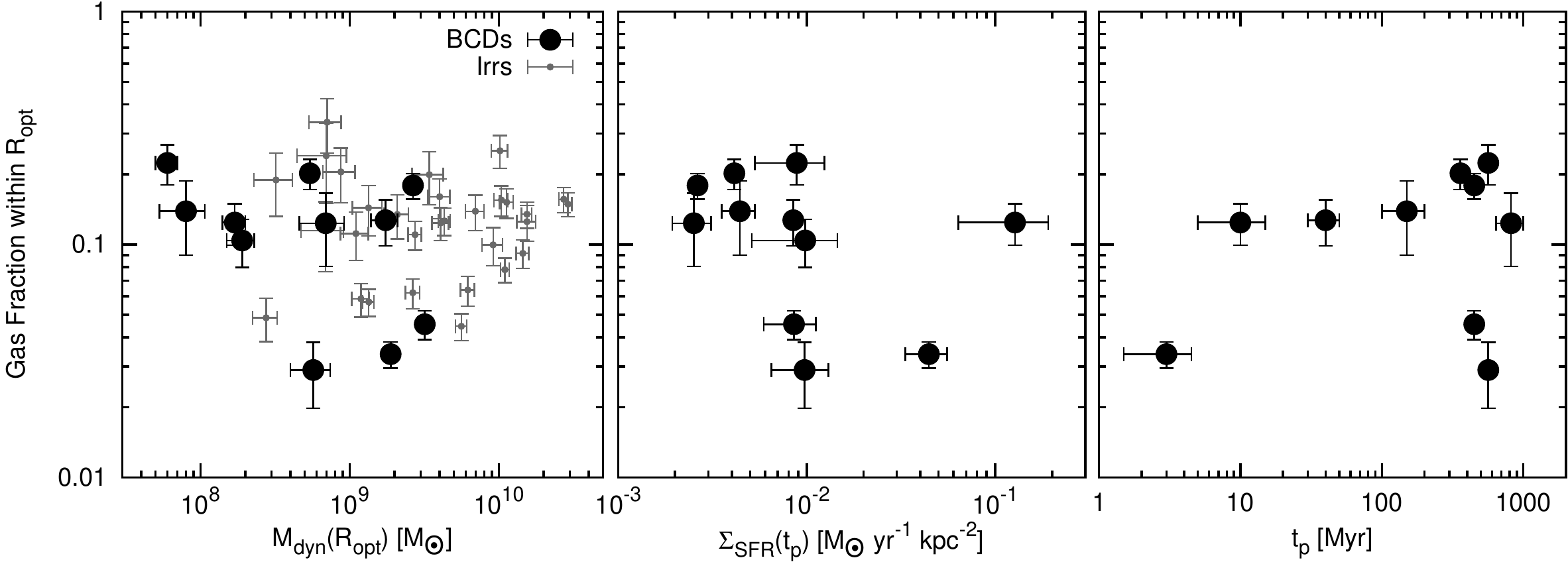}
\caption{\textit{Left}: the atomic gas fractions $f_{\rm gas} = 1.33
\, M_{\hi}/M_{\rm dyn}$ versus $M_{\rm dyn}$ calculated within the
optical radius $R_{\rm opt}$. Large, black dots and small, grey
dots indicate BCDs (this work) and Irrs \citep{Swaters2009},
respectively. \textit{Middle}: $f_{\rm gas}$ versus the SFR
surface density $\Sigma_{\rm SFR}(t_{\rm p}) = {\rm SFR_{p}}/
(\pi R_{\rm opt}^{2})$ where SFR$_{\rm p}$ is the peak SFR over
the last 1 Gyr (see Table~\ref{tab:sample}). \textit{Right}:
$f_{\rm gas}$ versus the look-back time $t_{\rm p}$ at
SFR$_{\rm p}$.}
\label{fig:gasFrac}
\end{figure*}
\subsection{Stellar feedback and gas outflows \label{sec:outflows}}

Simulations of galaxy formation in a $\Lambda$CDM cosmology
indicate that low-mass galaxies must experience massive gas outflows
at high redshifts: this is necessary to reproduce the observed stellar
and DM properties of dwarf galaxies at $z\simeq0$ \citep[e.g.,][]
{Governato2010, Governato2012, Sawala2012} as well as their number
density \citep[e.g.,][]{Okamoto2010, Sawala2013}. BCDs in the Local
Universe can be considered as good analogues of dwarf galaxies
at high redshifts, since they have small potential wells (with $20
\lesssim V_{\rm circ }\lesssim80$ km~s$^{-1}$) and are currently
experiencing intense star formation, as indicated by their recent
SFHs ($\lesssim$1~Gyr) and high values of the birthrate parameter
$b$ (see Table~\ref{tab:sample} and \citealt{McQuinn2010a}). Moreover,
starbursting dwarfs at $z\simeq0$ have clumpy morphologies that are
remarkably similar to those of more massive star-forming galaxies
at $z\simeq1-2$ (see e.g., Fig.~20 of \citealt{Elmegreen2009}),
further suggesting a similarity with high-$z$ star-forming objects.
While starburst galaxies are common at high redshifts, BCDs are
relatively rare in the Local Universe: \citet{Lee2009a} analyzed
H$\alpha$ observations from the 11HUGS survey \citep{Kennicutt2008}
and estimated that BCDs constitute only $\sim$6$\%$ of the population
of star-forming dwarfs at $z\simeq0$ (but see \citealt{McQuinn2010a}
regarding the limitations of H$\alpha$ observations to identify
starbursting dwarfs). BCDs, therefore, offer us the opportunity
to study in detail the actual importance of stellar feedback and
gas outflows in low-mass DM halos.

We computed atomic gas fractions $f_{\rm gas} = 1.33 \, M_{\hi}
/M_{\rm dyn}$ within $R_{\rm opt}$ for the 11 BCDs with accurate
estimates of $M_{\rm dyn}$ (see Sect.~\ref{sec:prelimi}) and for
the 30 Irrs considered in Sect.~\ref{sec:compa} \citep[from][]
{Swaters2009}. Fig.~\ref{fig:gasFrac} (left panel) shows that both
Irrs and BCDs have relatively high gas fractions ($0.1 \lesssim
f_{\rm gas} \lesssim 0.3$), and only a few objects show gas fractions
$\lesssim 5\%$. Moreover, the gas fractions of BCDs are similar to
those of Irrs. This suggests that either i) BCDs did \textit{not}
expell a large amount of gas out of their potential wells, or ii)
their gas fractions were much higher at the beginning of the
starburst, or iii) the gas expelled has been replenished by
gas inflows. These three hypothesis are discussed in the following.

The possibility that massive gas inflows replenish the presumed
outflowing gas seems unlikely, as we found evidence for radial
motions in only four galaxies (see Sect.~\ref{sec:DistRotBCD} and
\citealt{Lelli2012a, Lelli2012b}). We do not know the direction
of these radial motions, but if we interpret them as inflows, the
inferred gas inflow rates would be \textit{only} $\sim$1 order
of magnitude higher than the current SFRs and, thus, consistent
with a typical efficiency of $\sim$10$\%$ in converting gas
into stars. Moreover, it is likely that these radial motions
are recent and short-lived phenomena, as their typical timescales
are comparable with the orbital times \citep[][]{Lelli2012b}.
For the other BCDs with a regularly rotating disk, any radial
motion must be $\lesssim 5$ km~s$^{-1}$, which gives us a
firm upper-limit to the mean gas inflow rate of $\sim$0.3
$M_{\odot}$~yr$^{-1}$. This seems adequate to feed the current
star formation and build up the central concentration of gas
observed in BCDs (see Fig.~\ref{fig:hiprof}), but leave little
room for massive gas outflows.

In Fig.~\ref{fig:gasFrac} (middle panel), we plot $f_{\rm gas}$
versus the SFR surface density $\Sigma_{\rm SFR}(t_{\rm p}) =
{\rm SFR_{p}}/(\pi R_{\rm opt}^{2})$, where SFR$_{\rm p}$ is the
peak SFR over the past 1 Gyr (see Table~\ref{tab:sample}). The
SFHs of five galaxies (NGC~2366, NGC~4068, UGC~4483, UGC~9128, and
SBS~1415+437) show two distinct peaks with similar SFRs (consistent
within 1$\sigma$). Here we consider the older peak as this is the
one that formed more stars, given that the SFR is averaged over a
larger time-bin (typically a factor of $\sim$4, see \citealt[]
[]{McQuinn2010a}). One may expect that BCDs with higher values
of $\Sigma_{\rm SFR}(t_{\rm p})$ expell a higher fraction of
gas out of their potential well and, thus, might have lower
values of $f_{\rm gas}$. This is \textit{not} observed for the
11 objects considered here. However, in some BCDs the starburst
started only $\sim$10 Myr ago whereas in other ones it started
more than $\sim$500 Myr ago \citep[see e.g.,][]{McQuinn2010a}.
In Fig.~\ref{fig:gasFrac} (right panel), we also plot $f_{\rm gas}$
versus the look-back time $t_{\rm p}$ at SFR$_{\rm p}$, which can be
considered as the typical ``age'' of the starburst. There is also no
clear correlation between these two quantities. Similarly, we found
no clear correlation between $f_{\rm gas}$ and the product $\Sigma_{\rm SFR}
\times t_{\rm p}$ or other parameters that quantify the relative
strength of the burst, such as SFR$_{\rm p}/M_{\rm dyn}$, SFR$_{\rm p}/M_{*}$,
and the birthrate parameter $b$ (see Table~\ref{tab:sample}).
The lack of \textit{any} significant trend between $f_{\rm gas}$
and the starburst properties suggests that BCDs did \textit{not}
eject a large quantity of gas out of their potential wells.

To further investigate the possible effects of outflows, we estimated
the gas masses that might have been expelled from the potential wells
of BCDs by using i) the escape velocities derived from the \hi kinematics,
ii) the burst energies derived from the observed SFHs, and iii) the
feedback efficiencies derived by \citet{MacLow1999} and \citet{Ferrara2000}
using both analytical calculations and high-resolution hydrodynamical
simulations. Adopting Eq.~28 of \citet{Ferrara2000}, the mass-loss
rate $\dot{M}_{\rm out}$ due to stellar feedback is given by
\begin{equation}
\small
\begin{split}
 \dot{M}_{\rm out}=2 \, \xi \, \eta \, E \, \gamma / V_{\rm out}^{2},
\end{split}
\end{equation}
where $\eta$ and $\xi$ are parameters related to the feedback efficiency,
$E$ is the energy produced by supernovae and stellar winds, $\gamma$ is
the energy production rate, and $V_{\rm out}^{2}$ is the velocity of the
outflowing gas that must be higher than the escape velocity $V_{\rm esc}$.
The parameter $\eta$ represents the fraction of injected energy that is
converted into kinetic energy; for radiative bubbles $\eta\lesssim0.03$
\citep{Koo1992, Ferrara2000}. Since part of the kinetic energy accelerates
material in the equatorial plane of the bubble, the parameter $\xi$ corresponds
to the fraction of gas that is expelled from the galaxy almost perpendicular
to the disk. Using hydrodynamical simulations, \citet{MacLow1999} found
that $\xi$ is usually $\lesssim$7$\%$. Following \citet{Ferrara2000},
we here assume that $\eta = 0.03$ and $\xi = 0.07$, but we warn that the
actual values of these parameters are uncertain. Maximum mass-loss rates
occur when $V_{\rm out}=V_{\rm esc}$. Assuming that $V_{\rm esc}$ does
not significantly change during the burst, the maximum mass that can be
expelled from the galaxy is given by
\begin{equation}\label{eq:outmassEnergy}
\small
\begin{split}
M^{\rm{max}}_{\rm{out}} = \int_{t_{i}}^{t_{f}} \dot{M}_{\rm{out}} dt =
2 \, \xi \, \eta \, V_{\rm{esc}}^{-2} \int_{t_{i}}^{t_{f}} E \, \gamma dt
= 2 \, \xi \, \eta \, V_{\rm{esc}}^{-2} \, E_{\rm{burst}},
\end{split}
\end{equation}
where $t_{i}$ and $t_{f}$ are the initial and final times of the burst,
and $E_{\rm burst}$ is the total energy created during the burst.
In Table \ref{tab:outflow}, we list $M^{\rm max}_{\rm out}$ for seven
galaxies in our sample that have both good estimates of $E_{\rm burst}$
and $V_{\rm esc}$. We adopt the values of $E_{\rm burst}$ calculated
by \citet{McQuinn2010b} using the observed SFHs and the evolutionary
synthesis code STARBURST99, which can simulate the energy produced by
supernovae and stellar winds for a given SFR. We calculate $V_{\rm esc}$
as $\sqrt{2} V_{\rm circ}$ at the optical radius. Table~\ref{tab:outflow} 
shows that, for $\eta \times \xi \simeq 2 \times 10^{-3}$,
$M^{\rm max}_{\rm out}$ is very small, less than 10$\%$ of the
current atomic gas mass. Exceptions are UGC~4483 and UGC~9128,
which are among the lowest mass galaxies in our sample, with
rotation velocities of only $\sim$20 km~s$^{-1}$. Only if
one assumes that $\eta \times \xi$ is a few times $10^{-2}$, the
starbursting dwarfs in our sample could have expelled a gas mass
comparable to the current, atomic gas mass.

\begin{table}[t]
\caption{Ouflowing gas masses from Eq.~\ref{eq:outmassEnergy}.}
\centering
\footnotesize
\setlength{\tabcolsep}{4pt}
\begin{tabular}{l c c c c}
\hline
\hline
Galaxy       & $\log(E_{\rm{burst}})$ & $M^{\rm{max}}_{\rm{out}}$ & $<\dot{M}^{\rm max}_{\rm out}>$& $M^{\rm{max}}_{\rm{out}}$/$M_{\rm gas}$\\
             & (erg)  & (10$^{6}$ M$_{\odot}$) & (M$_{\odot}$ yr$^{-1}$) & \\
\hline
NGC~625      & 56.0 & 11.2$\pm$1.9         & 0.025                 & 0.09\\
NGC~1569     & 56.6 & 16.0$\pm$1.6         & 0.036                 & 0.04\\
NGC~2366     & 56.5 & 12.2$\pm$0.5         & 0.027                 & 0.01\\
NGC~4068     & 56.0 & 7.8$\pm$0.4          & 0.017                 & 0.04\\
NGC~4214     & 56.7 & 8.1$\pm$0.4          & 0.010                 & 0.01\\
NGC~6789     & 55.5 & 0.9$\pm$0.1          & 0.002                 & 0.04\\
UGC~4483     & 55.4 & 5.7$\pm$0.5          & 0.007                 & 0.14\\
UGC~9128     & 55.5 & 5.5$\pm$0.9          & 0.004                 & 0.32\\
\hline
\hline
\end{tabular}
\tablefoot{The burst energies are taken from \citet{McQuinn2010b}.
$M_{\rm out}^{\rm max}$ is calculated assuming a feedback efficiency
$\xi \times \eta \simeq 2 \times 10^{-3}$. See Sect.~\ref{sec:outflows}
for details.}
\label{tab:outflow}
\end{table}

Relatively low values of $M_{\rm out}/M_{\rm gas}$ are in line with the
results of both optical and X-ray observations. Studies of the H$\alpha$
kinematics have shown that galactic winds are common in BCDs, but the
velocities of the H$\alpha$ gas are usually smaller than the escape
velocities \citep{Martin1996, Martin1998, vanEymeren2009b, vanEymeren2009a,
vanEymeren2010}, implying that the ionized gas is gravitationally bound
to the galaxy. Similar results have been found by studies of the Na~D
absorption doublet \citep{Schwartz2004}. X-ray observations have revealed
that several starbursting dwarfs have a diffuse coronae of hot gas at
$T\simeq10^{6}$~K, which are likely due to outflows but have low masses,
$\sim$1$\%$ of the current \hi mass \citep{Ott2005a,Ott2005b}. The
observational evidence, therefore, suggests that galactic winds are
common in nearby dwarf galaxies, but they do not expell a significant
fraction of the gas mass out of the potential well.

\section{Summary and conclusions}

We presented a systematic study of the \hi content of 18 starbursting
dwarf galaxies, using both new and archival observations. We only selected
nearby galaxies that have been resolved into single stars by HST,
thus providing information on their total stellar masses. According to
their \hi distribution and kinematics, we classified starbursting dwarfs
into three main families: i) galaxies with a regularly rotating \hi disk
($\sim$50$\%$), ii) galaxies with a kinematically disturbed \hi disk
($\sim$40$\%$), and iii) galaxies with unsettled \hi distributions
($\sim$10$\%$). For galaxies with a regularly rotating \hi disk,
we derived rotation curves by building 3D kinematic models. For
galaxies with a kinematically disturbed \hi disk, we obtained
estimates of the rotation velocities in the outer parts.
Our main results can be summarized as follows:
\begin{enumerate}
 \item We firmly establish that the \hi surface density profiles
of starbursting dwarfs are different from those of typical Irrs.
On average, starbursting dwarfs have central \hi densities a
factor of $\sim$2 higher than typical Irrs.
 \item The average ratio of the \hi radius to the optical radius
(defined as 3.2 exponential scale lengths) is 1.7$\pm$0.5, similar
to the values found for gas-rich spiral and irregular galaxies.
 \item Disturbed \hi kinematics are much more common in starbursting
dwarfs ($\sim$50$\%$) than in typical Irrs ($\sim$10$\%$, see
\citealt{Swaters2009}). This may be related to stellar feedback
and/or to the starburst trigger (interactions/mergers or disk
instabilities).
\item Two galaxies (NGC~5253 and UGC~6456) show a velocity
gradient along the \hi minor axis. We modelled the \hi emission
by a disk dominated by \textit{radial} motions and derived
inflow/outflow timescales of $\sim$100-200 Myr. For NGC~5253,
the radial motions appear to be an inflow and would imply a
gas inflow rate $\sim$1 order of magnitude higher than the
current SFR.
 \item For 11 galaxies with accurate estimates of the circular
velocities, we calculated the baryonic fraction $f_{\rm bar}$
within the optical radius, using the stellar masses from the
HST studies of the resolved stellar populations. We found that,
on average, $f_{\rm bar} \simeq 0.3$ for a Kroupa IMF and
$f_{\rm bar} \simeq 0.4$ for a Salpeter IMF. If molecular
gas is also taken into account, the mean baryonic fraction
may increase up to $\sim$0.5.
 \item For four galaxies with a regularly rotating \hi disk
centered on the stellar component, we decomposed the rotation
curves into mass components. We found that baryons (both
stars and gas) are generally not sufficient to explain the
inner rise of the rotation curve, although they constitute
$\sim20-40\%$ of the total mass at $\sim$2.2 exponential
scale lengths.
\item Despite the star formation having injected $\sim$10$^{56}$~ergs
into the ISM in the last $\sim$500 Myr \citep{McQuinn2010b}, these
starbursting dwarfs have gas fractions comparable with those
of typical Irrs. This suggests that \textit{either} starbursting
dwarfs do not expell a large amount of gas out of their potential
wells, \textit{or} their gas fractions must have been much higher
at the beginning of the burst. The lack of any correlation
between the observed gas fractions and the starburst properties
favors the former scenario.
\end{enumerate}

\begin{acknowledgements}
We are grateful to Renzo Sancisi for sharing insights and ideas
that fueled this work. We thank Eline Tolstoy for stimulating
discussions. We also thank Ed Elson and Angel R. Lopez-Sanchez
for providing us with the \hi datacubes of NGC~1705 and NGC~5253,
respectively, and Polychronis Papaderos for the H$\alpha$-subtracted
HST image of I~Zw~18. We finally thank the members of the WHISP,
THINGS, and LITTLE-THINGS projects for having made the \hi
datacubes publicly available. FL acknowledges the Ubbo Emmius
bursary program of the University of Groningen and the Leids
Kerkhoven-Bosscha Fund. FF acknowledges financial support
from PRIN MIUR 2010-2011, project ``The Chemical and Dynamical
Evolution of the Milky Way and Local Group Galaxies'', prot.
2010LY5N2T.
\end{acknowledgements}

\begin{appendix}

\section{Notes on individual galaxies \label{app:indi}}

\subsection{Galaxies with a regularly rotating \hi disk}

\textbf{NGC~1705} has a strongly warped \hi disk. Our rotation curve rises
more steeply than those of \citet{Meurer1998} and \citet{Elson2013}
because we applied a beam-smearing correction to the inner velocity-points
using 3D disk models (see Fig.~\ref{fig:n1705}). \citet{Meurer1998} and
\citet{Elson2013} decomposed their rotation curves into mass components
and found that DM dominates the gravitational potential at all radii. 
We did \textit{not} build a detailed mass model because the optical
and kinematic centers differ by $\sim$550 pc, while PA$_{\rm opt}$
and PA$_{\rm kin}$ differ by $\sim$45$^{\circ}$.\\
\textbf{NGC~2366} has an extended \hi disk with a strong kinematic distortion
to the North-West (see its velocity field in Appendix~\ref{app:Atlas}). Our
rotation curve is in overall agreement with previous results \citep{Hunter2001,
Thuan2004, Oh2008, Swaters2009, vanEymeren2009b}, but we do not confirm the
declining part of the rotation curve found by \citet{Hunter2001} and
\citet{vanEymeren2009b} at $R\gtrsim5'$. This latter result appears to be
caused by an anomalous \hi cloud that lies at $V_{\rm l.o.s}\simeq130$
km~s$^{-1}$ along the major axis ($\sim7'$ from the galaxy center to the
North, see the PV-diagram in Fig.~\ref{fig:PVmodels} and Appendix~\ref{app:Atlas}).\\
\textbf{NGC~4068} has a \hi distribution characterized by a central depression and
several shell-like structures. The \hi kinematics is slightly lopsided. Our rotation
curve agrees with the one of \citet{Swaters2009} within the errors.\\
\textbf{NGC~4214} has a \hi disk with multiple spiral arms. Intriguingly, the optical
and \hi spiral arms wind in opposite directions (clockwise and counter-clockwise,
respectively). The \hi disk is close to face-on and strongly warped, thus the
rotation curve is uncertain. In the inner parts, our rotation curve rises more
steeply than the one derived by \citet{Swaters2009}; the difference seems to be
due to a different choice of the dynamical center (see the PV-diagram in
\citealt{Swaters2009}). \\
\textbf{NGC~6789} has a compact \hi disk that extends out to only $\sim$3.5 optical
scale lengths. The inclination is uncertain: we derived $i = 43^{\circ} \pm 7^{\circ}$
using 3D disk models. \\
\textbf{UGC~4483} has been studied in \citet{Lelli2012b}.\\
\textbf{I~Zw~18} has been studied in \citet{Lelli2012a}.\\
\textbf{I~Zw~36} has an extended and asymmetric \hi distribution \citep[see][]{Ashley2013},
but in the central parts the \hi forms a compact, rotating disk. The optical and
kinematic centers are offset by $\sim$12$''$ ($\sim$340~pc), while PA$_{\rm opt}$
and PA$_{\rm kin}$ differ by $\sim$36$^{\circ}$.\\
\textbf{SBS~1415+437} is a prototype ``cometary'' BCD, as the starburst region 
is located at the edge of an elongated stellar body. Remarkably, the kinematic
center does not coincide with the optical one but with the starburst region to
the South (see Appendix~\ref{app:Atlas}; the object at R.A.$\simeq$14$^{\rm h}$
17$^{\rm m}$ 00$^{s}$ and Dec.$\simeq$43$^{\circ}$ 29$'$ 45$''$ is a foreground
star). The lopsided \hi distribution and kinematics may be due to a pattern
of elliptical orbits centered on the starburst region \citep[cf.][]{Baldwin1980}.

\subsection{Galaxies with a kinematically disturbed \hi disk}

\textbf{NGC~625} has been previously studied by \citet{Cote2000} and \citet{Cannon2004}.
\citet{Cote2000} suggested that the complex \hi kinematics is due to an interaction/merger,
whereas \citet{Cannon2004} argued that it is best described by a gaseous outflow
superimposed on a rotating disk. We find it difficult to distinguish between these
two possibilities. It is clear, however, that the galaxy has a inner, rotating
disk with $V_{\rm rot}\simeq30$ km~s$^{-1}$ (see PV-diagram in Appendix~\ref{app:Atlas}).\\
\textbf{NGC~1569} has been previously studied by \citet{Stil2002} and \citet{Johnson2012}.
Both studies derived a rotation curve by fitting the \hi velocity field with a tilted-ring
model. The PV-diagram along the major axis, however, does not show any sign of rotation
in the inner parts (R$\lesssim$1$'$, see Appendix~\ref{app:Atlas}). Moreover, the \hi
line-profiles are very broad and asymmetric, likely due to strong non-circular motions.
For these reasons, we restrict our analysis to the rotation velocity in the outer parts
($\sim$50 km~s$^{-1}$).\\
\textbf{NGC~4163} shows a very small velocity gradient of $\sim$10 km~s$^{-1}$. The
complex \hi kinematics may be due to the low $V_{\rm{rot}}/\sigma_{\hi}$ ratio. The
PA of the stellar body and of the \hi disk significantly differ by $\sim$40$^{\circ}$.\\
\textbf{NGC~4449} has been previously studied by \citet{Hunter1998, Hunter1999b}, who
found that the \hi distribution forms 2 counter-rotating systems. For the inner \hi
disk, we find a rotation velocity of $\sim$35 km~s$^{-1}$. It is unclear whether the
outer gas system is really a counter-rotating disk or is formed by two or three \hi
tails wrapping around the inner disk (similarly to I~Zw~18, see \citealt{Lelli2012a}).\\
\textbf{NGC~5253} has been previously studied by \citet{Kobulnicky2008} and
\citet{LopezSanchez2012}, who discussed the possibility of gas inflows/outflows
along the minor axis of the galaxy. The data are, indeed, consistent with a \hi
disk with $V_{\rm rot} < 5$ km~s$^{-1}$ and $V_{\rm rad} \simeq 25$ km~s$^{-1}$
(see Fig.~\ref{fig:n5253}). Shadowing of the X-ray emission indicates that the
southern side of the galaxy is the nearest one to the observer \citep{Ott2005a},
suggesting that the radial motions are an inflow.\\
\textbf{UGC~6456} has been previously studied by \citet{Thuan2004} and \citet{Simpson2011}.
\citet{Simpson2011} derived a rotation curve using low-resolution (C+D array)
observations. They assumed different values of the PA for the approaching and
receding sides, which would imply an unusual, asymmetric warp starting within
the stellar component (see their Fig.~13). Our 3D models show that the \hi
kinematics may be simply explained by a disk with $V_{\rm rot} \simeq V_{\rm rad}
\simeq \sigma_{\hi} \simeq10$~km~s$^{-1}$ (see Fig~\ref{fig:u6456}).\\
\textbf{UGC~9128} has a \hi disk that rotates at $\sim$25 km~s$^{-1}$, but
the VF is very irregular and the \hi line profiles are broad and asymmetric,
possibly due to non-circular motions. The optical and kinematic PA differ
by $\sim$30$^{\circ}$.

\subsection{Galaxies with unsettled \hi distribution}

\textbf{UGC~6541} has a very asymmetric \hi distribution. Gas emission is
detected only in the northern half of the galaxy. This may be the remnant
of a disrupted disk.\\
\textbf{UGCA~290} has a \hi distribution that is offset with respect to the
stellar component. The kinematics is irregular and dominated by a few distinct
\hi clouds.

\section{Tables \label{app:tables}}

\begin{table*}[t]
\centering
\caption{Properties of the \hi datacubes.}
\setlength{\tabcolsep}{3pt}
\resizebox{18.5cm}{!}{
\begin{tabular}{l c c c c c c c c c c}
\hline
\hline
Name         & Telescope & \multicolumn{2}{c}{Original Beam}& Original $\Delta$V & \multicolumn{2}{c}{Final Beam}  & Final $\Delta$V & $R_{\hi}$/beam & Rms Noise & Source \\
             &           &(asec$\times$asec)&(pc$\times$pc)&(km s$^{-1}$)        &(asec$\times$asec)&(pc$\times$pc)&(km s$^{-1}$)    &                &(mJy/beam) & \\
(1)          & (2)       & (3)              & (4)            & (5)               & (6)              &(7)           & (8)      & (9)   & (10)  & (11)   \\
\hline
NGC 625      & VLA       & 18.9$\times$11.7 & 357$\times$221 & 2.6               & 30.0$\times$30.0 & 567$\times$567 & 5.2    & 4.6 & 1.80  & a \\
NGC 1569     & VLA       & 5.8$\times$5.0   & 96$\times$82   & 2.6               & 10.0$\times$10.0 & 165$\times$165 & 5.2    & 23.6& 0.46  & b \\
NGC 1705     & ATCA      & 16.7$\times$14.5 & 413$\times$358 & 4.0               & 16.7$\times$14.5 & 413$\times$358 & 7.0    & 5.5 & 0.40  & c \\
NGC 2366     & VLA       & 6.9$\times$5.9   & 107$\times$91  & 2.6               & 15.0$\times$15.0 & 233$\times$233 & 5.2    & 29.2& 0.66  & b \\
NGC 4068     & WSRT      & 14.8$\times$11.5 & 308$\times$240 & 2.5               & 20.0$\times$20.0 & 417$\times$417 & 6.1    & 7.4 & 2.00  & d \\
NGC 4163     & VLA       & 9.7$\times$5.9   & 141$\times$86  & 1.3               & 10.0$\times$10.0 & 145$\times$145 & 5.2    & 7.6 & 0.43  & b \\
NGC 4214     & VLA       & 7.6$\times$6.4   & 99$\times$84   & 1.3               & 30.0$\times$30.0 & 393$\times$393 & 5.2    & 14.0& 2.20  & b \\
NGC 4449     & VLA       & 13.7$\times$12.5 & 279$\times$254 & 5.2               & 20.0$\times$20.0 & 407$\times$407 & 10.4   & 21.9& 0.80  & e \\
NGC 5253     & ATCA      & 13.6$\times$7.5  & 231$\times$127 & 4.0               & 20.0$\times$20.0 & 339$\times$339 & 9.0    & 9.1 & 0.95  & f \\
NGC 6789     & WSRT      & 13.7$\times$12.7 & 239$\times$222 & 2.5               & 13.7$\times$12.7 & 239$\times$222 & 6.1    & 4.3 & 0.75  & a \\
UGC 4483     & VLA       & 5.7$\times$4.5   & 88$\times$70   & 2.6               & 10.0$\times$10.0 & 155$\times$155 & 5.2    & 9.0 & 0.66  & g \\
UGC 6456     & VLA       & 5.7$\times$4.8   & 119$\times$100 & 2.6               & 15.0$\times$15.0 & 313$\times$313 & 5.2    & 5.7 & 0.90  & a \\
UGC 6541     & VLA       & 6.2$\times$5.5   & 126$\times$112 & 1.3               & 10.0$\times$10.0 & 204$\times$204 & 5.2    & ... & 0.44  & b \\
UGC 9128     & VLA       & 6.2$\times$5.5   & 66$\times$59   & 2.6               & 15.0$\times$15.0 & 160$\times$160 & 5.2    & 5.6 & 0.80  & b \\
UGCA 290     & VLA       & 5.4$\times$4.2   & 175$\times$136 & 1.9               & 10.0$\times$10.0 & 325$\times$325 & 4.9    & ... & 0.56  & a \\
I Zw 18      & VLA       & 1.5$\times$1.4   & 132$\times$123 & 1.3               & 5.0$\times$5.0   & 441$\times$441 & 5.2    & 7.5 & 0.16  & h \\
I Zw 36      & VLA       & 6.8$\times$5.5   & 194$\times$157 & 2.6               & 6.8$\times$5.5   & 194$\times$157 & 5.2    & 10.9& 0.34  & b \\
SBS 1415+437 & VLA       & 4.6$\times$4.3   & 303$\times$283 & 1.9               & 10.0$\times$10.0 & 659$\times$659 & 4.9    & 6.5 & 0.60  & a \\
\hline
\hline
\end{tabular}
}
\tablebib{(a)~This work; (b)~\citet{Hunter2012}; (c)~\citet{Elson2013}; (d)~\citet{Swaters2002a}; (e)~\citet{Walter2008};
(f)~\citet{LopezSanchez2012}; (g)~\citet{Lelli2012b}; (h)~\citet{Lelli2012a}.}
\label{tab:data}
\end{table*}
\begin{table*}[t]
\caption{Optical and \hi orientation parameters.}
\centering
\resizebox{18.5cm}{!}{
\begin{tabular}{l c c c c c c c c c c c}
\hline
\hline
Name     & RA$_{\rm opt}$  & Dec$_{\rm opt}$  & $\epsilon_{\rm{opt}}$ & $i_{\rm{opt}}$ & PA$_{\rm{opt}}$ & RA$_{\rm{kin}}$ & Dec$_{\rm{kin}}$ & $V_{\rm{sys}}$ & $i_{\rm{kin}}$  & PA$_{\rm{kin}}$ & $\Delta \rm{c}$\\
         & (J2000)         & (J2000)          &            & ($^{\circ}$)   & ($^{\circ}$)    & (J2000)         & (J2000)          & (km s$^{-1}$)  & ($^{\circ}$)    & ($^{\circ}$)   & (pc)             \\
(1)      & (2)             & (3)              & (4)        & (5)            & (6)             & (7)             & (8)              & (9)            & (10)            & (11)& (12) \\
\hline
\multicolumn{12}{l}{\textit{Galaxies with a regularly rotating \hi disk}}\\
NGC 1705       & 04 54 13.9 & -53 21 25 & 0.28 & 47$\pm$2 & 55$\pm$3 & 04 54 16.1 & -53 21 35 & 635$\pm$2 & 45:85    & 10$\pm$5  & 552$\pm$164 \\
NGC 2366       & 07 28 51.9 & +69 12 34 & 0.66 & 80$\pm$2 & 29$\pm$4 & 07 28 53.3 & +69 12 43 & 103$\pm$1 & 68$\pm$5 & 42$\pm$2  & 150$\pm$99 \\
NGC 4068       & 12 04 02.7 & +52 35 28 & 0.38 & 56$\pm$4 & 31$\pm$4 & 12 04 03.0 & +52 35 30 & 206$\pm$2 & 44$\pm$6 & 24$\pm$3  & 0 \\
NGC 4214       & 12 15 38.8 & +36 19 39 & 0.09 & 26$\pm$5 & 40$\pm$20& 12 15 36.9 & +36 19 59 & 291$\pm$1 & 30:$-1$  & 65:84     & 393$\pm$167 \\
NGC 6789       & 19 16 41.9 & +63 58 17 & 0.15 & 34$\pm$2 & 86$\pm$4 & 19 16 41.9 & +63 58 17 &-151$\pm$2 & 43$\pm$7 & 82$\pm$5  & 0 \\
UGC 4483       & 08 37 03.4 & +69 46 31 & 0.47 & 63$\pm$3 &-13$\pm$5 & 08 37 03.4 & +69 46 31 & 158$\pm$2 & 58$\pm$3 & 0$\pm$5   & 0 \\
I Zw 18        & 09 34 02.0 & +55 14 25 & 0.50 & 65$\pm$5 &135$\pm$1 & 09 34 02.0 & +55 14 25 & 767$\pm$4 & 70$\pm$4 & 145$\pm$5 & 0 \\
I Zw 36        & 12 26 16.8 & +48 29 39 & 0.30 & 49$\pm$2 & 80$\pm$3 & 12 26 18.0 & +48 29 41 & 277$\pm$2 & 67$\pm$3 & 44$\pm$3  & 340$\pm$74 \\
SBS 1415+437   & 14 17 02.1 & +43 30 19 & 0.66 & 80$\pm$3 & 30$\pm$5 & 14 17 01.7 & +43 30 07 & 616$\pm$2 & 66$\pm$3 & 23$\pm$3  & 824$\pm$280 \\
\multicolumn{12}{l}{\textit{Galaxies with a kinematically disturbed \hi disk}}\\
NGC 625        & 01 35 04.3 & -41 26 15 & 0.64 & 78$\pm$2 & 94$\pm$1 & 01 35 06.3 & -41 26 17 &398$\pm$6  & ...      & 120$\pm$10 & 0 \\
NGC 1569       & 04 30 49.0 & +64 50 53 & 0.54 & 69$\pm$2 & 118$\pm$3& 04 30 51.9 & +64 50 56 &-80$\pm$10 & ...      & 115$\pm$10 & 310$\pm$70 \\
NGC 4163       & 12 12 09.0 & +36 10 11 & 0.30 & 49$\pm$2 & 14$\pm$2 & 12 12 09.0 & +36 10 16 &158$\pm$4  & ...      & -25$\pm$10 & 72$\pm$62  \\
NGC 4449       & 12 28 10.8 & +44 05 37 & 0.40 & 57$\pm$3 & 55$\pm$3 & 12 28 11.3 & +44 05 58 &210$\pm$5  & ...      & 60$\pm$5   & 444$\pm$173 \\
NGC 5253       & 13 39 56.0 & -31 38 31 & 0.53 & 68$\pm$2 & 43$\pm$2 & 13 39 56.0 & -31 38 31 &410$\pm$10 & ...      &  40$\pm$5  & 0 \\
UGC 6456       & 11 27 57.2 & +78 59 48 & 0.50 & 65$\pm$5 &-10$\pm$5 & 11 27 58.8 & +78 59 51 &-102$\pm$4 & ...      &  0$\pm$5   & 0 \\
UGC 9128       & 14 15 56.8 & +23 03 22 & 0.28 & 47$\pm$6 & 33$\pm$6 & 14 15 57.6 & +23 03 08 & 150$\pm$4 & ...      & 0$\pm$10   & 181$\pm$68 \\
\multicolumn{12}{l}{\textit{Galaxies with unsettled \hi distribution}}\\
UGC 6541       & 11 33 28.9 & +49 14 22 & 0.50 & 65$\pm$4 &129$\pm$2 & ...        & ...       & 250$\pm$2 & ...      & ...        & ... \\
UGCA 290       & 12 37 22.1 & +38 44 41 & 0.50 & 65$\pm$3 & 47$\pm$3 & ...        & ...       & 468$\pm$5 & ...      & ...        & ... \\
\hline
\hline
\end{tabular}
}
\label{tab:param}
\end{table*}
\begin{table*}[hbt]
\caption{Mass budget within the optical radius.}
\centering
\resizebox{18.5cm}{!}{
\begin{tabular}{l c c c c c c c c c c c}
\hline
\hline
Name     & $M^{\rm{Sal}}_{*}$ & $M_{\rm{mol}}$ & $M_{\hi}(R_{\rm{opt}})$ & $M^{\rm{Kr}}_{\rm{bar}}$ & $M^{\rm{Sal}}_{\rm{bar}}$ & $M^{\rm{mol}}_{\rm{bar}}$ & $V_{\rm{circ}}(R_{\rm opt})$ & $M_{\rm{dyn}}(R_{\rm{opt}})$& $f^{\rm{Kr}}_{\rm{bar}}$ & $f^{\rm{Sal}}_{\rm{bar}}$ & $f^{\rm{mol}}_{\rm{bar}}$\\
         & \multicolumn{6}{c}{-------------------------------- (10$^{7}$ M$_{\odot}$) --------------------------------} & (km s$^{-1}$)    & (10$^{7}$ M$_{\odot}$) &    &   & \\
(1)      & (2)         & (3)         & (4)         & (5)        & (6)       &(7)        & (8)      & (9)        & (10)    & (11)    & (12)\\
\hline
\multicolumn{12}{l}{\textit{Galaxies with a regularly rotating \hi disk}}\\
NGC 1705 & $>$20       & 60$\pm$18   & 4.8$\pm$0.5 & $>$19       & $>$26     & $>$86     & 73$\pm$3 & 185$\pm$36 & $>$0.10       & $>$0.14       & $>$0.47 \\
NGC 2366 & 26$\pm$3    &  8$\pm$2    & 36$\pm$4    & 64$\pm$5    & 74$\pm$6  & 82$\pm$6  & 51$\pm$2 & 263$\pm$28 & 0.24$\pm$0.03 & 0.28$\pm$0.04 & 0.31$\pm$0.04 \\
NGC 4068 & 22$\pm$3    & 5.9$\pm$1.8 & 8.2$\pm$0.8 & 25$\pm$2    & 33$\pm$3  & 39$\pm$4  & 36$\pm$2 & 54$\pm$13  & 0.46$\pm$0.12 & 0.61$\pm$0.16 & 0.72$\pm$0.19\\
NGC 4214 & $>$28       & 12$\pm$4    & 11$\pm$1    & $>$32       & $>$42     & $>$54     & 79$\pm$4 & 325$\pm$104& $>$0.10       & $>$0.13       & $>$0.17\\
NGC 6789 & 7$\pm$2     & 0.6$\pm$0.2 & 1.2$\pm$0.1 & 6.0$\pm$1.3 &8.6$\pm$2.0&9.2$\pm$2.0& 59$\pm$9 & 60$\pm$24  & 0.10$\pm$0.05 & 0.14$\pm$0.07 & 0.15$\pm$0.07\\
UGC 4483 & 1.0$\pm$0.2 & 2.1$\pm$0.6 & 1.0$\pm$0.1 & 2.0$\pm$0.2 &2.3$\pm$0.2&4.4$\pm$0.6& 21$\pm$2 & 6.6$\pm$1.3& 0.30$\pm$0.07 & 0.36$\pm$0.08 & 0.68$\pm$0.17\\
I Zw 18  & $>$1.7      & 7$\pm$2     & 1.6$\pm$0.2 & $>$3.2      &$>$3.8     &$>$11      & 38$\pm$4 & 16$\pm$3   & $>$0.20       & $>$0.24       & $>$0.67 \\
I Zw 36  & $>$0.8      & 4.7$\pm$1.4 & 1.5$\pm$0.1 & $>$2.5      &$>$2.8     &$>$7.5     & 30$\pm$3 & 19$\pm$4   & $>$0.13       & $>$0.14       & $>$0.38 \\
SBS 1415+437&17$\pm$3  & 7.6$\pm$2.3 & 6.8$\pm$0.7 & 20$\pm$2    & 26$\pm$3  & 34$\pm$4  & 22$\pm$2 & 27$\pm$5   & 0.74$\pm$0.16 & 0.98$\pm$0.22 & \textit{1.26$\pm$0.28}\\
\multicolumn{12}{l}{\textit{Galaxies with a kinematically disturbed \hi disk}}\\
NGC 625  & 26$\pm$10   & 0.8$\pm$0.2 & 6.4$\pm$0.6 & 25$\pm$6    & 34$\pm$10 & 35$\pm$10 & 30$\pm$5 & 70$\pm$23  & 0.36$\pm$0.15  & 0.50$\pm$0.22& 0.51$\pm$0.22\\
NGC 1569 & 70$\pm$7    & 15$\pm$5    & 17$\pm$2    & 66$\pm$5    & 92$\pm$7  & 107$\pm$9 & 50$\pm$5 & 177$\pm$36 & 0.37$\pm$0.08  & 0.52$\pm$0.11& 0.61$\pm$0.13 \\
NGC 4163 & 10$\pm$3    & 1.0$\pm$0.3 & 1.1$\pm$0.1 & 8$\pm$2     & 11$\pm$3  & 12$\pm$3  & 10$\pm$4 & 2.4$\pm$1.9& \textit{3.3$\pm$2.7} & \textit{4.8$\pm$4.1} & \textit{5.2$\pm$4.4}\\
NGC 4449 & 210$\pm$35  & 184$\pm$55  & 32$\pm$3    & 174$\pm$22  &252$\pm$35 & 436$\pm$65& 35$\pm$5 & 94$\pm$28  & \textit{1.9$\pm$0.6} & \textit{2.7$\pm$0.9} & \textit{4.6$\pm$1.5}\\
NGC 5253 & 154$\pm$21  & 31$\pm$9    & 6.8$\pm$0.7 & 106$\pm$13  &163$\pm$21 & 194$\pm$23& $<$5     & ...        & ...            & ...          & ... \\
UGC 6456 & 5$\pm$2     & 4.3$\pm$1.3 & 2.6$\pm$0.3 & 6.6$\pm$1.3 &8.5$\pm$2.0& 13$\pm$2  & 10$\pm$5 & ...        & ...            & ...          & ...  \\
UGC 9128 & 1.3$\pm$0.2 &0.13$\pm$0.04& 0.8$\pm$0.1 & 2.0$\pm$0.2 &2.4$\pm$0.2&2.5$\pm$0.2& 24$\pm$4 & 7.7$\pm$3.0& 0.25$\pm$0.10  & 0.31$\pm$0.12&0.33$\pm$0.13  \\
\multicolumn{12}{l}{\textit{Galaxies with unsettled \hi distribution}}\\
UGC 6541 & $>$0.8      & 0.6$\pm$0.2 & 1.2$\pm$0.1 & $>$2.1      & $>$2.4    & $>$2.7     & ...      & ...        & ...           & ...          & ...  \\
UGCA 290 & $>$1        & 2.1$\pm$0.6 & 1.4$\pm$0.2 & $>$2.5      & $>$2.9    & $>$4.9     & ...      &...         & ...           & ...          & ...  \\
\hline
\hline
\end{tabular}
}
\label{tab:dynamics}
\end{table*}

\subsection{Properties of the \hi datacubes}

\textit{Column} (1) gives the galaxy name, following the ordering NGC, UGC, UGCA, Zwicky, SBS. \\
\textit{Column} (2) gives the radio interferometer used for the 21cm-line observations.\\
\textit{Column} (3), (4), and (5) give the spatial and spectral resolutions of the original 
cube. This cube is typically obtained using a Robust parameter $\Re\simeq0$.\\
\textit{Column} (6), (7), and (8) give the spatial and spectral resolutions of the cube
after Gaussian smoothing.\\
\textit{Column} (9) gives the ratio of the \hi radius $R_{\hi}$ (given in
Table~\ref{tab:extent}) to the final \hi beam.\\
\textit{Column} (10) gives the noise in the final cube.\\
\textit{Column} (11) provides the reference for the original cube.

\subsection{Optical and \hi orientation parameters}

\textit{Column} (1) gives the galaxy name.\\
\textit{Column} (2), (3), (4), (5) and (6) give the optical center, ellipticity,
inclination, and position angle. These values are derived by interactively
fitting ellipses to the outer isophotes. The inclination is calculated
assuming an oblate spheroid with intrinsic thickness $q_{0}=0.3$.\\
\textit{Column} (7), (8), (9), (10) and (11) give the kinematical center,
systemic velocity, inclination, and position angle. These values are
derived using \hi velocity fields, channel maps, PV-diagrams, and
building 3D disk models.\\
\textit{Column} (12) gives the projected offset between the optical
and kinematical centers. This is calculated as $\sqrt{(\alpha_{\rm{opt}}-
\alpha_{\rm{kin}})^{2} -(\delta_{\rm{opt}}-\delta_{\rm{kin}})^{2}}$,
assuming the galaxy distance given in Table \ref{tab:sample}. The error
is estimated as FWHM/2.35, where FWHM is the beam of the smoothed \hi
datacube (see Table~\ref{tab:data}). Projected distances smaller than
FWHM/2.35 are assumed to be zero.

\subsection{Mass budget within the optical radius.}

\textit{Column} (1) gives the galaxy name.\\
\textit{Column} (2) gives the stellar mass. This is calculated by integrating the galaxy SFH
and assuming a gas-recycling efficiency of 30$\%$. The SFHs were derived by fitting the CMDs
of the resolved stellar populations and assuming a Salpeter IMF from 0.1 to 100 M$_{\odot}$.\\
\textit{Column} (3) gives the molecular mass. This is indirectly estimated using Eq.~\ref{eq:mol},
which assumes that the star-formation efficiency in dwarfs is the same as in spirals.\\
\textit{Column} (4) gives the \hi mass $M_{\hi}$ within $R_{\rm{opt}}$.\\
\textit{Column} (5), (6), and (7) give the baryonic mass within $R_{\rm{opt}}$ assuming, respectively,
a Kroupa IMF, a Salpeter IMF, and a Salpeter IMF plus the possible contribution of molecules.\\
\textit{Column} (8) gives the circular velocity at $R_{\rm{opt}}$. \\
\textit{Column} (9) gives the dynamical mass within $R_{\rm{opt}}$ calculated as
$M_{\rm{dyn}} = V_{\rm{circ}}^{2}(R_{\rm opt})\times R_{\rm{opt}}/G$.\\
\textit{Column} (10, 11, 12) gives the baryonic fraction within $R_{\rm{opt}}$ assuming,
respectively, a Kroupa IMF, a Salpeter IMF, and a Salpeter IMF plus molecules.
\textit{Italics} indicate unphysical values $>$1.

\section{\label{app:Atlas}Atlas}

In the following, we present overview figures for the 18 starbursting dwarfs in our sample.
For each galaxy, we show six panels including both optical and \hi data.\\
\textbf{Top-left}: a sky-subtracted optical image in the $R$ or $V$ band.
The cross shows the optical center.\\
\textbf{Bottom-left}: an isophotal map (black contours) overlaid with a set of concentric
ellipses (white contours). The value of the outermost isophote $\mu_{\rm out}$ is given
in the note; the isophotes increase in steps of 1 mag~arcsec$^{-2}$. The orientation
parameters for the ellipses ($\epsilon_{\rm opt}$ and PA$_{\rm opt}$) are given in
Table~\ref{tab:param}. The cross shows the optical center. For I~Zw~18, the isophotal
map was derived from a $R$-band HST image after the subtraction of the H$\alpha$ emission,
as the nebular emission dominates the galaxy morphology (see \citealt{Papaderos2002}).\\
\textbf{Top-middle}: the total \hi map. The contour levels are at 1, 2, 4, 8, ... $\times$
$N_{\hi}(3\sigma)$, where $N_{\hi}(3\sigma)$ is the pseudo-3$\sigma$ contour, calculated
following \citet{Verheijen2001}. The value of $N_{\hi}(3\sigma)$ is given in the note.
The cross shows the optical center. The ellipse shows the beam.\\
\textbf{Bottom-middle}: the \hi surface density profile, derived by azimuthally averaging
over the entire \hi disk (black line) and over the approaching and receding sides separately
(filled and open circles, respectively). In UGC~6541 and UGCA~290, \hi emission is detected
only on one side of the galaxy, thus the \hi surface density profile was derived using
the optical orientation parameters and averaging over a single side.\\
\textbf{Top-right}: the \hi velocity field. Light and dark shading indicate approaching
and receding velocities, respectively. The thick, black line shows the systemic velocity.
The velocity interval between approaching (black) and receding (white) contours is given
in the note. The cross shows the optical center, while the circle shows the kinematic center.
The dashed line indicates the kinematic position angle. The ellipse shows the beam.\\
\textbf{Bottom-right}: Position-Velocity diagram taken through the kinematic center and along
the kinematic major axis. Contours are at -3, -1.5 (dashed), 1.5, 3, 6, 12, ... $\times$ $\sigma$.
The value of $\sigma$ is given in Table~\ref{tab:data}. The vertical and horizontal lines
show the kinematic center and the systemic velocity, respectively. For galaxies with a
regularly rotating \hi disk, squares show the rotation curve as derived in Sect.~\ref{sec:rotcur},
projected along the line of sight. For galaxies with a kinematically disturbed \hi disk,
arrows show the estimated value of $V_{\rm{rot}}$, projected along the line of sight.

\renewcommand\bibname{{References}}
\bibliographystyle{aa}
\bibliography{bibliography.bib}

\newpage

\begin{figure*}
\centering
\includegraphics[width=17.5 cm]{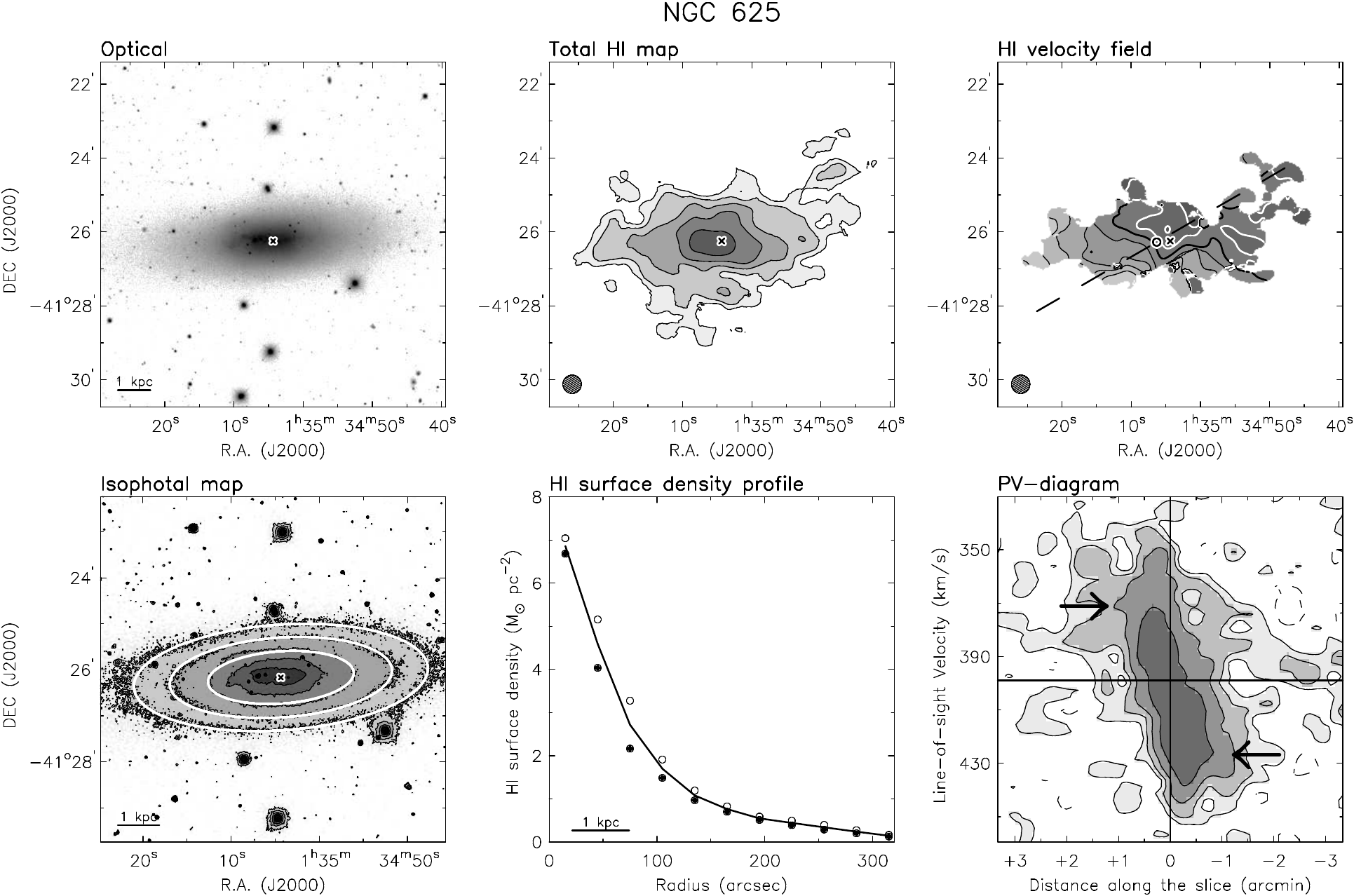}
\caption{\textbf{Contours:} $\mu_{\rm{out}} = 24.5$ R mag arcsec$^{-2}$; $N_{\hi}(3\sigma) = 1.1 \times 10^{20}$ atoms cm$^{-2}$; $V_{\rm l.o.s} = 398 \pm 10$ km s$^{-1}$.}
\vspace{0.3 cm}
\centering
\includegraphics[width=17.5 cm]{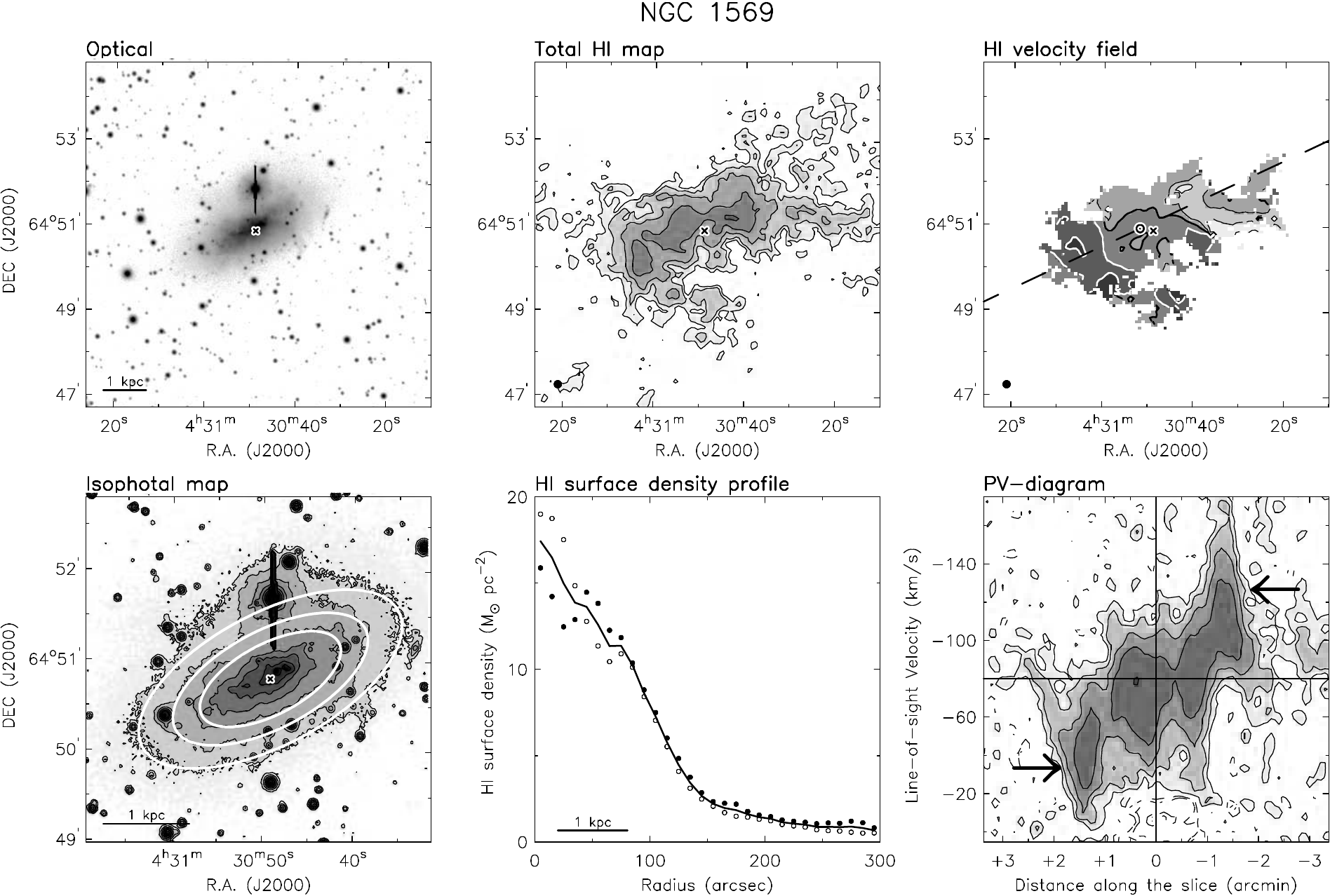}
\caption{\textbf{Contours:} $\mu_{\rm{out}} = 24$ V mag arcsec$^{-2}$; $N_{\hi}(3\sigma) = 4.3 \times 10^{20}$ atoms cm$^{-2}$; $V_{\rm l.o.s} = -80 \pm 20$ km s$^{-1}$.}
\end{figure*}

\begin{figure*}
\centering
\includegraphics[width=17.5 cm]{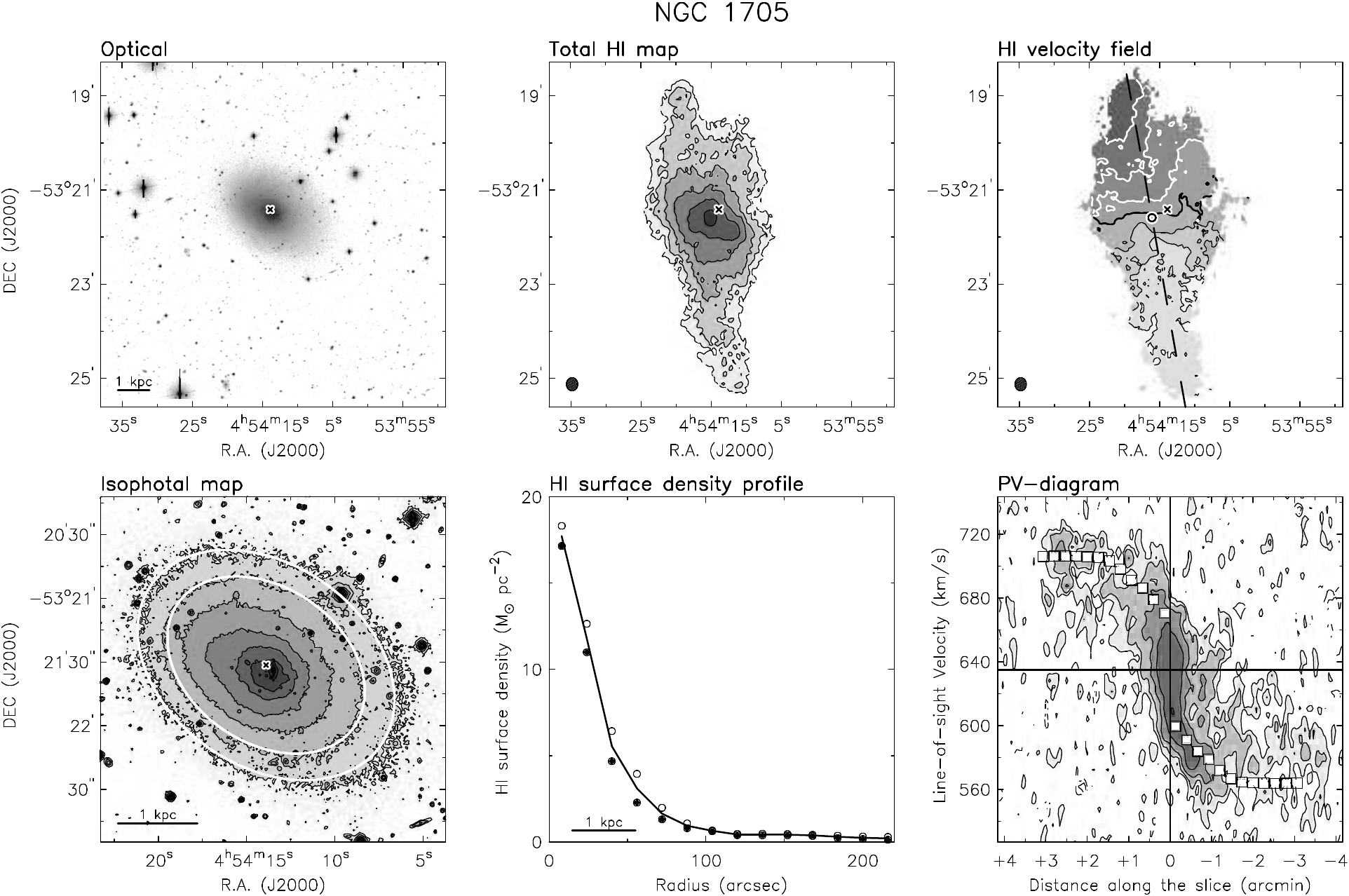}
\caption{\textbf{Contours:} $\mu_{\rm{out}} = 25.5$ R mag arcsec$^{-2}$; $N_{\hi}(3\sigma) = 1.1 \times 10^{20}$ atoms cm$^{-2}$; $V_{\rm l.o.s} = 635 \pm 20$ km s$^{-1}$.}
\vspace{0.3 cm}
\centering
\includegraphics[width=17.5 cm]{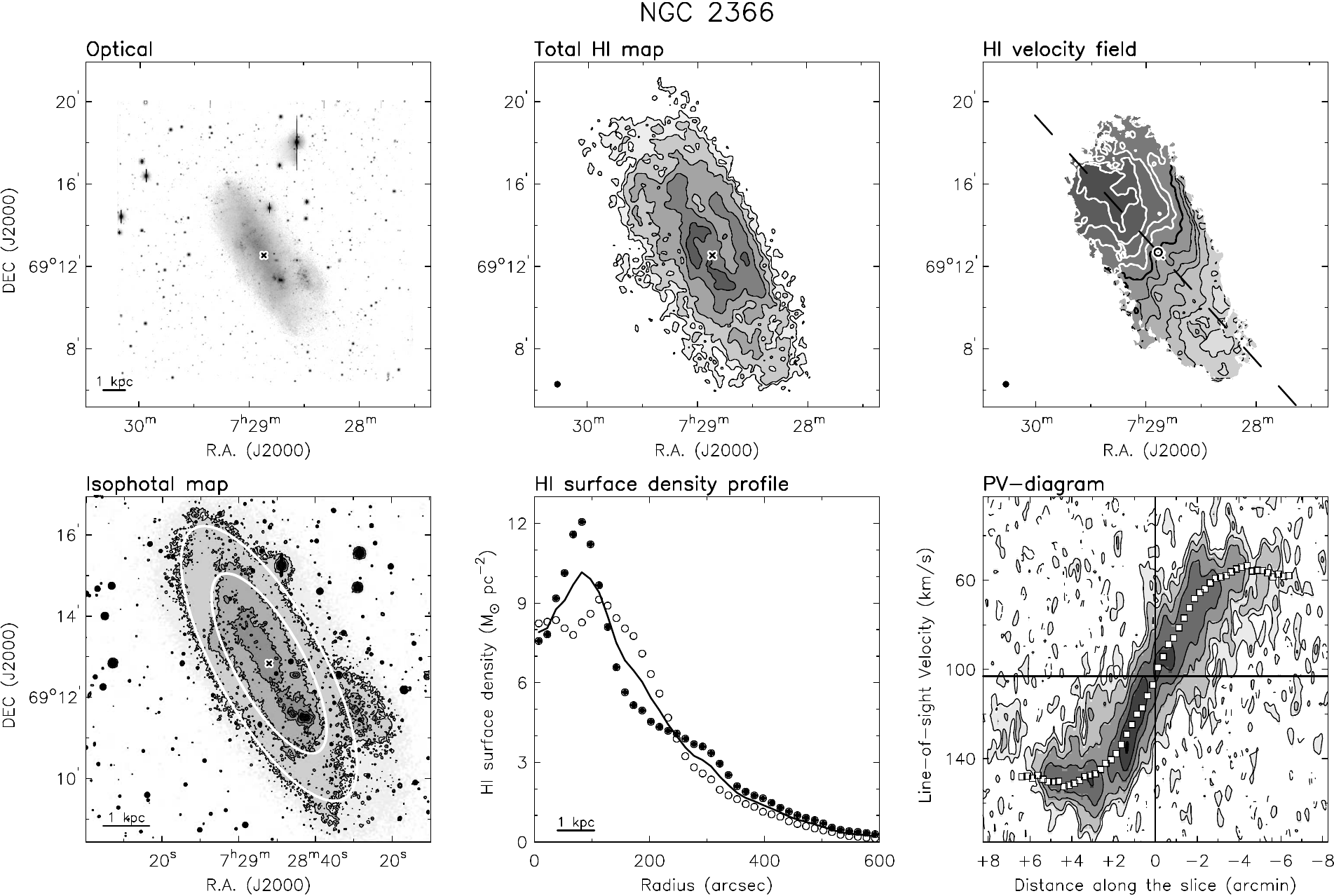}
\caption{\textbf{Contours:} $\mu_{\rm{out}} = 24.5$ V mag arcsec$^{-2}$; $N_{\hi}(3\sigma) = 2.3 \times 10^{20}$ atoms cm$^{-2}$; $V_{\rm l.o.s} = 103 \pm 10$ km s$^{-1}$.}
\end{figure*}

\begin{figure*}
\centering
\includegraphics[width=17.5 cm]{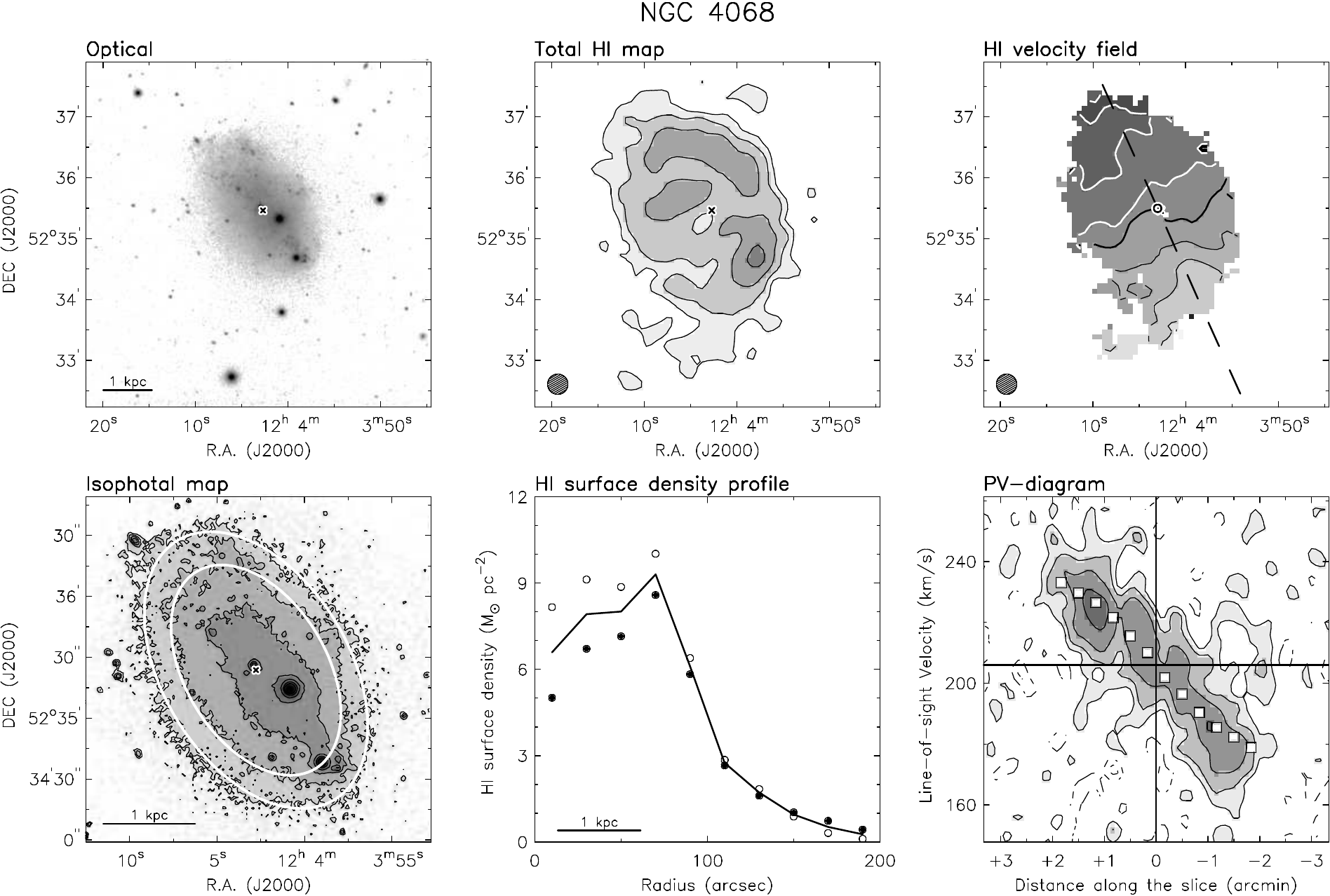}
\caption{\textbf{Contours:} $\mu_{\rm{out}} = 24.5$ R mag arcsec$^{-2}$; $N_{\hi}(3\sigma) = 3.6 \times 10^{20}$ atoms cm$^{-2}$; $V_{\rm l.o.s} = 206 \pm 10$ km s$^{-1}$.}
\vspace{0.3 cm}
\centering
\includegraphics[width=17.5 cm]{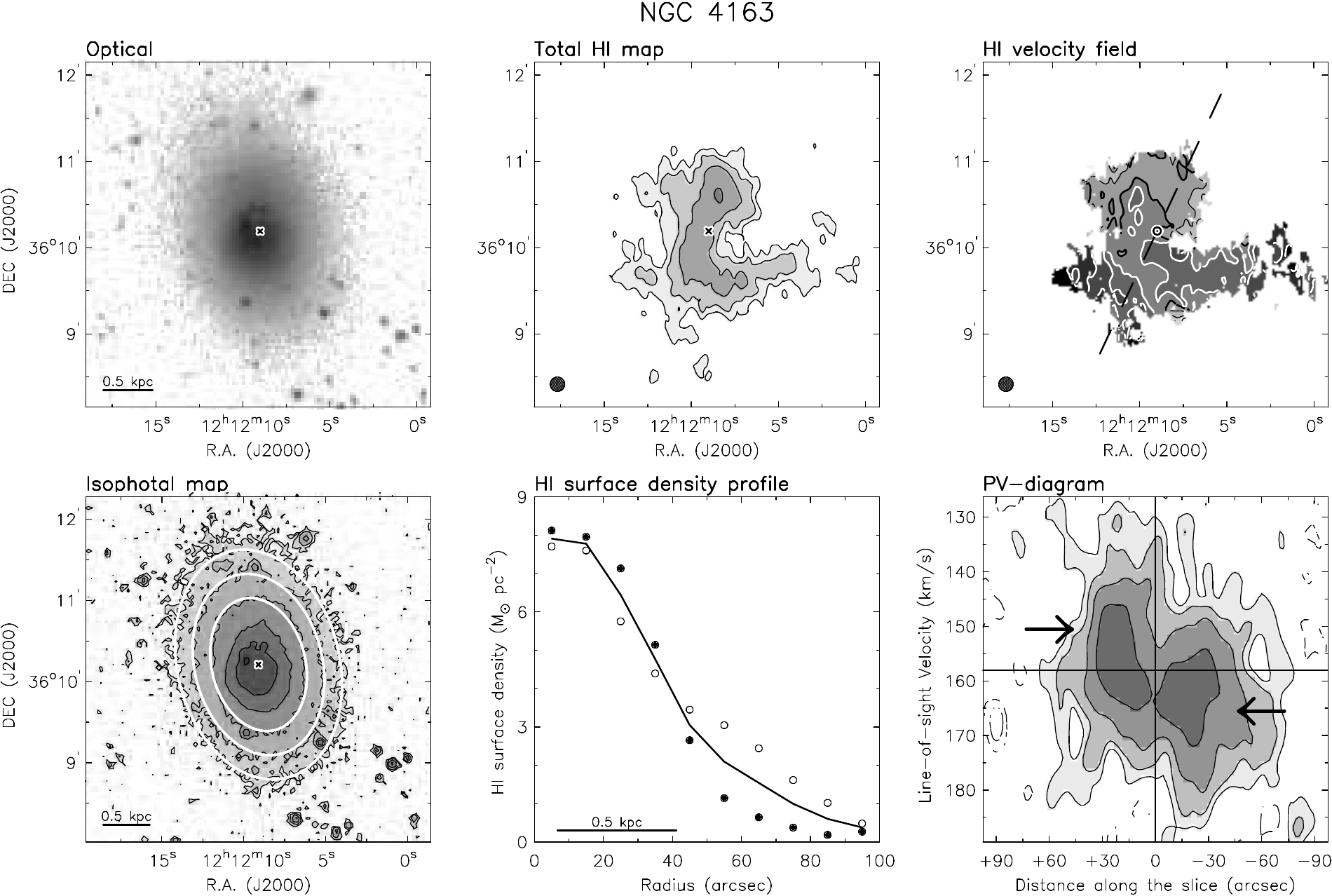}
\caption{\textbf{Contours:} $\mu_{\rm{out}} = 26.5$ V mag arcsec$^{-2}$; $N_{\hi}(3\sigma) = 2.7 \times 10^{20}$ atoms cm$^{-2}$; $V_{\rm l.o.s} = 158 \pm 5$ km s$^{-1}$.}
\end{figure*}

\begin{figure*}
\centering
\includegraphics[width=17.5 cm]{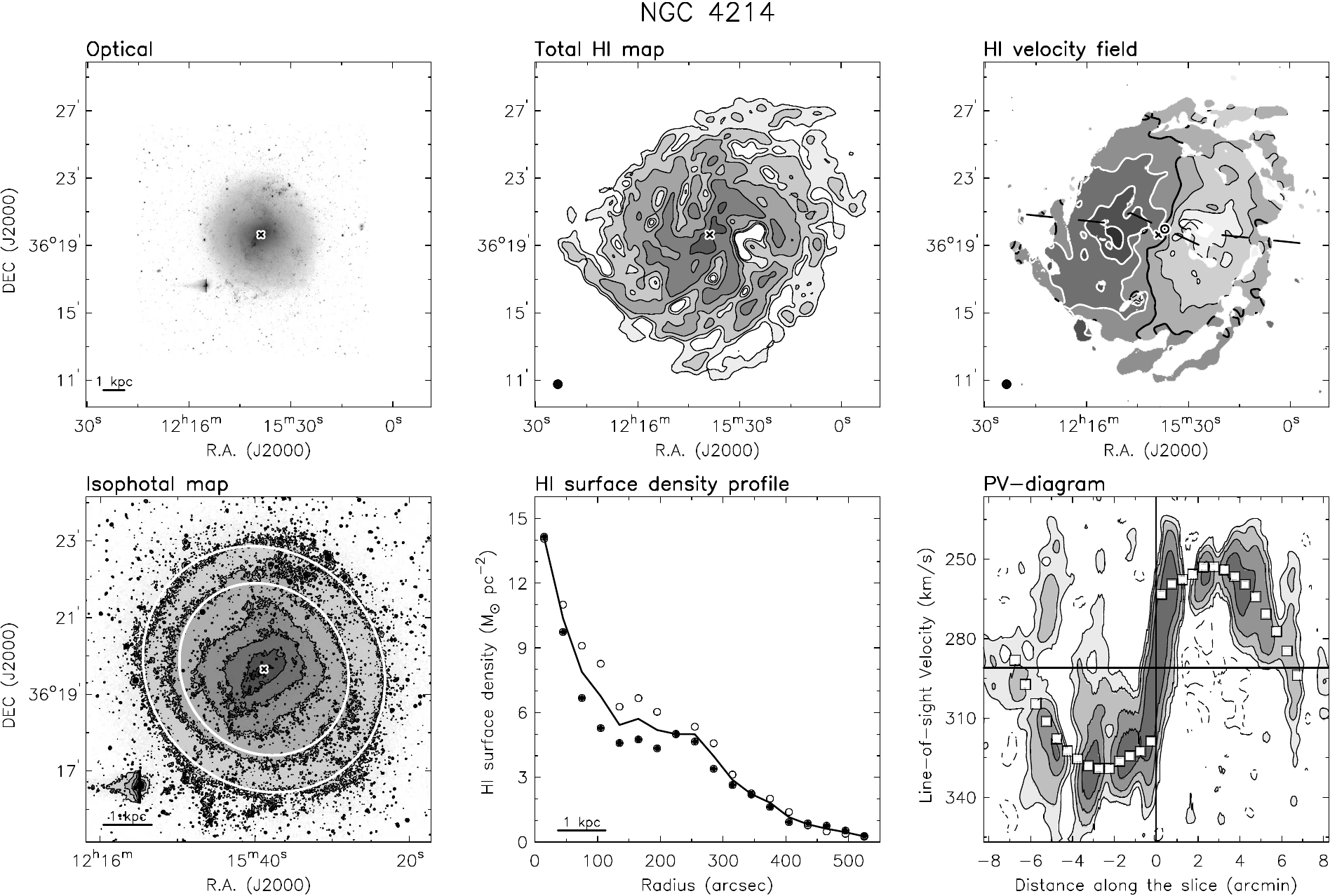}
\caption{\textbf{Contours:} $\mu_{\rm{out}} = 24.5$ V mag arcsec$^{-2}$; $N_{\hi}(3\sigma) = 1.2 \times 10^{20}$ atoms cm$^{-2}$; $V_{\rm l.o.s} = 291 \pm 15$ km s$^{-1}$.}
\vspace{0.3 cm}
\centering
\includegraphics[width=17.5 cm]{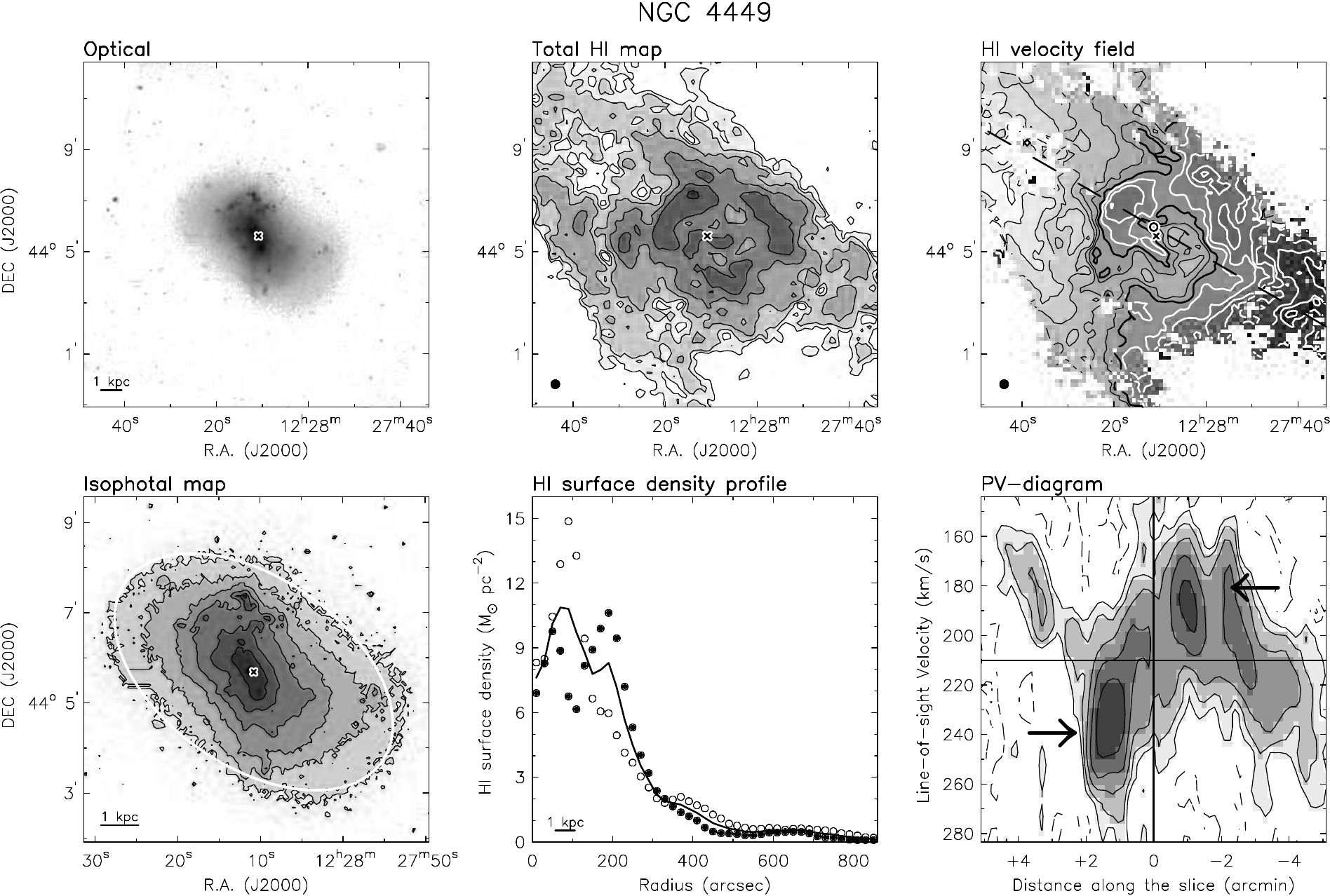}
\caption{\textbf{Contours:} $\mu_{\rm{out}} = 25$ V mag arcsec$^{-2}$; $N_{\hi}(3\sigma) = 1.4 \times 10^{20}$ atoms cm$^{-2}$; $V_{\rm l.o.s} = 210 \pm 10$ km s$^{-1}$.}
\end{figure*}

\begin{figure*}
\centering
\includegraphics[width=17.5 cm]{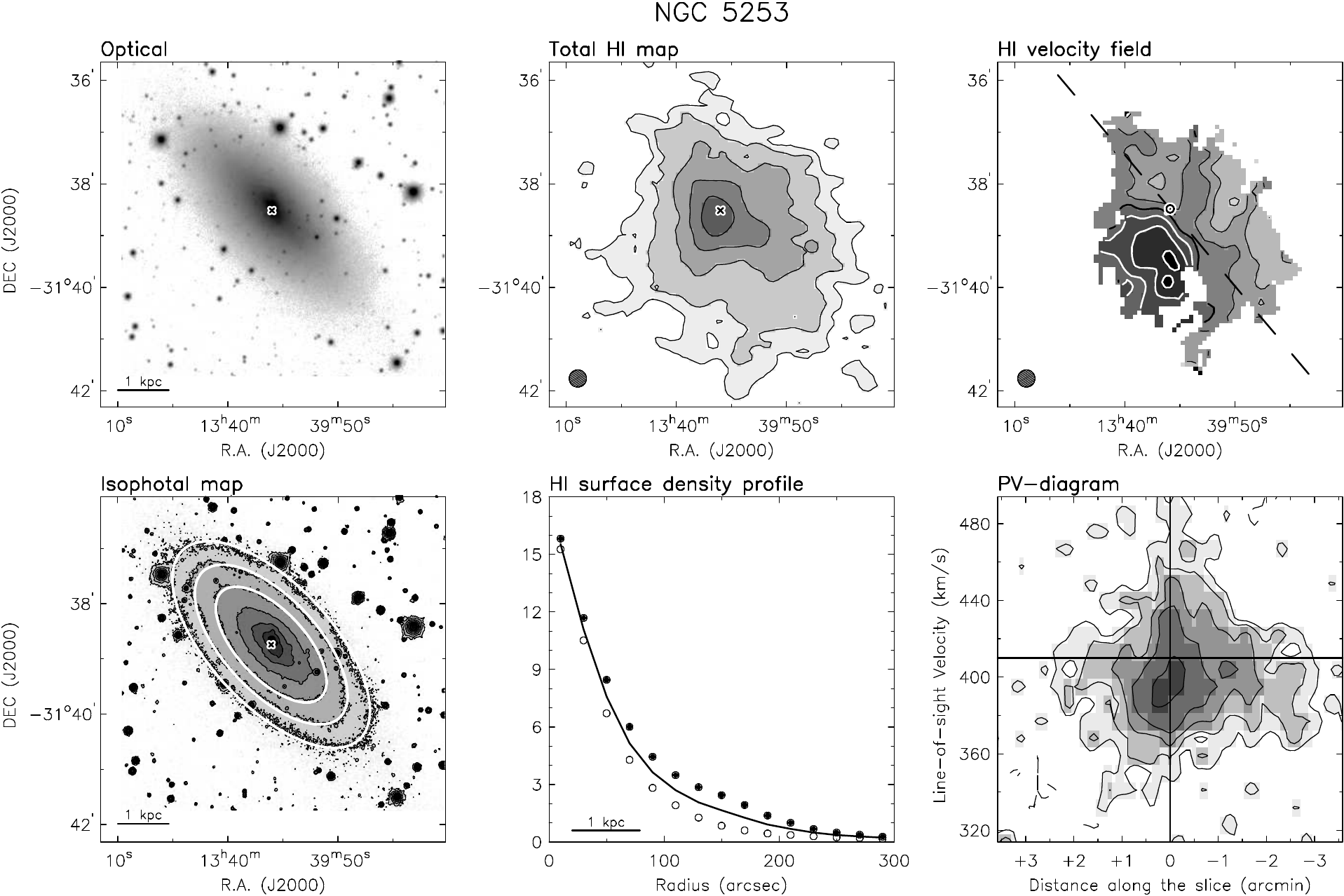}
\caption{\textbf{Contours:} $\mu_{\rm{out}} = 24$ R mag arcsec$^{-2}$; $N_{\hi}(3\sigma) = 2.1 \times 10^{20}$ atoms cm$^{-2}$; $V_{\rm l.o.s} = 410 \pm 10$ km s$^{-1}$.}
\centering
\vspace{0.3 cm}
\includegraphics[width=17.5 cm]{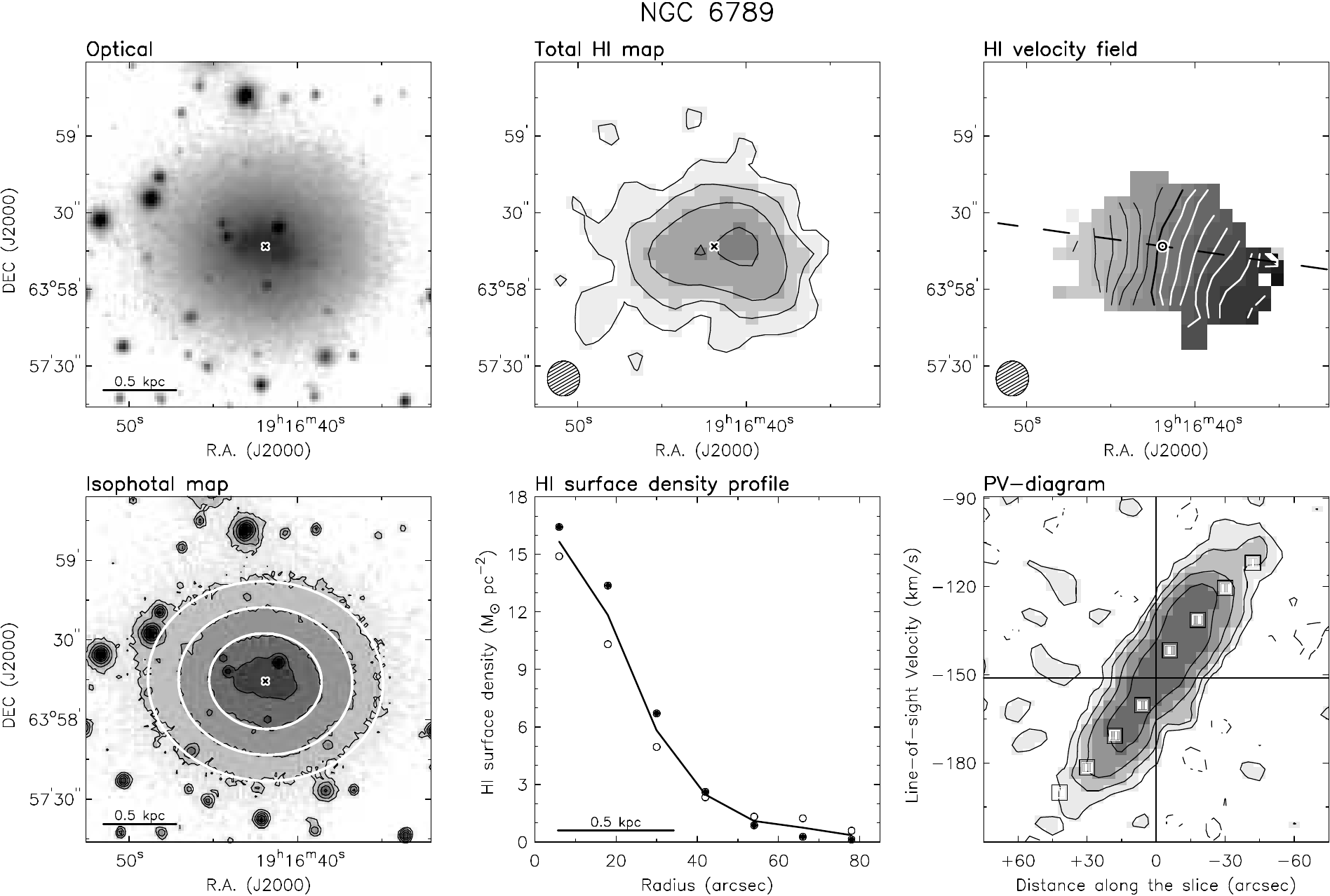}
\caption{\textbf{Contours:} $\mu_{\rm{out}} = 24.5$ R mag arcsec$^{-2}$; $N_{\hi}(3\sigma) = 3.4 \times 10^{20}$ atoms cm$^{-2}$; $V_{\rm l.o.s} = -151 \pm 5$ km s$^{-1}$.}
\vspace{0.3 cm}
\end{figure*}

\begin{figure*}
\centering
\includegraphics[width=17.5 cm]{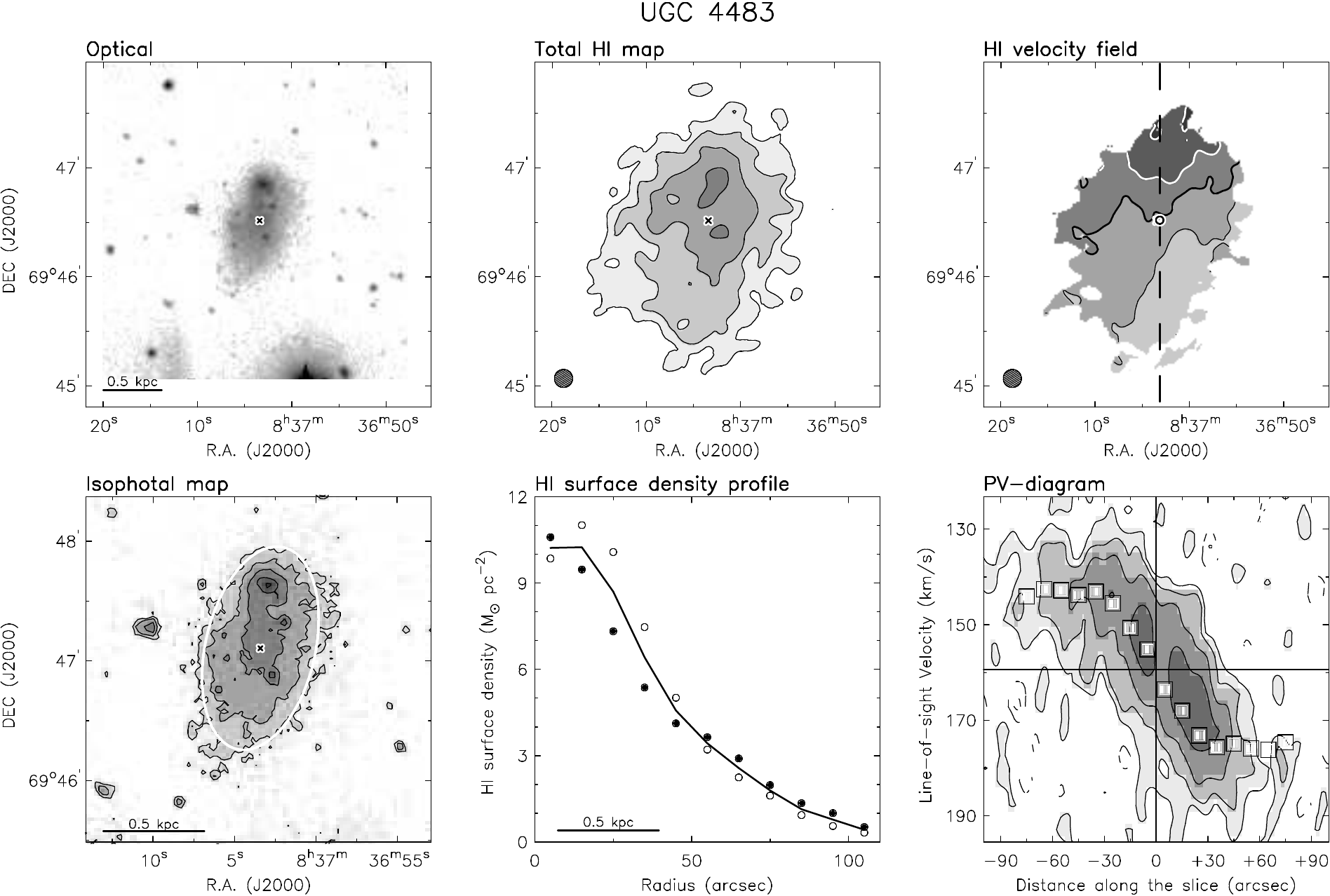}
\caption{\textbf{Contours:} $\mu_{\rm{out}} = 25$ R mag arcsec$^{-2}$; $N_{\hi}(3\sigma) = 3.4 \times 10^{20}$ atoms cm$^{-2}$; $V_{\rm l.o.s} = 158 \pm 10$ km s$^{-1}$.}
\vspace{0.3 cm}
\centering
\includegraphics[width=17.5 cm]{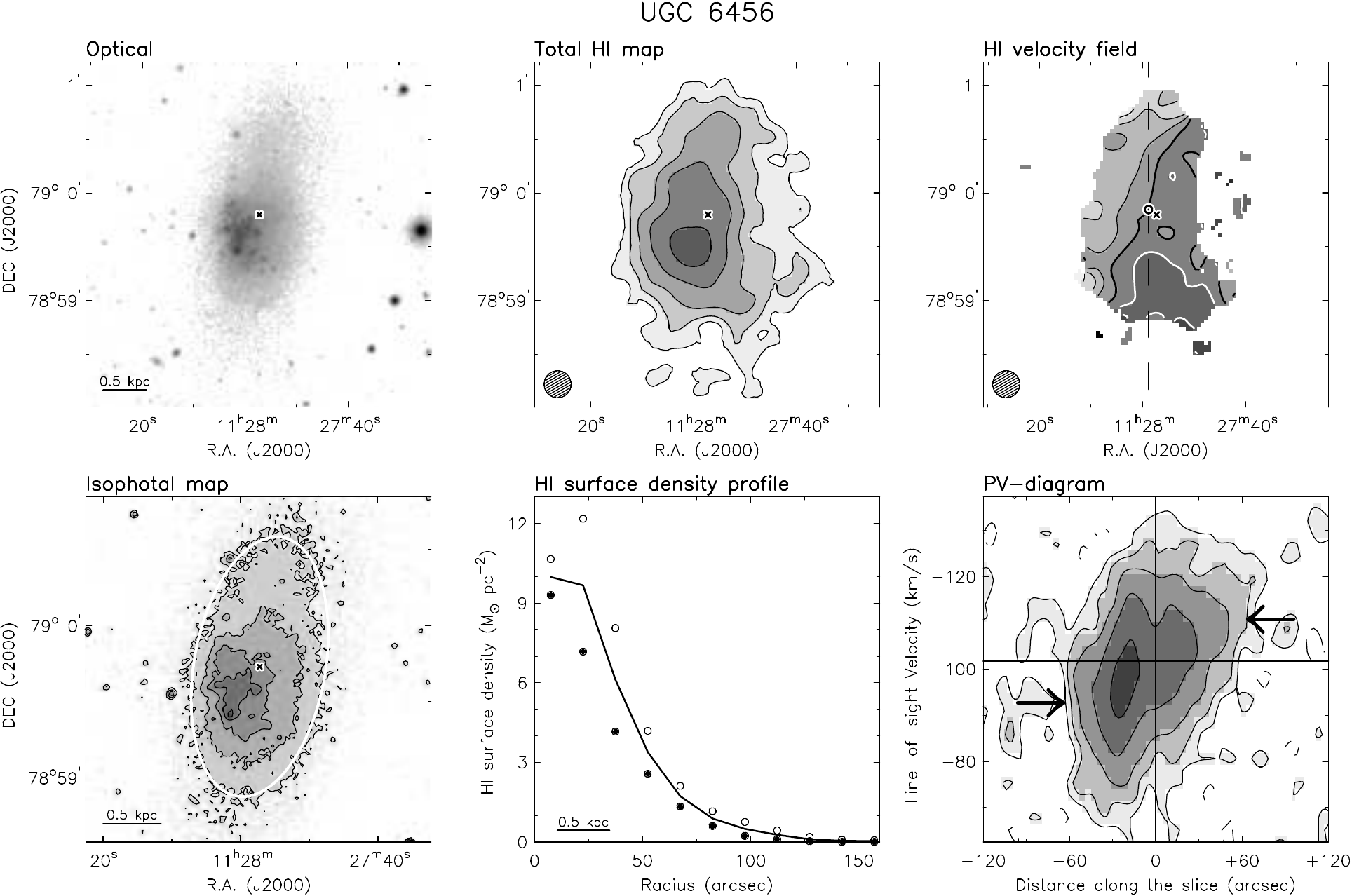}
\caption{\textbf{Contours:} $\mu_{\rm{out}} = 25$ R mag arcsec$^{-2}$; $N_{\hi}(3\sigma) = 1.8 \times 10^{20}$ atoms cm$^{-2}$; $V_{\rm l.o.s} = -102 \pm 5$ km s$^{-1}$.}
\end{figure*}

\begin{figure*}
\centering
\includegraphics[width=17.5 cm]{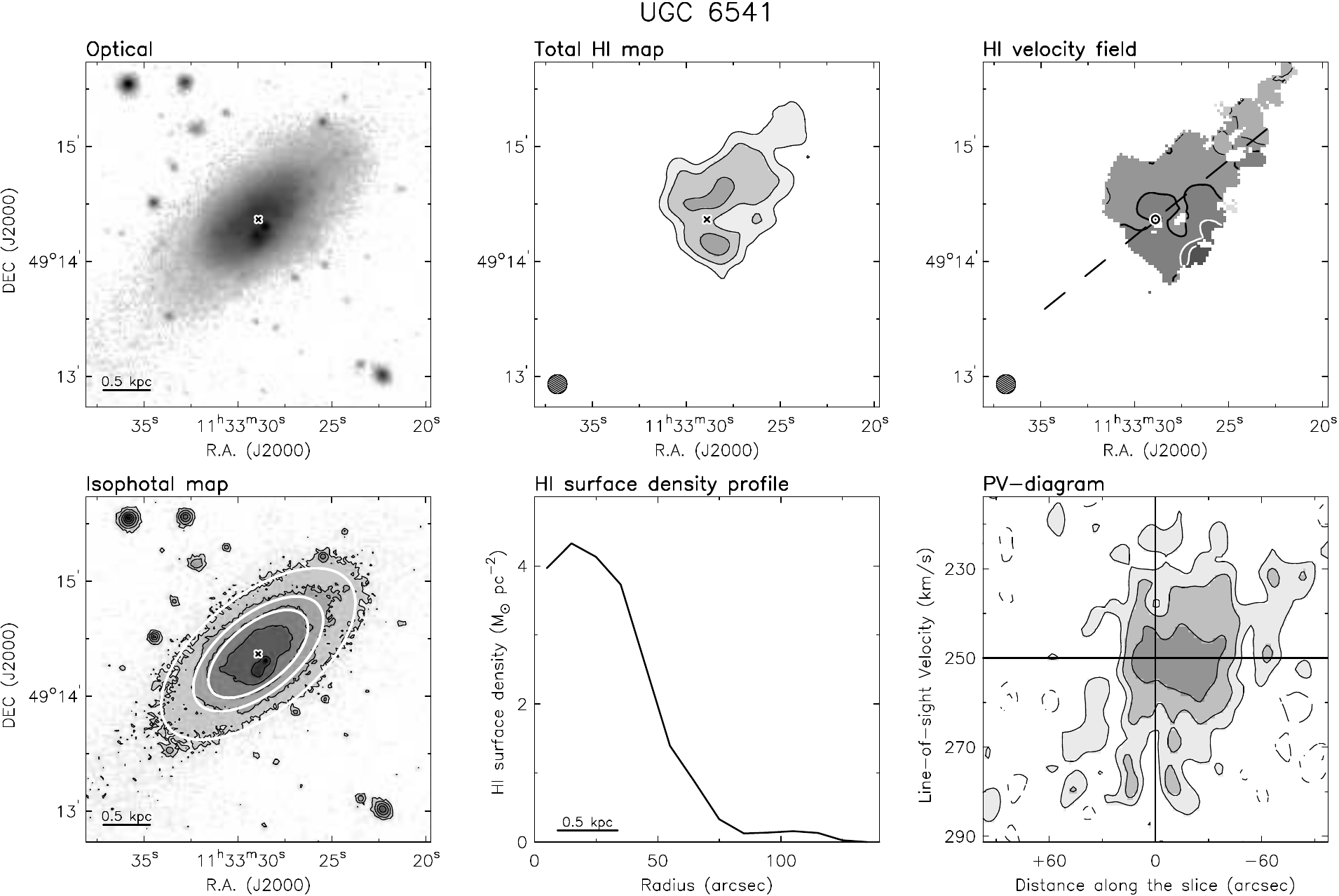}
\caption{\textbf{Contours:} $\mu_{\rm{out}} = 25.5$ R mag arcsec$^{-2}$; $N_{\hi}(3\sigma) = 3.4 \times 10^{20}$ atoms cm$^{-2}$; $V_{\rm l.o.s} = 250 \pm 10$ km s$^{-1}$.}
\vspace{0.3 cm}
\centering
\includegraphics[width=17.5 cm]{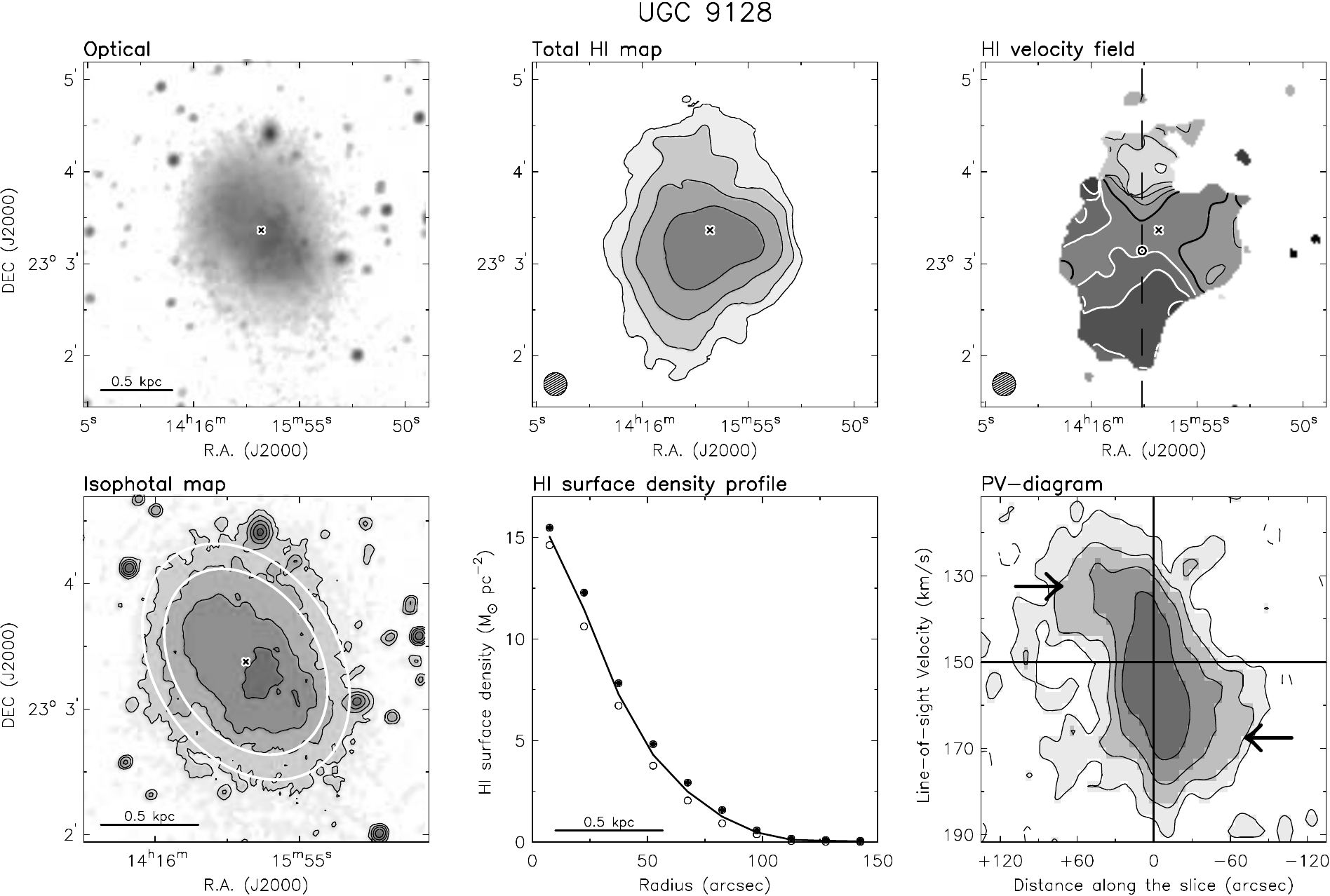}
\caption{\textbf{Contours:} $\mu_{\rm{out}} = 26.5$ V mag arcsec$^{-2}$; $N_{\hi}(3\sigma) = 2.0 \times 10^{20}$ atoms cm$^{-2}$; $V_{\rm l.o.s} = 150 \pm 5$ km s$^{-1}$.}
\end{figure*}

\begin{figure*}
\centering
\includegraphics[width=17.5 cm]{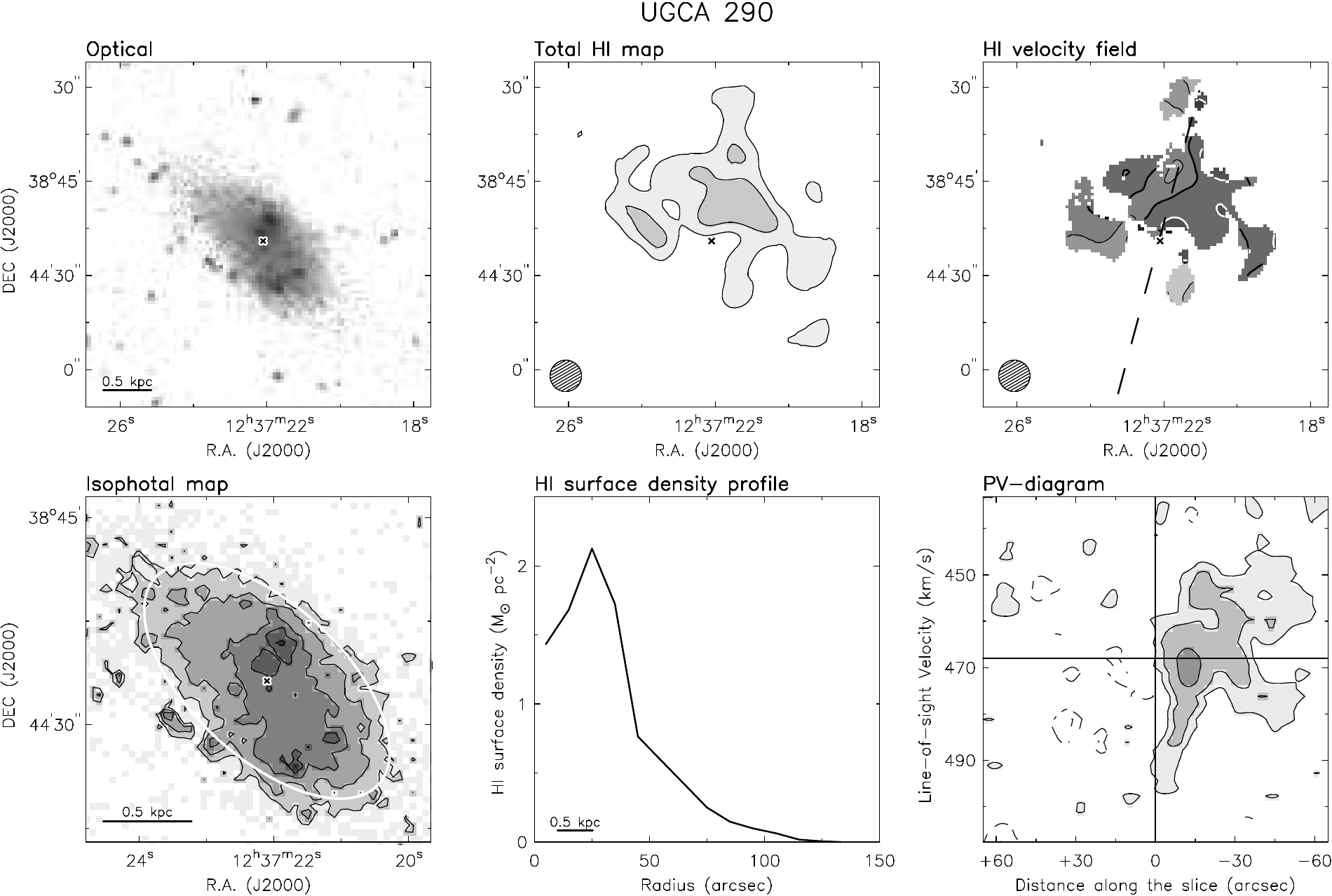}
\caption{\textbf{Contours:} $\mu_{\rm{out}} = 25$ R mag arcsec$^{-2}$; $N_{\hi}(3\sigma) = 3.1 \times 10^{20}$ atoms cm$^{-2}$; $V_{\rm l.o.s} = 468 \pm 5$ km s$^{-1}$.}
\vspace{0.3 cm}
\centering
\includegraphics[width=17.5 cm]{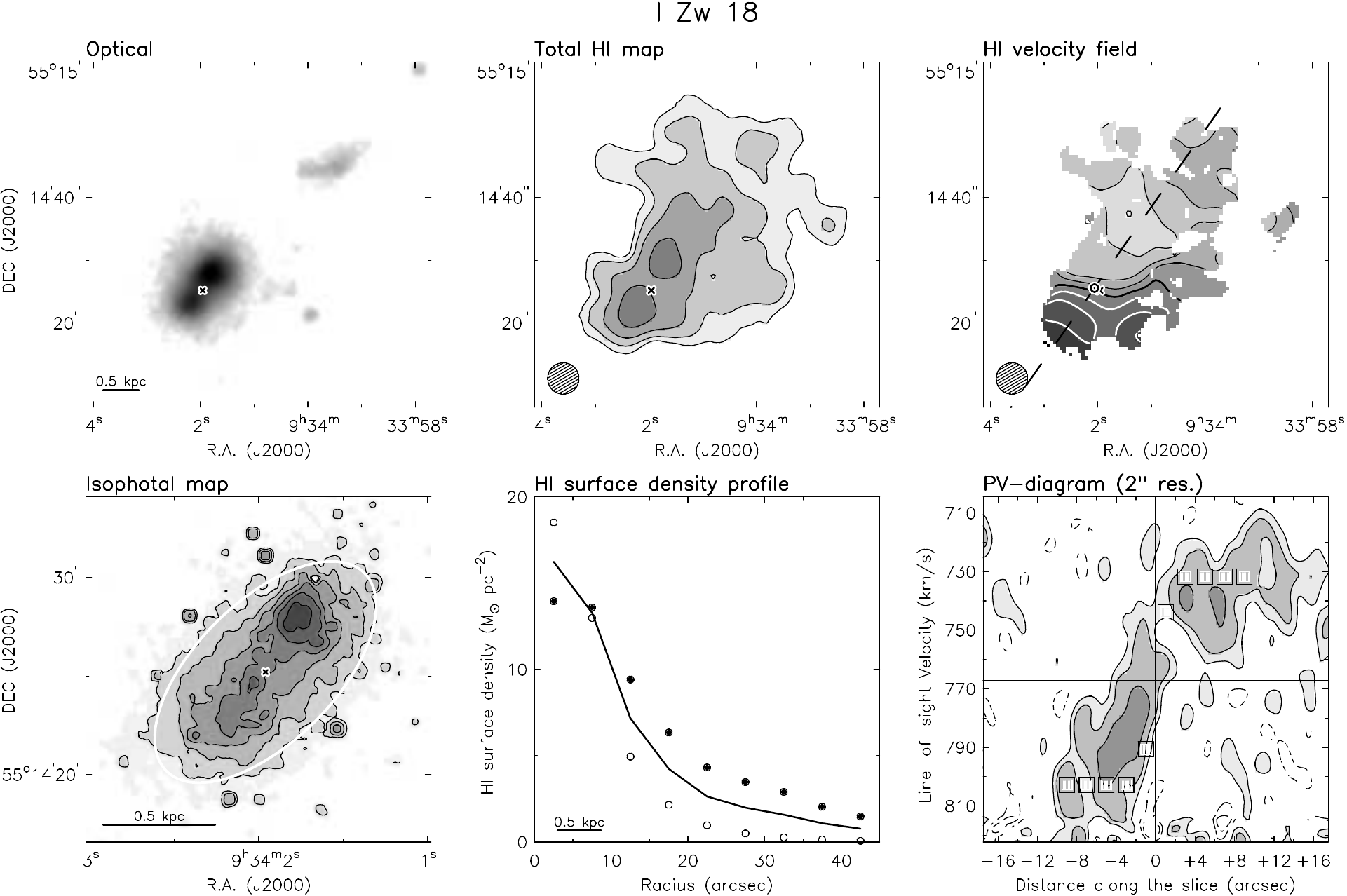}
\caption{\textbf{Contours:} $\mu_{\rm{out}} = 24$ R mag arcsec$^{-2}$; $N_{\hi}(3\sigma) = 6.3 \times 10^{20}$ atoms cm$^{-2}$; $V_{\rm l.o.s} = 767 \pm 10$ km s$^{-1}$.}
\end{figure*}

\begin{figure*}
\centering
\includegraphics[width=17.5 cm]{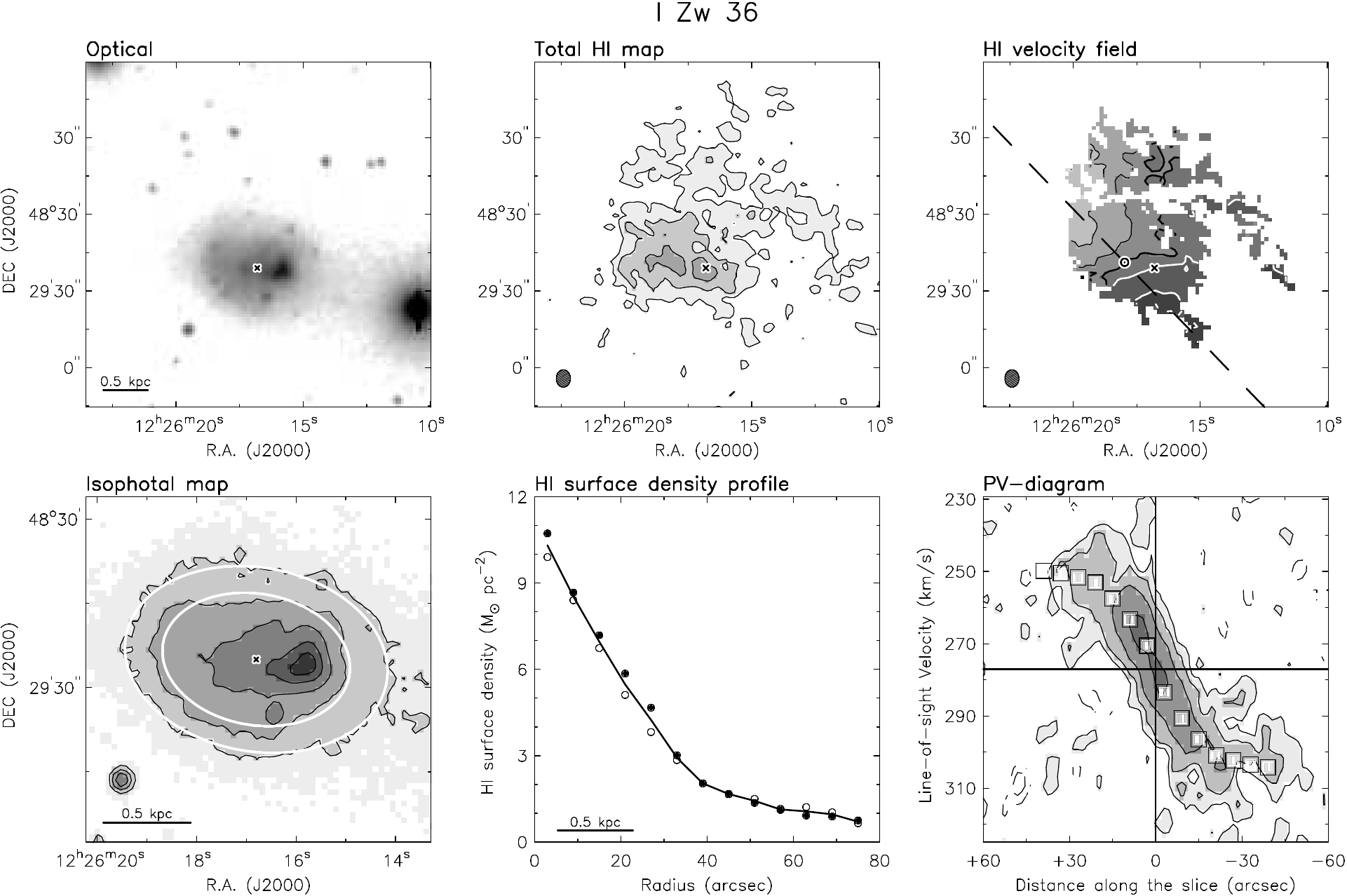}
\caption{\textbf{Contours:} $\mu_{\rm{out}} = 24.5$ R mag arcsec$^{-2}$; $N_{\hi}(3\sigma) = 7.4 \times 10^{20}$ atoms cm$^{-2}$; $V_{\rm l.o.s} = 277 \pm 10$ km s$^{-1}$.}
\vspace{0.3 cm}
\centering
\includegraphics[width=17.5 cm]{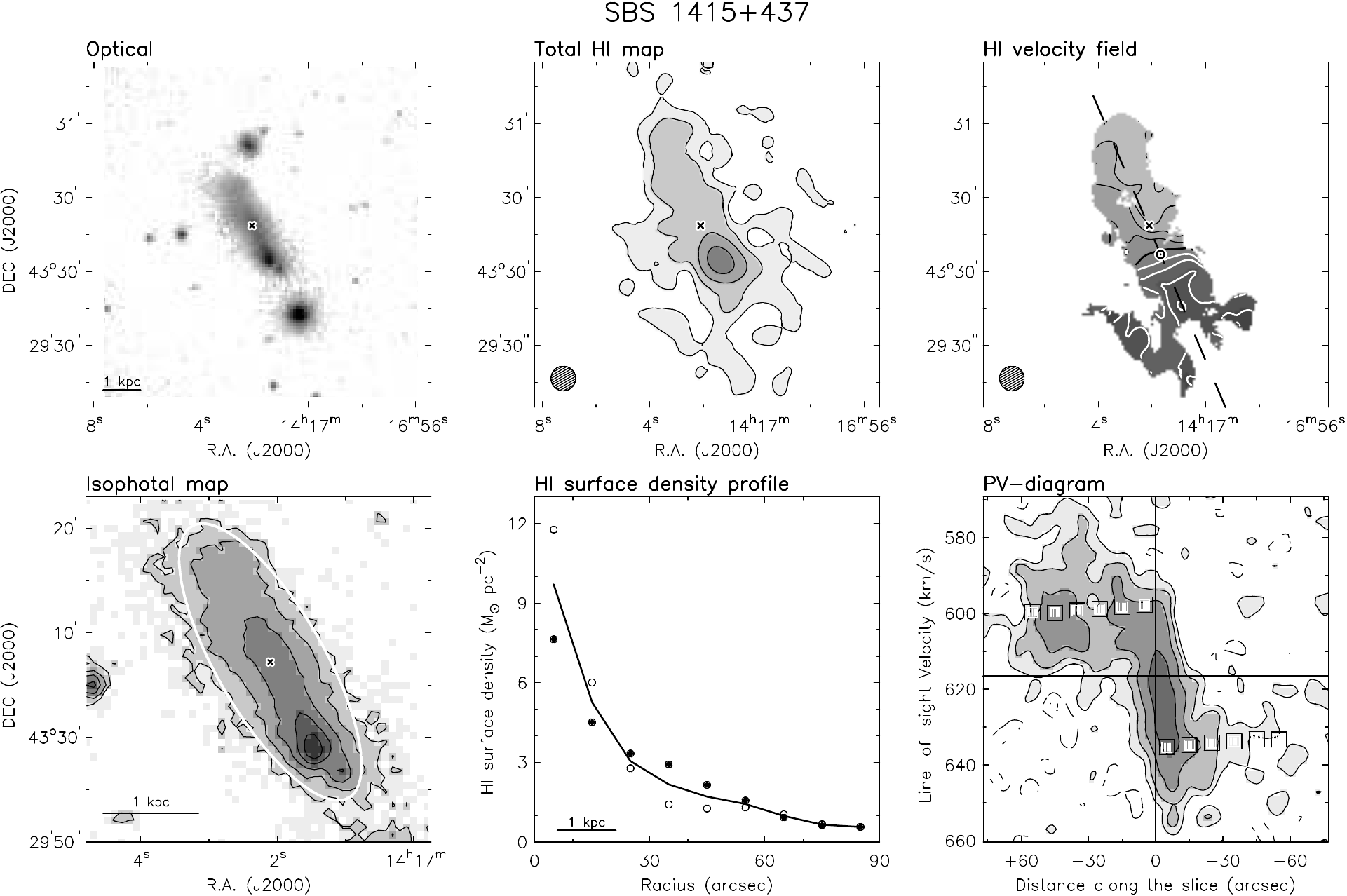}
\caption{\textbf{Contours:} $\mu_{\rm{out}} = 24.5$ R mag arcsec$^{-2}$; $N_{\hi}(3\sigma) = 3.7 \times 10^{20}$ atoms cm$^{-2}$; $V_{\rm l.o.s} = 616 \pm 5$ km s$^{-1}$.}
\end{figure*}

\end{appendix}

\end{document}